\documentclass[aps,prl,twocolumn,reprint,amsmath,amssymb,superscriptaddress,floatfix,footinbib,longbibliography]{revtex4-2}

\usepackage{graphicx}
\usepackage{epstopdf}

\usepackage[T1]{fontenc}
\usepackage[applemac]{inputenc}
\usepackage{lmodern}
\usepackage{amsmath}
\usepackage{amssymb}
\usepackage[english]{babel}
\usepackage{natbib}
\usepackage{ae}
\usepackage{units}
\usepackage{ulem}
\usepackage{amsmath,amssymb,natbib,bm}
\usepackage{psfrag}
\usepackage{subfigure}
\usepackage{amsthm}
\usepackage{booktabs}
\usepackage{slashed}

\usepackage[americaninductors]{circuitikz}
\usepackage{tikz}
\usetikzlibrary{arrows}
\usepackage{ulem}
\usepackage{color}
\usepackage{url}

\usepackage[colorlinks]{hyperref}
\hypersetup{%
        plainpages=true,
        breaklinks=true,
        hypertexnames=false,
        pageanchor=true,
        colorlinks=true,
        linkcolor={blue},
        citecolor={magenta},
        urlcolor={blue},
        anchorcolor={black}
      }

\usepackage{mleftright} 

\newcommand{\figureref}[1]{\mbox{Figure~\ref{#1}}}
\newcommand{\figref}[1]{\mbox{Fig.~\ref{#1}}}

\renewcommand{\eqref}[1]{\mbox{Eq.~(\ref{#1})}}
\newcommand{\figpanel}[2]{Fig.~\hyperref[#1]{\ref*{#1}(#2)}} 
\newcommand{\figpanelsand}[3]{Figs.~\hyperref[#1]{\ref*{#1}(#2)} and \hyperref[#1]{\ref*{#1}(#3)}} 
\newcommand{\figpanels}[3]{Figs.~\hyperref[#1]{\ref*{#1}(#2)--(#3)}} 
\newcommand{\figpanelNoPrefix}[2]{\hyperref[#1]{\ref*{#1}(#2)}} 
\newcommand{\figpanelsNoPrefix}[3]{\hyperref[#1]{\ref*{#1}(#2)-(#3)}} 

\newcommand{\ket}[1]{|#1\rangle}

\newcommand{\be}{\begin{equation}}
\newcommand{\ee}{\end{equation}}
\newcommand{\bea}{\begin{eqnarray}}
\newcommand{\eea}{\end{eqnarray}}

\AtBeginDocument{%
    \newwrite\bibnotes
    \def\bibnotesext{Notes.bib}
    \immediate\openout\bibnotes=\jobname\bibnotesext
    \immediate\write\bibnotes{@CONTROL{REVTEX42Control}}
    \immediate\write\bibnotes{@CONTROL{%
    apsrev42Control,author="08",editor="1",pages="0",title="0",year="1"}}
     \if@filesw
     \immediate\write\@auxout{\string\citation{apsrev42Control}}%
    \fi
}%

\begin{document}

\title{Cavity magnonics and bound states in the continuum with Bragg and anti-Bragg mirrors}

\author{B.-Y.~Wu}
\thanks{These authors contributed equally}
\affiliation{Department of Physics, City University of Hong Kong, Kowloon, Hong Kong SAR, China}

\author{K.-P.~Li}
\thanks{These authors contributed equally}
\affiliation{Center for Joint Quantum Studies and Department of Physics, School of Science, Tianjin University, Tianjin 300350, China}
\affiliation{Tianjin Key Laboratory of Low Dimensional Materials Physics and Preparing Technology, Tianjin University, Tianjin 300350, China}

\author{G.~Chen}
\affiliation{Department of Microtechnology and Nanoscience, Chalmers University of Technology, 41296 Gothenburg, Sweden
}

\author{W.-M.~Zhou}
\affiliation{Department of Physics, City University of Hong Kong, Kowloon, Hong Kong SAR, China}

\author{C.-X.~Run}
\affiliation{Department of Physics, City University of Hong Kong, Kowloon, Hong Kong SAR, China}

\author{K.-M.~Hsieh}
\affiliation{Department of Physics, City University of Hong Kong, Kowloon, Hong Kong SAR, China} 

\author{A.~F.~Kockum}
\affiliation{Department of Microtechnology and Nanoscience, Chalmers University of Technology, 41296 Gothenburg, Sweden
}

\author{Z.-R.~Lin}
\email[e-mail:]{zrlin@mail.sim.ac.cn}
\affiliation{Shanghai Institute of Microsystem and Information Technology, CAS, China}

\author{W.~Nie}
\email[e-mail:]{weinie@tju.edu.cn}
\affiliation{Center for Joint Quantum Studies and Department of Physics, School of Science, Tianjin University, Tianjin 300350, China}
\affiliation{Tianjin Key Laboratory of Low Dimensional Materials Physics and Preparing Technology, Tianjin University, Tianjin 300350, China}

\author{I.-C.~Hoi}
\email[e-mail:]{iochoi@cityu.edu.hk}
\affiliation{Department of Physics, City University of Hong Kong, Kowloon, Hong Kong SAR, China}


\begin{abstract}

A periodic resonant emitter array coupled to a one-dimensional waveguide can act as a mirror. 
A typical example is a Bragg mirror, where emitters are spaced by multiples of half wavelengths and collectively enhance light reflection. 
Two such mirrors form an effective cavity that hosts bound states in the continuum (BICs). 
However, generating and detecting the spatial profiles of such BICs has proven challenging.
Here, we experimentally demonstrate BICs in cavity magnonics using two periodic ferrimagnetic-sphere arrays and a probe sphere in a dual-open-waveguide architecture.
We realize both Bragg and anti-Bragg cavities, whose mirrors have different lattice constants. 
In the Bragg cavity, there are degenerate supermodes formed by mirror spheres. 
We show that a single dark supermode coherently couples to the probe magnon in the cavity region, forming two polaritons whose splitting scales with both the probe-sphere size and the number of mirror spheres.
By contrast, the anti-Bragg cavity has bright and dark supermodes in a bandgap, substantially changing the magnon-cavity interaction.
Moreover, by moving the probe sphere, we perform position-dependent detection of the cavity field, highlighting the role of BICs in these cavities. 
Our flexible experimental setup with a scanning probe opens possibilities to detect other exotic states created by light-matter interaction, to interface with superconducting circuits in hybrid quantum networks, and to study non-Hermitian physics with Bragg and anti-Bragg cavities.

\end{abstract}


\date{\today}

\maketitle


Optical cavities, e.g., Fabry-P\'erot cavities~\cite{Kavokin2017Microcavities}, whispering-gallery-mode microcavities~\cite{Vahala2003,5d16d972da5e4fa7a5b682efbe3d2e51}, and photonic crystal cavities~\cite{painter1999twodimensional, akahane2003highq,Yoshie2004}, play a central role in studying light-matter interaction.
In the pursuit of fundamental and diverse examples of such interactions, a new paradigm for cavities has been developed~\cite{ZhouPhysRevA.78.063827,HetetPhysRevLett2011,Chang_2012,FratiniPhysRevLett2014, Albrecht_2019,Mirhosseini2019,GuimondPhysRevLett2019}, where cavity mirrors consist of single or multiple atoms~\cite{ShenPhysRevLett2005,Chang2007,ZhouPhysRevLett2008,ChangPhysRevA.83.013825,Rui2020,Srakaew2023}.
This development, which builds on the Dicke model~\cite{Dicke1954Coherence, ScullyPRL2009} and the Purcell effect~\cite{Purcell1946SpontaneousEmission}, is promising since periodic arrays of emitters on the subwavelength scale not only exhibit photonic bandgaps~\cite{Tsoi_2008,Brehm2021Waveguide,niu2025}, but also host novel collective phenomena~\cite{RevModPhys.90.031002,Bekenstein2020,Solntsev2021,2023SheremetRevModPhys.95.015002,Gonzlez-Tudela2024}, including superradiance and subradiance~\cite{BettlesPhysRevLett2016,ShahmoonPhysRevLett2017,AsenjoPhysRevX2017,ZhangPhysRevLett2019}.
In one-dimensional (1D) waveguide quantum electrodynamics (QED), photon-mediated long-range interactions enhance collective quantum effects~\cite{vanLoo2013PhotonMediatedIB,GobanPhysRevLett2015,WenPRL2019,Lin2019Scientific,Corzo2019,Zanner2022,Tiranov2023} and dramatically alter light-matter interaction in non-Hermitian cavity QED~\cite{Nie_2023}, which differs from conventional optical cavities.

In waveguide QED systems, placing atoms in a Bragg array with a lattice constant $d_1 = n \lambda_0 / 2$ ($n$ integer, $\lambda_0$ the resonant wavelength) enhances light reflection~\cite{Corzo2016LargeBragg,CoherentPhysRevLett.117.133604,niu2025}, forming an efficient mirror that can be used to construct a cavity~\cite{Chang_2012}.
In the first experiment of atomic cavities, two superconducting artificial atoms separated by $\lambda_0/2$ and a probe atom were coupled to a waveguide to realize cavity QED~\cite{Mirhosseini2019}.
A collective dark state of two mirror atoms acts as an effective cavity mode and the bright state is decoupled from the probe atom placed in the center of the cavity.
More generally, two Bragg arrays separated by $n \lambda_0 / 2$ form a cavity, and the atom-cavity coupling is proportional to the square root of the number of mirror atoms~\cite{Chang_2012}.
An alternative mirror setup also merits attention for anti-Bragg arrays ($d_1 = \lambda_0 / 4 + n \lambda_0 / 2$)~\cite{PoddubnyPhysRevA2022,Wang2026UnidirectionalEP}, which facilitate purely exchange-type interactions between adjacent atoms~\cite{vanLoo2013PhotonMediatedIB} and produce broad photonic bandgaps~\cite{niu2025}.
A cavity formed by two anti-Bragg arrays has been predicted to possess a cavity mode with long lifetime, i.e., the dark state, while the other degenerate superradiant state is important for non-Hermitian cavity QED~\cite{Nie_2023}.

Light-matter interaction in atomic cavities significantly relies on the position of the probe atom~\cite{Nie_2023}, which has non-Hermitian couplings with collective cavity modes. 
A crucial step towards understanding atomic cavity QED is to uncover the role played by localized optical fields of dark modes as bound states in the continuum (BICs)~\cite{Hsu2016,CalaPhysRevLett2019,FacchiPhysRevA2019,Dinc2019exactmarkoviannon,Kang2023}.
A Fabry-P\'erot BIC arises from the destructive interference of resonantly scattered photons from two emitters in the waveguide~\cite{Hsu2016}. 
However, due to the lack of effective detection methods, this interpretation has remained phenomenological, and experimental detection of BICs in waveguide QED systems has not been reported.

\begin{figure*}
\includegraphics[width=\linewidth]{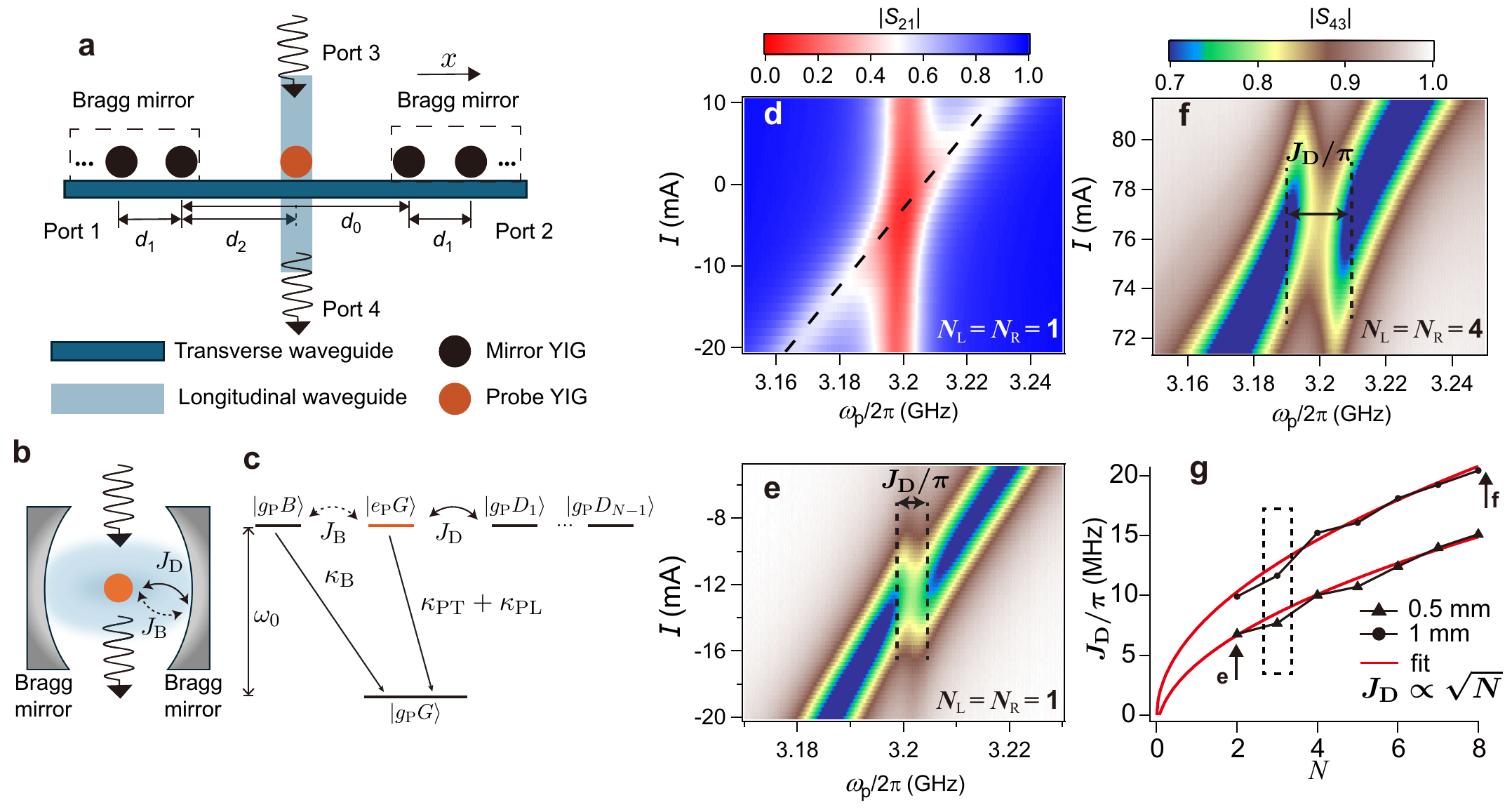}
\caption{Bragg cavity and the interaction with a probe magnon (PM).
(a) Schematic illustration of the dual-waveguide setup with $N$ mirror YIG spheres (diameter \unit[1.2]{mm}, black spheres) and a moveable probe YIG sphere (orange) on the transverse waveguide, and the longitudinal waveguide positioned above the probe sphere.
The drawn spheres are enlarged relative to the resonance wavelength $\lambda_0$ for clarity. Here $d_1 = \lambda_0/2$ denotes the spacing between adjacent YIG spheres, $d_2$ denotes the spacing between the probe sphere and the left mirror sphere, and $d_0$ denotes the cavity region.
A local external magnetic field (not shown) is applied to each sphere to tune its magnon resonance frequency $\omega_0$.
See Supplementary Section 2 for more information.
(b) The setup in (a) is analogous to a cavity QED system. 
(c) The energy-level diagram of $N$ supermodes and the PM in the cavity region between two mirrors.
Here, $J_\mathrm{D}$ ($J_\mathrm{B}$) denotes real (complex) couplings between the PM and the dark (bright) supermode; $\kappa_\mathrm{B}$ is the decay rate of the bright supermode; $\kappa_\mathrm{PT}$ and $\kappa_\mathrm{PL}$ are decay rates of the PM to the transverse and longitudinal waveguides, respectively.
The remaining degenerate $N - 2$ dark supermodes are completely decoupled from the PM.
(d) Amplitude of the transmission coefficient for a weak coherent probe from Port 1 to Port 2, $|S_{21}|$, as a function of the applied current $I$ (which tunes the PM resonance frequency, shown by the dashed curve) and the probe frequency $\omega_\mathrm{p}$.
Here, $N = 2$ with $N_\mathrm{L} = N_\mathrm{R} = 1$; $N_\mathrm{L}$ and $N_\mathrm{R}$ denote the numbers of YIG spheres in the left and right mirrors, respectively.
Two MMs are set to $\omega_0 / 2 \pi \approx \unit[3.2]{GHz}$.
The diameter of the probe YIG sphere is \unit[0.5]{mm}.
(e) Amplitude of the transmission coefficient measured from Port 3 to Port 4, $|S_{43}|$, as a function of $I$ and $\omega_\mathrm{p}$, in the same configuration as in (d).
(f) Measured $|S_{43}|$ as a function of $I$ and $\omega_\mathrm{p}$ for $N = 8$ ($N_\mathrm{L} = N_\mathrm{R} = 4$), with eight MMs at $\omega_0 / 2 \pi \approx \unit[3.2]{GHz}$. The probe-YIG-sphere diameter is \unit[1.0]{mm}.
(g) Coherent coupling $J_\mathrm{D}$ as a function of $N$ for two probe-YIG-sphere diameters (\unit[0.5]{mm} and \unit[1.0]{mm}). 
}
\label{fig:cavity_formed_by_Bragg_mirror}
\end{figure*}

In this article, to overcome experimental challenges for atomic systems, we make use of spin ensembles instead and explore light-matter interaction in dual-waveguide Bragg and anti-Bragg cavities. 
In the limit of weak driving, concepts from atomic systems can be extended to such spin systems, as magnon (spin-wave) excitations behave analogously to atomic or qubit excitations; this has been demonstrated, e.g., when exploring giant-atom physics~\cite{Anton_2018, Kannan2020, Wang2022}.
In the Bragg cavity, the coherent coupling strength exhibits the expected square-root scaling. 
However, this scaling law breaks down in the anti-Bragg cavity.
Through spatial displacement of the probe spin ensemble, we scan standing-wave patterns from magnon-cavity interactions, indicating the presence of BICs.
Detailed differences and advantages of our dual-waveguide system compared to other cavity-based and free-space platforms are summarized in Supplementary Section 5.


\textbf{Photon-mediated dissipative and coherent couplings between YIG spheres.}
The preferred spin ensemble for carrying magnons is an yttrium iron garnet (YIG, Y$_3$Fe$_5$O$_{12}$) sphere, a ferrimagnetic insulator.
The size of a YIG sphere can be much smaller than the wavelength of microwave photons it interacts with, allowing it to be treated as a point-like emitter.
Moreover, a YIG sphere has high spin density~\cite{Gilleo_1958} and its Kittel mode~\cite{KittelPhysRev}, i.e., the uniform-precession magnon mode, enables strong magnetic-dipole coupling to both confined photons in cavities~\cite{Li_2020_JAP, Harder2021,ZARERAMESHTI20221} and propagating photons in waveguides~\cite{Wang2022, Qian2023, wu2024microwaveinterferencespinensemble, run2025realizingondemandalltoallselective} under weak microwave excitation.

In \figpanel{fig:cavity_formed_by_Bragg_mirror}{a}, we show two periodic arrays with $N$ YIG spheres coupled to a coplanar waveguide (the transverse waveguide).
The Kittel mode of each YIG sphere is referred to as a mirror magnon mode (MM).
For a magnon in the $j$th mirror YIG sphere, its annihilation (creation) operator is denoted by $m_{\mathrm{M},j}$ ($m_{\mathrm{M},j}^\dag$); the MMs are numbered from left to right.
Each MM resonates at $\omega_0 / 2\pi = \unit[3.2]{GHz}$, corresponding to a wavelength $\lambda_0 = v / (\omega_0 / 2\pi) = \unit[60]{mm}$, where $v = \unit[1.92 \times 10^8]{m/s}$ is the microwave propagation speed in the transverse waveguide~\cite{wu2024microwaveinterferencespinensemble}.
In the weak-excitation limit and the Born-Markov approximation~\cite{Zhan2022, Wang2022, wu2024microwaveinterferencespinensemble,en1}, the system is described by an effective non-Hermitian Hamiltonian
\begin{equation}
\begin{split}
H_\mathrm{eff} &= \hbar\sum_{j = 1}^N \mleft( \omega_0 - i\frac{\alpha_{\mathrm{M},j}}{2} \mright) m_{\mathrm{M},j}^\dag m_{\mathrm{M},j} \\
&\quad + \hbar\sum_{j, l = 1}^N \mleft( J_{j, l} -i \gamma_{j, l} \mright) m_{\mathrm{M},j}^\dag m_{\mathrm{M},l} .
\end{split}
\label{Effective_H}
\end{equation}
Here, $\hbar$ is the reduced Planck constant, $\omega_0$ is the bare magnon resonance frequency, $\alpha_{\mathrm{M},j}$ is the intrinsic loss rate of the $j$th MM. 
The photon-mediated coupling through the waveguide between the MMs includes both a coherent part ($J$) and a dissipative part ($i\gamma$).
The coherent coupling between MMs $j$ and $l$, known as the exchange interaction, is~\cite{RevModPhys.90.031002,2023SheremetRevModPhys.95.015002,Gonzlez-Tudela2024}
\begin{equation}
J_{j, l} =\frac{1}{2} \sqrt{\kappa_{\mathrm{M}, j} \kappa_{\mathrm{M}, l}} \sin(\phi_{j, l}),
\label{coherent coupling}
\end{equation}
where $\kappa_{\mathrm{M}, j}$ denotes the radiative decay rate of the $j$th mirror YIG sphere into the waveguide, and the phase $\phi_{j, l} = 2 \pi |x_j - x_l| / \lambda_0$ is accumulated as photons travel between mirror YIG spheres located at $x_j$ and $x_l$.
Note that the variation of $J_{j, l}$ with inter-emitter distance is purely sinusoidal. This coupling thus does not attenuate with distance and provides a platform for realizing long-range exchange interactions. 
The waveguide-mediated correlated dissipation between MMs $j$ and $l$ is
\begin{equation}
\gamma_{j, l} =\frac{1}{2} \sqrt{\kappa_{\mathrm{M}, j} \kappa_{\mathrm{M}, l}} \cos(\phi_{j, l}),
\label{corelated decay}
\end{equation}
which is responsible for novel collective effects~\cite{2023SheremetRevModPhys.95.015002,Gonzlez-Tudela2024}.

\begin{figure}
\includegraphics[width=\linewidth]{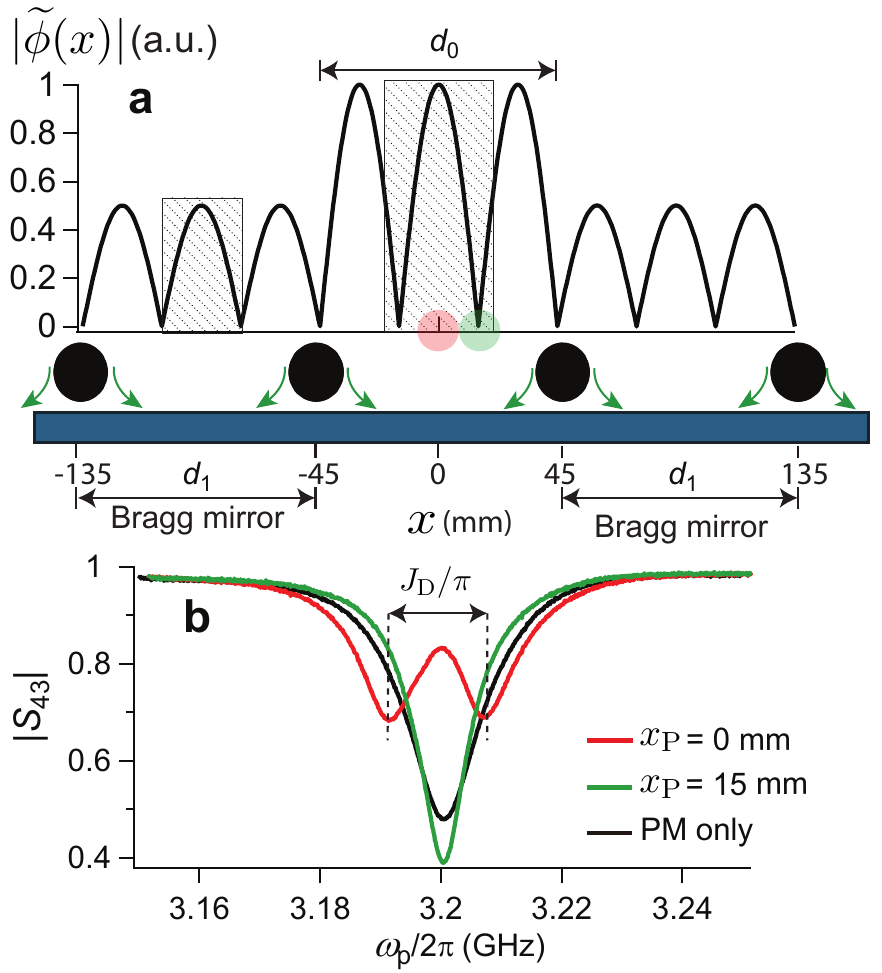}
\caption{
Field distribution of the BIC $\ket{D_1}$ in a Bragg cavity and transmission spectra of the PM.
(a) Rescaled optical field $|\widetilde{\phi}(x)|$ of the BIC in the Bragg cavity with $N = 4$ mirror YIG spheres ($d_1 = 3 \lambda_0 / 2$).
Here, $|\widetilde{\phi}(x)|$ denotes the photonic spatial profile \(|\phi(x)|\) normalized to its maximum value.
The dashed regions indicate the experimental measurements in \figpanel{fig:mapping the anti_Bragg cavity}{c}.
(b) Measured $|S_{43}|$ for a \unit[1.0]{mm} probe YIG sphere [lightly shaded in (a)] at two positions: $x_\mathrm{P} = \unit[0]{mm}$ (red curve, coherent coupling $J_\text{D}$ with the BIC $\ket{D_1}$) and 
$x_\mathrm{P} = \unit[15]{mm}$ [green curve, dissipative coupling $J_B$ with the bright supermode, as shown in \figpanel{fig:cavity_formed_by_Bragg_mirror}{c}].
The black curve is a reference with far-detuned MMs.}
\label{fig:BIC}
\end{figure}
\begin{figure*}
\includegraphics[width=\linewidth]{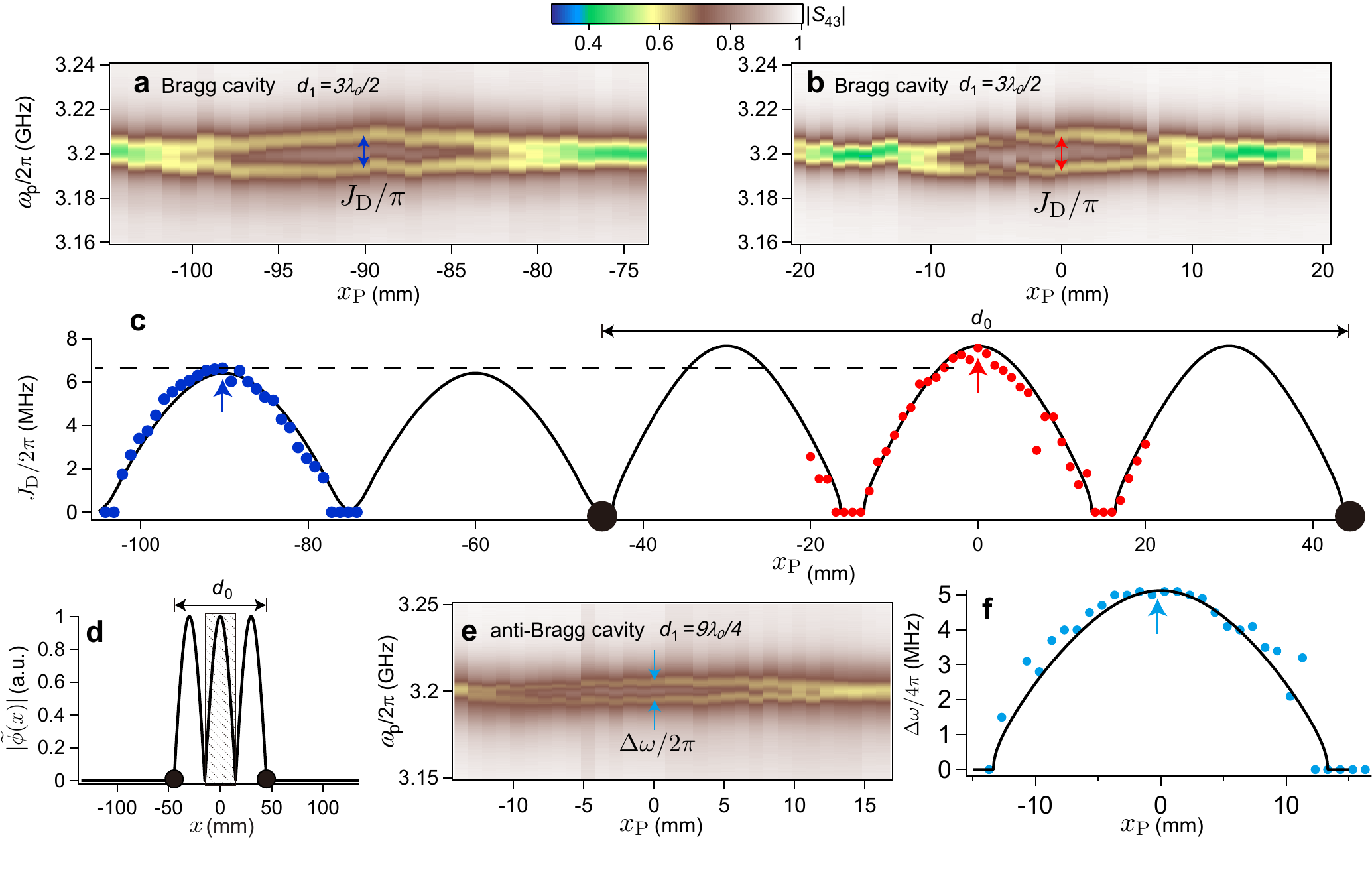}
\caption{
Probe-position field detection in Bragg and anti-Bragg cavities.
(a, b) Measured $|S_{43}|$ spectra with the PM tuned into resonance with the Bragg MMs ($N_\mathrm{L} = N_\mathrm{R} = 2, d_1 = 3 \lambda_0 / 2$) at \unit[3.2]{GHz} [\figpanel{fig:BIC}{a}], recorded for a series of probe YIG sphere (\unit[1.0]{mm}) displacements, $x_\mathrm{P}$.
(c) Coupling strength $J_\mathrm{D}$ between the PM and the Bragg cavity extracted from panels (a, b), as a function of $x_\mathrm{P}$. 
The bold black spheres indicate the locations of mirror spheres.
(d) Rescaled optical field $|\widetilde{\phi}(x)|$ of the single anti-Bragg BIC [zoom in of Supplementary Fig. S2(a)], which is confined within the cavity region $d_0$ ($\unit[-45]{mm} < x < \unit[45]{mm}$) for $N_\mathrm{L} = N_\mathrm{R} = 2$ and $d_1 = 9 \lambda_0 / 4$.
The dashed region indicates where the field is detected in \figpanel{fig:mapping the anti_Bragg cavity}{f}.
(e) Measured $|S_{43}|$ for a \unit[1.2]{mm} probe sphere tuned into resonance with the configuration shown at the bottom of (d) from $x_\mathrm{P} = 0$ to $\pm \unit[15]{mm}$.
(f) Extracted $\Delta \omega / 4 \pi$ as a function of $x_\mathrm{P}$.
The solid black curves in panels (c) and (f) are taken from Supplementary Figs.~S4(e) and S7(c), respectively.
}
\label{fig:mapping the anti_Bragg cavity}
\end{figure*}

\textbf{Magnonic cavity formed by Bragg mirrors.}
\begin{figure*}
\centering
\includegraphics[width=\linewidth]{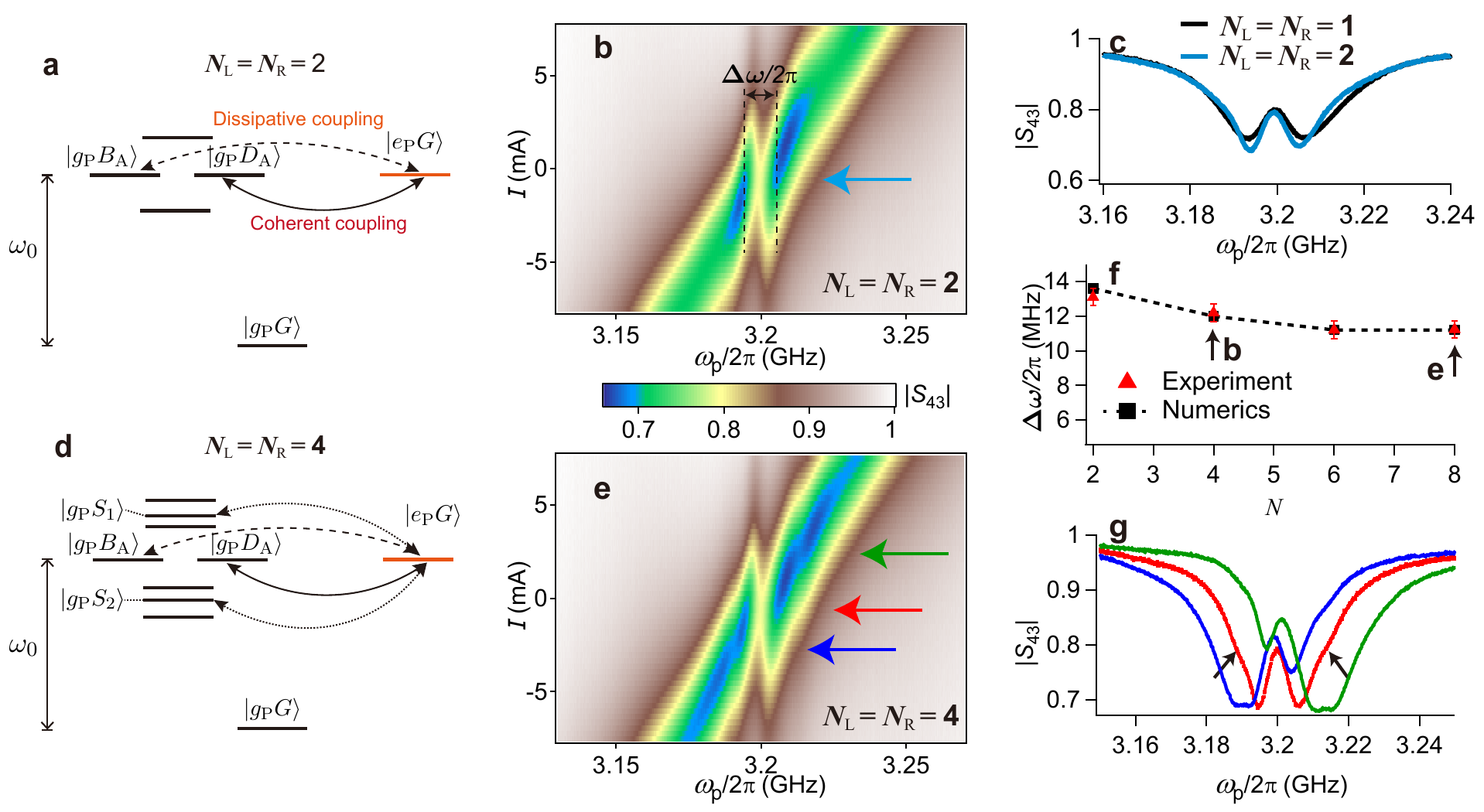}
\caption{
Anti-Bragg cavity QED. 
Here, we consider $d_1 = 3\lambda_0/4$, $d_0 = 3\lambda_0/2$ and use a probe YIG sphere with a diameter of \unit[1.2]{mm}.
(a) The energy-level diagram of four supermodes formed in an anti-Bragg cavity with $N_\mathrm{L} = N_\mathrm{R} = 2$.
The PM $\ket{e_\mathrm{P}}$ couples dissipatively and coherently to two degenerate bright and dark supermodes, respectively.
(b) Measured $|S_{43}|$ as a function of $I$ and $\omega_\mathrm{p}$ with four MMs set to $\omega_0 / 2 \pi \approx \unit[3.2]{GHz}$.
(c) Representative line cut from (b), compared with the resonance case $N_\mathrm{L} = N_\mathrm{R} = 1$.
(d) Same as (a), but for $N_\mathrm{L} = N_\mathrm{R} = 4$, showing the PM interacting with a denser set of collective supermodes.
The dashed and solid arrows indicate both dissipative and coherent couplings.
(e) Measured $|S_{43}|$ as a function of $I$ and $\omega_\mathrm{p}$ with eight MMs set to $\omega_0 / 2 \pi \approx \unit[3.2]{GHz}$.
(f) Extracted frequency gap $\Delta \omega$ as a function of $N$. The error bars on the experimental data reflect the measurement resolution and frequency uncertainty of the two absorption-dip minima.
The numerics are based on Eq.~(S25) in Supplementary Section 1 using parameters: $\kappa_\mathrm{PT} / 2 \pi = \unit[8]{MHz}$, $\kappa_\mathrm{PL} / 2 \pi = \unit[10]{MHz}$, $\kappa_\mathrm{M} / 2 \pi = \unit[10]{MHz}$, $\alpha_\mathrm{M} / 2 \pi = \unit[6.7]{MHz}$, and $\alpha_\mathrm{P} / 2 \pi = \unit[13.1]{MHz}$.
(g) Representative line cuts from (e), indicated by the corresponding colored arrows.
The black arrows indicate additional dips that are barely visible, which are more prominent in the lossless case in Supplementary Fig.~S9.
}
\label{fig:anti_bragg_cavity}
\end{figure*}
%
In a Bragg mirror, neighboring YIG spheres are separated by $d_1 = n \lambda_0 / 2$; we use $d_1 = \lambda_0 / 2$ for the measurements in \figref{fig:cavity_formed_by_Bragg_mirror} and $d_1 = 3 \lambda_0 / 2$ for those in Figs.~\ref{fig:BIC} and \ref{fig:mapping the anti_Bragg cavity}.
The coherent coupling between all MMs is zero, while the strength of dissipative coupling $|\gamma_{j, l}|$ is maximized.
The Bragg mirror can enhance light reflection~\cite{niu2025}.
We characterize the Bragg mirror through waveguide transmission spectroscopy from port 1 to port 2 under a weak coherent drive of $\unit[-30]{dBm}$ using a vector network analyzer and demonstrate a broadened Lorentzian line shape in the transmission spectrum $|S_{21}|$ due to constructive interference~\cite{Chang_2012}. See Supplementary Section 3 for more information.

We use two Bragg mirrors, separated by $d_0=3 \lambda_0 / 2$, to form a cavity with each mirror containing $N/2$ YIG spheres.
The Bragg cavity can be described by the effective Hamiltonian in Eq.~(\ref{Effective_H}).
We diagonalize the cavity Hamiltonian in the single-excitation subspace and find multiple collective states, including a single superradiant state and $N-1$ dark states. 
For $N=2$, a dark state is shown to play the role of an effective cavity mode~\cite{Mirhosseini2019}.
However, the localized optical field of the cavity is not characterized, because the photonic degrees of freedom are traced out while deriving the effective Hamiltonian.
To study optical fields in the waveguide, we consider a full Hamiltonian of the system~\cite{zhang2025}, including Hamiltonians of two magnonic mirrors, the transverse waveguide, and their interactions.
A general form of cavity supermodes can be expressed as~\cite{CalaPhysRevLett2019,Dinc2019exactmarkoviannon,FacchiPhysRevA2019} 
\begin{equation}
\ket{\Psi} = \mleft(\sum_{\alpha = \mathrm{F,B}} \int dx \, \phi_\alpha(x) a_\alpha^\dag(x) + \sum_{j = 1}^N c_j m_{\mathrm{M}, j}^\dag\mright) \ket{\mathrm{vac}, G} ,
\label{cavity supermodes}
\end{equation}
where $\phi_\alpha(x)$ ($\alpha = \mathrm{F,B}$) and $a_\alpha^\dag(x)$ are the spatial amplitude and creation operator of a forward- or backward-propagating photon at position $x$ of the transverse waveguide, $c_j$ denotes the excitation amplitude of the $j$th MM, $\ket{\mathrm{vac}}$ represents the vacuum state of the waveguide, and $\ket{G}$ is the reference ground state of the supermode.
Substituting \eqref{cavity supermodes} into the Schr\"odinger equation of the full system, we can derive the photonic components of supermodes.
See Supplementary Section 1 for details.
Magnonic components of supermodes are obtained from the effective cavity Hamiltonian.
Here, $N-1$ dark supermodes are BICs, because their photonic components vanish outside the Bragg mirrors.
Particularly, one dark supermode has a nonzero photonic component in both cavity and mirror regions [Supplementary Fig.~S4(g)], and its magnonic part is $\ket{D_1} = (1/\sqrt{N})\sum_{j=1}^{N/2} (-1)^{j-1} \big( m_{\mathrm{M}, j}^\dagger + m_{\mathrm{M}, N+1-j}^\dagger\big) \ket{G}$.
Other dark supermodes have localized fields only in the mirror regions. 
However, the superradiant state of MMs has field distribution delocalized outside the Bragg mirrors [Supplementary Fig.~S4(h)] with the form $\ket{B} = (1/\sqrt{N}) \sum^N_{j = 1} (-1)^j m_{\mathrm{M}, j}^\dag \ket{G}$, producing the bright supermode.
For theoretical details, see Supplementary Figure S4.

To detect the cavity, we consider a probe magnon (PM) coupled to the transverse waveguide with radiative decay rate $\kappa_\mathrm{PT}$.
\figureref{fig:cavity_formed_by_Bragg_mirror}(b) shows a schematic of a Bragg cavity with a probe YIG sphere.
The PM-MM interaction is $H_{\mathrm{int}}=\sum_{j=1}^{N}g_j(m_{\mathrm{M},j}^\dagger m_{\mathrm{P}}+\mathrm{H.c.})$, where $m_{\mathrm{P}}$ is the annihilation operator of the PM and the complex coupling strength $g_j$ between the PM and the $j$th MM can be described by Eqs.~(\ref{coherent coupling}) and (\ref{corelated decay}).
We define a coupling vector $\vec{g} = (g_1, g_2, \ldots, g_N)^\mathrm{T}$, which depends on the position of the probe sphere.
Projecting this vector into the cavity supermode basis, we obtain the effective PM-cavity couplings~\cite{Nie_2023}.
As shown in \figpanel{fig:cavity_formed_by_Bragg_mirror}{c}, the PM is coupled to the dark and bright supermodes via $J_\mathrm{D}$ and $J_\mathrm{B}$, respectively.
Here, $J_\mathrm{D}$ is real whereas $J_\mathrm{B}$ is complex, and both depend on the position of the probe sphere.
When the probe sphere is located at $d_2=3\lambda_0/4$ from the nearest mirror sphere on the left, the coupling vector $\vec{g}$ is parallel to the dark supermode $\ket{D_1}$ and orthogonal to the other supermodes.
Hence, the PM only couples coherently to $\ket{D_1}$ with coupling strength $J_\mathrm{D} = \sqrt{N \kappa_\mathrm{M} \kappa_\mathrm{PT}}/2$ (we assume $\kappa_\mathrm{M}=\kappa_{\mathrm{M},j}$).
The PM-cavity interaction exhibits cooperative enhancement as the number of mirror spheres increases, which differs from the atom-cavity coupling enhanced by atoms in an optical cavity~\cite{RaizenPhysRevLett1989,ThompsonPhysRevLett1992}.
In \figpanel{fig:cavity_formed_by_Bragg_mirror}{d}, detection via the transverse waveguide does not reveal the expected energy splitting when the PM is tuned into resonance with two MMs (dashed line).
The reason is that magnonic mirrors strongly reflect incident light and hinder the detection of PM-cavity interaction~\cite{Mirhosseini2019}.

To circumvent this challenge, the longitudinal waveguide in our dual-waveguide setup is implemented vertically on top of the probe YIG sphere and can eliminate the spectral influence of the bright supermode.
By directly exciting the PM at \unit[-30]{dBm} from port 3 to port 4 [\figpanel{fig:cavity_formed_by_Bragg_mirror}{a}], we observe two well-resolved absorption dips in $|S_{43}|$ [\figpanel{fig:cavity_formed_by_Bragg_mirror}{e}].
These two dips indicate two polaritons~\cite{WeisbuchPRL1992,FranciscoScience2021}, i.e., the hybridized (dressed) eigenstates formed by the coherent coupling of the PM with the dark supermode of $N=2$ MMs.
Fitting the data shown in Supplementary Fig.~S14 yields a coherent coupling of $2 J_\mathrm{D} / 2 \pi = \unit[6.8]{MHz}$, a total PM decay rate of $\Gamma_\mathrm{PM} / 2 \pi = \unit[5.5]{MHz}$, and a total cavity decay rate of $\Gamma_\mathrm{C} / 2 \pi = \unit[2.7]{MHz}$.
The cooperativity for this system is thus $C = (2 J_\mathrm{D})^2 / (\Gamma_\mathrm{C} \Gamma_\mathrm{PM}) = 3.1$~\cite{Mirhosseini2019}.
To further increase the PM-waveguide coupling, we can use a larger probe YIG sphere with more spins.
For $N = 8$, the probe sphere with a diameter \unit[1.0]{mm} enhances the polariton splitting [see \figpanel{fig:cavity_formed_by_Bragg_mirror}{f}] with $2 J_\mathrm{D} / 2 \pi = \unit[20.4]{MHz}$, $\Gamma_\mathrm{PM} / 2 \pi = \unit[10.1]{MHz}$, and $\Gamma_\mathrm{C} / 2 \pi = \unit[4.7]{MHz}$, giving $C = 8.8$, which enters the regime of strong coupling ($ J_\mathrm{D} > \{ \Gamma_\mathrm{PM} , \Gamma_\mathrm{C} \}$).
In Bragg cavities, we confirm the $\sqrt{N}$ scaling of $J_\mathrm{D}$ using two probe YIG spheres with different sizes [\figpanel{fig:cavity_formed_by_Bragg_mirror}{g}]. 
The fitted $J_\mathrm{D}/(\sqrt{N}\pi)$ is \unit[4.8]{MHz} for the \unit[0.5]{mm} probe sphere and \unit[7.2]{MHz} for the \unit[1.0]{mm} probe sphere, reflecting the increased $\kappa_{\mathrm{PT}}$ from the PM-waveguide coupling.

So far, the Bragg cavities have the same number of YIG spheres in the left and right mirrors.
If we add an additional mirror sphere to the left or right mirror, i.e., $|N_\mathrm{L}-N_\mathrm{R}|=1$ with odd $N$, the cavity becomes asymmetric.
As a result, the only dark supermode that couples to the PM is modified and its magnonic part is $\ket{D'_1} = \big(\sqrt{N}\,\ket{D_1}_{N-1}\otimes\ket{g_N} - \ket{B}_{N-1}\otimes\ket{g_N}/\sqrt{N} + \sqrt{1-1/N}\,\ket{G}_{N-1}\otimes\ket{e_N}\big)/\sqrt{N+1}$.
Here, $\ket{D_1}_{N-1}$ ($\ket{B}_{N-1}$) is the dark (bright) supermode of the $N-1$ MMs in the symmetrical case, and $\ket{g_N}$ ($\ket{e_N}$) is the ground (excited) state of the added MM.
The coupling vector $\vec{g}$ is no longer parallel to $\ket{D'_1}$, and the PM coherently couples to the bright supermode.
The coherent coupling strength reduces to $J_\mathrm{D}=\sqrt{N\kappa_\mathrm{M}\kappa_\mathrm{PT}(1-1/N^2)}/2$, see Supplementary S1D.
Accordingly, $J_\mathrm{D}$ falls below the $\sqrt{N}$ scaling in the $N=3$ case [dashed box in \figpanel{fig:cavity_formed_by_Bragg_mirror}{g}], and this deviation diminishes as $N$ increases.

\textbf{Magnonic cavity formed by anti-Bragg mirrors.}
In an anti-Bragg mirror, adjacent mirror YIG spheres are separated by $d_1 = 3 \lambda_0 / 4$ for the cavities in \figref{fig:anti_bragg_cavity} and by $d_1 = 9 \lambda_0 / 4$ for the one in \figref{fig:mapping the anti_Bragg cavity}.
The corresponding phase difference is $\phi_{j, j+1} = 3 \pi / 2$, which maximizes the coherent coupling $J_{j, j+1}$ according to \eqref{coherent coupling}.
The anti-Bragg mirror can produce a photonic bandgap~\cite{niu2025}.
We characterize it and confirm a non-Lorentzian line shape in $|S_{21}|$ shown in Supplementary Figs.~S12 and S13.
We consider an anti-Bragg cavity with four MMs ($N=4$), which has four supermodes [\figpanel{fig:anti_bragg_cavity}{a}]. 
Similar to Bragg cavities, the bright and dark supermodes are degenerate.
However, the two remaining supermodes are subradiant and their eigenfrequencies are symmetric about the degenerate frequency.
In contrast to the Bragg cavity, the anti-Bragg cavity has a single dark supermode (BIC).
Intriguingly, the optical field of this BIC is localized in the cavity region $d_0$ and its magnonic part is $\ket{D_\mathrm{A}} = (1/\sqrt{2}) \mleft(m^\dag_{\mathrm{M},N / 2} + m^\dag_{\mathrm{M},N / 2 + 1} \mright) \ket{G}$ ($\mathrm{A}$ denotes anti-Bragg).
When the probe sphere is located at $d_2 = 3 \lambda_0 / 4$, the PM couples coherently to the dark supermode with $J_\mathrm{D} = \sqrt{2 \kappa_\mathrm{M} \kappa_\mathrm{PT}}/2$, which is independent of the cavity size.
Moreover, the PM couples dissipatively to the bright supermode.
The dissipative coupling induces level attraction~\cite{Wang_2020_JAP, Harder2021}.
In \figpanel{fig:anti_bragg_cavity}{b}, we observe an avoided-crossing gap of $\Delta\omega/2\pi=\unit[12.2]{MHz}$.
The gap arises from two polaritons formed by the dark supermode ($\ket{D_\mathrm{A}}$) and the PM, and is influenced by the dissipative coupling to the bright supermode.
We confirm this by comparing with the $N_\mathrm{L}=N_\mathrm{R}=1$ case, where only the coherent coupling is present, and find that the gap is reduced and the two absorption dips become deeper, as shown in \figpanel{fig:anti_bragg_cavity}{c}.

For a general even $N$, the cavity supports multiple supermodes separated by a photonic bandgap~\cite{Nie_2023}, where one bright, one dark, and $N-2$ subradiant supermodes are formed.
For $N_\mathrm{L}=N_\mathrm{R}=4$, the PM interacts with four supermodes, as shown in \figpanel{fig:anti_bragg_cavity}{d}.
In \figpanel{fig:anti_bragg_cavity}{e}, $\Delta\omega$ is reduced to $\unit[11.2]{MHz}$.
Numerical simulations in \figpanel{fig:anti_bragg_cavity}{f} agree well with the experimental results and the gap eventually reduces to a constant value as $N \to \infty$ (Supplementary Fig.~S6).
In addition, \figpanel{fig:anti_bragg_cavity}{e} reveals faint avoided crossings on both sides of the main avoided crossing, together with non-Lorentzian transmission dips away from resonance [line cuts in \figpanel{fig:anti_bragg_cavity}{g}].
These features stem from the coupling between the PM and the subradiant supermodes indicated by dotted arrows in \figpanel{fig:anti_bragg_cavity}{d}, as confirmed by lossless calculations that produce pronounced side anticrossings [Supplementary Fig.~S9].
The faint features are attributed primarily to the large intrinsic losses in our system.


\textbf{Probing the photonic field distribution of Bragg and anti-Bragg BICs.}
In the discussion above, we keep the probe sphere fixed for simplicity.
In our magnonic system, we can take advantage of the mobility of the probe sphere attached to the longitudinal waveguide and measure the spatial profiles of BICs along the transverse waveguide. 
In \figpanel{fig:BIC}{a}, we consider a Bragg cavity with $N_\mathrm{L} = N_\mathrm{R} = 2$, $d_0 = 3 \lambda_0 / 2$, and $d_1 = 3 \lambda_0 / 2$.
When the MMs are resonant at \unit[3.2]{GHz}, this configuration supports three Bragg BICs (dark supermodes).
The rescaled localized field $|\widetilde{\phi}(x)|$ of the Bragg BIC associated with $\ket{D_1}$ in \figpanel{fig:BIC}{a} has three peaks (antinodes) in the cavity region $d_0$, while its peaks are relatively weaker in the mirror region.
The polariton splitting produced by PM-cavity coupling is proportional to the localized field.
When the probe sphere is placed at $x_\mathrm{P} = \unit[0]{mm}$, a clear polariton splitting with $2 J_\mathrm{D} / 2 \pi = \unit[14.6]{MHz}$ is observed in \figpanel{fig:BIC}{b}.
At $x_\mathrm{P} = \unit[15]{mm}$, the splitting disappears because the BIC field is zero.
Interestingly, we observe a narrower linewidth than in the bare PM response [black curve in \figpanel{fig:BIC}{b}]. 
The reason is that the PM only couples dissipatively to the bright supermode with $J_\mathrm{B}=i \sqrt{N \kappa_\mathrm{M} \kappa_\mathrm{PT}}/2$ [Supplementary Fig.~S4(d), star marker on the dashed curve], forming two degenerate hybridized modes. 
One hybridized mode dominated by the PM has low dissipation via destructive interference and leads to a linewidth-reduced transmission spectrum.

To better understand Bragg BICs, we move the probe sphere in the cavity and the mirror regions [see \figpanelsand{fig:BIC}{a}{b}].
In \figpanel{fig:mapping the anti_Bragg cavity}{c}, 
the extracted $J_\mathrm{D}$ decreases as the probe sphere is moved away from the center toward $x_\mathrm{P} = \pm \unit[15]{mm}$, and then increases again.
This directly maps the standing wave of the BIC $\ket{D_1}$ in the cavity region.
In the mirror region, the maximum coupling at $x_\mathrm{P} = \unit[-90]{mm}$ is smaller than that at the cavity center, and the ratio between the maximum couplings in the two regions differs from the corresponding ratio of the maximum $|\widetilde{\phi}(x)|$ of the BIC $\ket{D_1}$.
This is because four supermodes ($\ket{D_1},\ket{D_2},\ket{D_3}$, $\ket{B}$) interact with the PM in the mirror region [Supplementary Figs.~S4(a-d)], thereby modifying the coherent coupling strength.
Theoretical results based on the splitting of the calculated eigenvalues [black curve in \figpanel{fig:mapping the anti_Bragg cavity}{c}] show good agreement with the experimental data.

For the single anti-Bragg BIC, the optical field is confined only within the cavity region, as shown in \figpanel{fig:mapping the anti_Bragg cavity}{d}, where the anti-Bragg cavity has four mirror spheres with $d_1 = 9 \lambda_0 / 4$.
At $x = \pm\unit[15]{mm}$, the field strength of the anti-Bragg BIC is zero.
As the probe sphere is moved from $x_\mathrm{P} = \unit[0]{mm}$ to $\pm\unit[15]{mm}$, the avoided-crossing gap decreases to zero, as shown in \figpanel{fig:mapping the anti_Bragg cavity}{e}.
At $x_\mathrm{P} = \pm\unit[15]{mm}$, the PM does not couple to the dark and bright supermodes, but couples dissipatively and coherently to the nondegenerate subradiant supermodes [see \figpanel{fig:anti_bragg_cavity}{a}].
Therefore, we also observe a linewidth-reduced spectrum [see Supplementary Fig.~S17(a)].
The extracted values of $\Delta \omega$ in \figpanel{fig:mapping the anti_Bragg cavity}{f} agree well with theory, confirming the field distribution in \figpanel{fig:mapping the anti_Bragg cavity}{d}.


\textbf{Conclusion and outlook.}
Our experiment demonstrates that mirrors made by periodic emitter arrays in a waveguide create novel BICs, and opens a way to understand how optical cavities form at a microscopic level. 
In the symmetric Bragg cavity, a dark supermode (a BIC) coherently couples to the central probe magnon, and their coupling follows the $\sqrt{N}$ relation, showing the collective enhancement of light-matter interaction.
In the asymmetric Bragg cavity, the probe magnon couples to a modified dark supermode with a slightly reduced interaction strength.
The anti-Bragg cavity hosts degenerate dark and bright supermodes that couple coherently and dissipatively to the probe magnon, reducing the avoided-crossing gap. 
In fact, Bragg and anti-Bragg cavities both have a single BIC in the cavity region.
By moving the probe magnon, we characterize distinct properties of BICs in Bragg and anti-Bragg cavities, which can be attributed to light reflection of mirrors~\cite{niu2025}.
Crucially, the role played by superradiant supermodes coexisting with BICs is pinpointed in non-Hermitian cavity QED, distinguishing our model from conventional optical cavities~\cite{Kavokin2017Microcavities, Vahala2003,5d16d972da5e4fa7a5b682efbe3d2e51,painter1999twodimensional, akahane2003highq,Yoshie2004}.

Looking to the future, our scheme can be applied to other waveguide-coupled emitters, e.g., superconducting circuits~\cite{Mirhosseini2019,niu2025,vanLoo2013PhotonMediatedIB,Zanner2022,WenPRL2019}, quantum dots~\cite{Tiranov2023}, and natural atoms~\cite{GobanPhysRevLett2015,Corzo2016LargeBragg,CoherentPhysRevLett.117.133604}.
The scanning-probe framework enables the investigation of various exotic states, such as emitter-photon bound states~\cite{Liu2017}. 
Furthermore, as magnon excitations in YIG spheres have been achieved at millikelvin temperatures~\cite{Tabuchi2015}, extending the capabilities of magnon arrays to interface with superconducting qubits offers a promising route toward scalable quantum networks~\cite{LiPRXQuantum} via the dual-waveguide setup~\cite{29th-t4ks}.


\acknowledgments{\textbf{Acknowledgments.}}
I.-C. H.~acknowledges financial support from City University of Hong Kong through the start-up project 9610569, from the Research Grants
Council of Hong Kong (Grant No. 11312322 and Grant No. 11307324), from Guangdong Provincial Quantum Science Strategic
Initiative (No. GDZX2303005, No. GDZX2203001, and No. GDZX2403001), and technical support from JX Quantum (www.jx-quantum.com). 
W.~N.~is supported by the National Natural Science Foundation of China (Grant No. 92476115).
A. F. K. acknowledges support from the Swedish Foundation for Strategic Research (Grants No. FFL21-0279 and No. FUS21-0063), 
the Horizon Europe programme HORIZON-CL4-2022-QUANTUM01-SGA via Project No. 101113946 OpenSuperQPlus100, and from the Knut and Alice Wallenberg Foundation through the Wallenberg Centre for Quantum Technology (WACQT). 

\normalem
\bibliography{References_main}

\begin{thebibliography}{74}
\bibitem [{en1()}]{en1}%
  \BibitemOpen
  \href@noop {} {}\bibinfo {note} {Radiative decay rates of magnon modes in YIG spheres at resonance frequency \unit[3.2]{GHz} vary slightly, centered around $\kappa_\mathrm{M}/2\pi\approx\unit[9]{MHz}$. Thus, non-Markovian effects are negligible in our work, because the magnon relaxation time ($1/\kappa\approx\unit[17.7]{ns}$) is longer than the photon propagation time along the transverse waveguide ($L/v\approx\unit[2.3]{ns}$).}\BibitemShut {Stop}%
\end{thebibliography}

\end{document}


\title{Supplementary Material for \\ ``Cavity magnonics and bound states in the continuum with Bragg and anti-Bragg mirrors"}

\author{B.-Y.~Wu}
\thanks{These authors contributed equally}
\affiliation{Department of Physics, City University of Hong Kong, Kowloon, Hong Kong SAR, China}

\author{K.-P.~Li}
\thanks{These authors contributed equally}
\affiliation{Center for Joint Quantum Studies and Department of Physics, School of Science, Tianjin University, Tianjin 300350, China}
\affiliation{Tianjin Key Laboratory of Low Dimensional Materials Physics and Preparing Technology, Tianjin University, Tianjin 300350, China}

\author{G.~Chen}
\affiliation{Department of Microtechnology and Nanoscience, Chalmers University of Technology, 41296 Gothenburg, Sweden
}

\author{W.-M.~Zhou}
\affiliation{Department of Physics, City University of Hong Kong, Kowloon, Hong Kong SAR, China}

\author{K.-M.~Hsieh}
\affiliation{Department of Physics, City University of Hong Kong, Kowloon, Hong Kong SAR, China} 

\author{C.-X.~Run}
\affiliation{Department of Physics, City University of Hong Kong, Kowloon, Hong Kong SAR, China}

\author{A.~F.~Kockum}
\affiliation{Department of Microtechnology and Nanoscience, Chalmers University of Technology, 41296 Gothenburg, Sweden
}

\author{Z.-R.~Lin}
\email[e-mail:]{zrlin@mail.sim.ac.cn}
\affiliation{Shanghai Institute of Microsystem and Information Technology, CAS, China}

\author{W.~Nie}
\email[e-mail:]{weinie@tju.edu.cn}
\affiliation{Center for Joint Quantum Studies and Department of Physics, School of Science, Tianjin University, Tianjin 300350, China}
\affiliation{Tianjin Key Laboratory of Low Dimensional Materials Physics and Preparing Technology, Tianjin University, Tianjin 300350, China}

\author{I.-C.~Hoi}
\email[e-mail:]{iochoi@cityu.edu.hk}
\affiliation{Department of Physics, City University of Hong Kong, Kowloon, Hong Kong SAR, China}
\date{\today}

\maketitle

\tableofcontents

\section{Theoretical framework}

In this section, we present the details of the theoretical framework we use to analyze our experimental results. First, we write down the Hamiltonian describing the full system of yttrium iron garnet (YIG) spheres with two waveguides and derive the master equation for the YIG spheres. Then, we show how scattering theory describes experiments probing the system with a weak coherent field. Finally, we derive the modes of the cavity formed by the YIG spheres and the bound states in the continuum (BICs) that arise, and show how a probe YIG sphere couples to these modes.


\subsection{Model of the magnonic cavity and the dual waveguides}

\begin{figure*}
\includegraphics[width=0.7\linewidth]{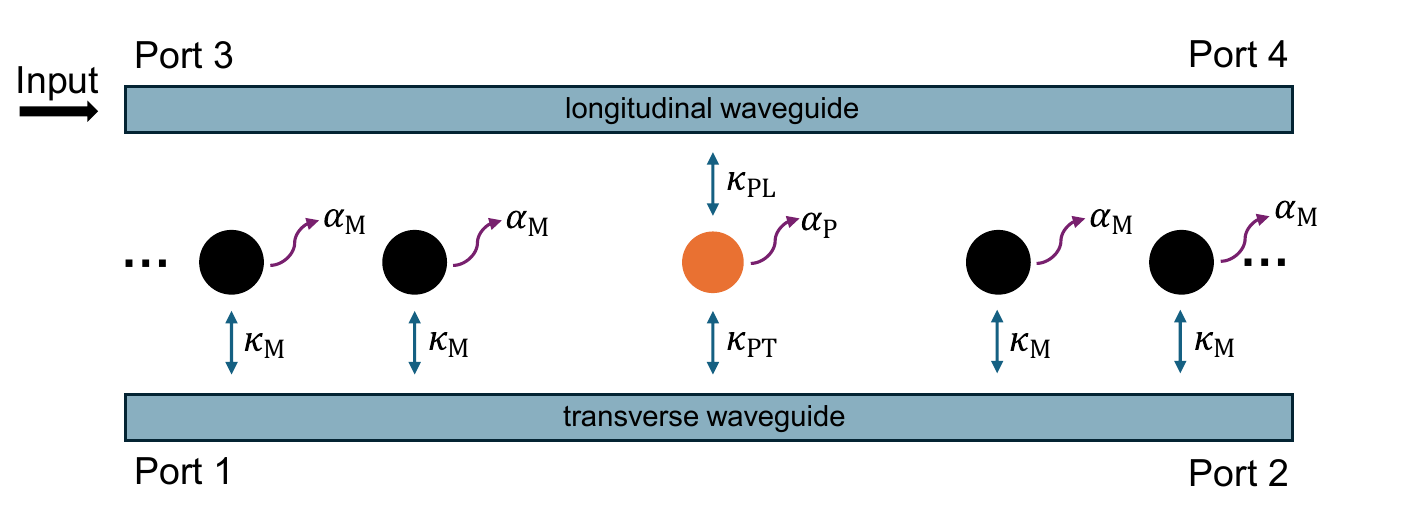}
\caption{Schematic illustration of the system with a dual-waveguide (teal) setup coupling to mirror YIG spheres (black) and a probe YIG sphere (orange). Note that for visual clarity and layout convenience, the longitudinal waveguide is depicted horizontally in this schematic. In this illustration, we assume uniform radiative decay rates $\kappa_\mathrm{M}$ and intrinsic magnon loss rates $\alpha_\mathrm{M}$ for all the mirror YIG spheres. For the probe YIG sphere, $\kappa_{\mathrm{PT}}$ ($\kappa_{\mathrm{PL}}$) represents its radiative decay rate into the transverse (longitudinal) waveguide, and $\alpha_\mathrm{P}$ is its intrinsic loss rate.}
\label{system}
\end{figure*}

In our experiment, we constructed a system consisting of multiple YIG spheres coupled to a dual-waveguide setup, as schematically illustrated in \figref{system}, and observed collective magnon-photon interaction. 
Here, we give the derivation of the master equation for this system, starting from the full system Hamiltonian. 


\subsubsection{Hamiltonian}   

The total Hamiltonian of the system is given by
\begin{equation}
H = H_0 + H_\text{I},
\end{equation}
where $H_0 = H_\mathrm{wg} + H_\mathrm{M}$ is the free Hamiltonian, consisting of a waveguide part $H_\mathrm{wg}$ and a magnon part $H_\mathrm{M}$ for the YIG spheres, and $H_\text{I}$ is the interaction Hamiltonian describing the interaction between photons in the waveguides and magnons in the YIG spheres.
 
The waveguide Hamiltonian can be written as
\begin{equation}
H_{\mathrm{wg}} = \sum_{\alpha = \text{F,B}} \int_0^\infty d\omega\, \hbar \omega \mleft[ a_\alpha^\dag(\omega) a_\alpha(\omega) + b_\alpha^\dag(\omega) b_\alpha(\omega) \mright].
\end{equation}
Here, $a_\alpha(\omega)$ and $b_\alpha(\omega)$ denote annihilation operators for propagating modes in transverse and longitudinal waveguides, respectively; $a_{\alpha}^{\dagger}(\omega)$ and $b_{\alpha}^{\dagger}(\omega)$ are the corresponding creation operators.
The index $\alpha = \text{F}$ (B) denotes forward (backward) propagation, which we take to be from the left to the right (the right to the left) in the transverse waveguide and from the top to the bottom (the bottom to the top) in the longitudinal waveguide.
		
The magnon Hamiltonian is so named due to the collective spin excitations (Kittel modes, KMs) in the YIG spheres, which are referred to as magnon modes. In the low-excitation limit~\cite{Zhan2022, Wang2022, wu2024microwaveinterferencespinensemble}, these collective dynamics can be effectively described by bosonic operators. Accordingly, the magnon Hamiltonian is given by
\begin{equation}
H_\mathrm{M} = \sum_{j=1}^N \hbar \omega_0 m_{\mathrm{M}, j}^\dag m_{\mathrm{M}, j} + \hbar \omega_0 m_\mathrm{P}^\dag m_\mathrm{P} ,
\end{equation}
where $m_{\mathrm{M}, j}^\dag$ ($m_{\mathrm{M}, j}$) is the magnon creation (annihilation) operator associated with the KM of the $j$th mirror YIG sphere in the system and $m_\text{P}^\dag$ ($m_\text{P}$) is the same for the probe YIG sphere.

Finally, the interaction Hamiltonian $H_\mathrm{I}$ takes the following form in the Schr\"odinger picture:
\begin{equation}
\begin{aligned}
H_\mathrm{I} &= i \hbar \sum_{\alpha = \text{F,B}} \sum_{j=1}^{N} \int_0^\infty d\omega \, g_\mathrm{M} \mleft[ a_{\alpha}^\dag(\omega) e^{-i \eta_\alpha \omega x_j / v} m_{\mathrm{M},j} - \mathrm{H.c.} \mright] \\
& \quad + i \hbar \sum_{\alpha = \text{F,B}} \int_0^\infty d\omega \, \mleft[ \mleft( g_\mathrm{PL} b_\alpha^\dag(\omega) e^{-i \eta_\alpha \omega y_\mathrm{P} / v} + g_\mathrm{PT} a_\alpha^\dag(\omega) e^{-i \eta_\alpha \omega x_\mathrm{P} / v} \mright) m_\mathrm{P} - \mathrm{H.c.} \mright] .
\end{aligned}
\label{Hint_Schrodinger}
\end{equation}
Here, $\eta_\alpha$ is a sign factor accounting for the propagation direction, with $\eta_\mathrm{F} = +1$ and $\eta_\mathrm{B} = -1$. The parameter $g_\mathrm{M}$ denotes the coupling strength between the mirror YIG spheres and the transverse waveguide, while $g_\mathrm{PT}$ and $g_\mathrm{PL}$ describe the coupling strengths of the probe YIG sphere to the transverse and longitudinal waveguides, respectively. The coordinates $x_j$ and $x_\mathrm{P}$ represent the positions of the $j$th mirror YIG sphere and the probe YIG sphere, respectively, along the transverse waveguide, while $y_\mathrm{P}$ denotes the position of the probe YIG sphere along the longitudinal waveguide. Note that by setting the spatial origin of the longitudinal waveguide at the probe YIG sphere ($y_\mathrm{P} = 0$), its corresponding phase factor reduces to unity.

Moving to the interaction picture via the unitary transformation $U(t) = \exp[-i H_0 t / \hbar]$, the field and magnon operators acquire free-evolution time dependence, which we denote with a tilde [e.g., $\tilde{a}_\alpha(\omega, t)$]. The interaction Hamiltonian thus becomes $\tilde{H}_\mathrm{I}(t)$, taking the form
\begin{equation}
\begin{aligned}
\tilde{H}_\mathrm{I}(t) &= i \hbar \sum_{\alpha = \text{F,B}}\sum_{j=1}^{N} \int_0^\infty d\omega \, g_\mathrm{M} \mleft[ \tilde{a}_\alpha^\dag(\omega,t) e^{-i \eta_\alpha \omega x_j / v} \tilde{m}_{\mathrm{M}, j}(t) - \mathrm{H.c.} \mright] \\
&\quad + i \hbar \sum_{\alpha = \text{F,B}} \int_0^\infty d\omega \, \mleft[ \mleft( g_\mathrm{PL} \tilde{b}_\alpha^\dag(\omega, t) e^{-i \eta_\alpha \omega y_\mathrm{P} / v} + g_\mathrm{PT} \tilde{a}_\alpha^\dag(\omega, t) e^{-i \eta_\alpha \omega x_\mathrm{P} / v} \mright) \tilde{m}_\mathrm{P}(t) -\mathrm{H.c.} \mright] .
\end{aligned}
\label{Hint}
\end{equation}
%


\subsubsection{Master equation}

In the interaction picture, the density matrix $\tilde{\rho}_\mathrm{tot}(t)$ of the total system satisfies the von Neumann equation
\begin{equation}
\frac{d\tilde{\rho}_\mathrm{tot}(t)}{dt} = -\frac{i}{\hbar} \mleft[ \tilde{H}_\mathrm{I}(t), \tilde{\rho}_\mathrm{tot}(t) \mright] .
\label{von}
\end{equation}
Integrating this equation over time, we obtain the formal solution
\begin{equation}
\tilde{\rho}_\mathrm{tot}(t) = \tilde{\rho}_\mathrm{tot}(0) - \frac{i}{\hbar} \int_0^t d\tau \, \mleft[ \tilde{H}_\mathrm{I}(\tau), \tilde{\rho}_\mathrm{tot}(\tau) \mright] .
\label{eq:FormalSolution}
\end{equation}
To obtain the evolution of the system's reduced density matrix for the YIG spheres, we trace out the continuum of waveguide modes. Substituting the formal solution from \eqref{eq:FormalSolution} into the right side of \eqref{von} and performing this partial trace, we obtain
\begin{equation}
	\frac{d\tilde{\rho}(t)}{dt} = -\frac{i}{\hbar}\, \mathrm{tr}_E \mleft( \mleft[ \tilde{H}_\mathrm{I}(t), \tilde{\rho}_\mathrm{tot}(0) \mright] \mright)
	- \frac{1}{\hbar^2} \int_0^t d\tau \, \mathrm{tr}_E \mleft( \mleft[ \tilde{H}_\mathrm{I}(t), \mleft[ \tilde{H}_\mathrm{I}(\tau), \tilde{\rho}_\mathrm{tot}(\tau) \mright] \mright] \mright) ,
	\label{rhofinal}
\end{equation} 
where $\tilde{\rho}(t) = \mathrm{tr}_E[\tilde{\rho}_\mathrm{tot}(t)]$ is the reduced density matrix of the system in the interaction picture and the subscript E denotes the environment formed by the waveguides. Although the waveguides are in a room-temperature thermal state, the system's bilinear Hamiltonian ensures that coherent signals and zero-mean thermal fluctuations evolve independently. Therefore, when evaluating the coherent transmission spectra, the thermal background serves as an effective vacuum state. This allows the first term on the right-hand side to vanish analytically [$\mathrm{tr}_E \mleft( \mleft[ \tilde{H}_I(t), \tilde{\rho}_\mathrm{tot}(0) \mright] \mright) = 0$].

Applying the standard Born--Markov approximation, we substitute the interaction Hamiltonian \eqref{Hint} into \eqref{rhofinal} above. Tracing out the waveguide modes and transforming back to the Schr\"odinger picture, we obtain the Markovian master equation for the system's (YIG spheres') density matrix $\rho(t)$:
\begin{equation}
\frac{d\rho}{dt} = -\frac{i}{\hbar} \mleft[ H_\mathrm{M} + H_\mathrm{coh}, \rho \mright] + \mathcal{D}[\rho].
\label{eq:final_master}
\end{equation}
In this equation, $H_\mathrm{coh}$ accounts for the waveguide-mediated coherent exchange coupling between the YIG spheres. It takes the form
\begin{equation}
\label{coherent_dis_1}
H_\mathrm{coh} = \hbar \sum_{j, l = 1}^N  J_{j, l}  m_{\mathrm{M}, j}^\dag m_{\mathrm{M}, l} + \hbar \sum_{j=1}^N J_{j, \mathrm{P}} \mleft( m_\mathrm{P}^\dag m_{\mathrm{M}, j} + \text{H.c.} \mright) ,
\end{equation}
with the coherent coupling strengths given by
\begin{equation}
\begin{aligned}
J_{j, l} &= \frac{1}{2} \sqrt{\kappa_{\mathrm{M}, j} \kappa_{\mathrm{M}, l}} \, \sin \mleft( \frac{2\pi |x_j-x_l|}{\lambda_0} \mright) , \\
J_{j, \mathrm{P}} &= \frac{1}{2} \sqrt{\kappa_{\mathrm{M}, j} \kappa_\mathrm{PT}} \, \sin \mleft( \frac{2\pi |x_j-x_\mathrm{P}|}{\lambda_0} \mright) .
\end{aligned}
\end{equation}
Here, $\lambda_0 = v / (\omega_0 / 2\pi)$ is the resonant wavelength with $v$ being the propagation velocity in the waveguide, and we have introduced the waveguide-induced radiative decay rates $\kappa_{\mathrm{M}, j} = 4\pi g_{\mathrm{M}, j}^2$, $\kappa_\mathrm{PL} = 4\pi g_\mathrm{PL}^2$, and $\kappa_\mathrm{PT} = 4\pi g_\mathrm{PT}^2$. These rates correspond to the decay of the $j$th mirror YIG sphere into the transverse waveguide, and the decay of the probe YIG sphere into the longitudinal and transverse waveguides, respectively.
Finally, $\mathcal{D}[\rho]$ is the Lindblad superoperator describing dissipation into the dual waveguides, which takes the form
\begin{equation}
\label{coherent_dis_2}
\begin{split}
\mathcal{D}[\rho] =& \sum_{j, l = 1}^N \gamma_{j, l} \mleft( 2 m_{\mathrm{M}, j} \rho m_{\mathrm{M}, l}^\dag - m_{\mathrm{M}, j}^\dag m_{\mathrm{M}, l} \rho - \rho m_{\mathrm{M}, j}^\dag m_{\mathrm{M}, l} \mright) \\
&+ \gamma_\mathrm{P} \mleft( 2 m_\mathrm{P} \rho m_\mathrm{P}^\dag - m_\mathrm{P}^\dag m_\mathrm{P} \rho - \rho m_\mathrm{P}^\dag m_\mathrm{P} \mright) \\
&+ \sum_{j = 1}^N \gamma_{j, \mathrm{P}} \mleft( 2 m_\mathrm{P} \rho m_{\mathrm{M}, j}^\dag - m_\mathrm{P}^\dag m_{\mathrm{M}, j} \rho - \rho m_\mathrm{P}^\dag m_{\mathrm{M}, j} \mright) ,
\end{split}
\end{equation}
where the waveguide-induced dissipation rates are given by
\begin{equation}
\label{dissipation_rates}
\begin{aligned}
\gamma_{j, l} &= \frac{1}{2} \sqrt{\kappa_{\mathrm{M}, j} \kappa_{\mathrm{M}, l}} \, \cos \mleft( \frac{2\pi |x_j-x_l|}{\lambda_0} \mright) , \\
\gamma_{j, \mathrm{P}} &= \frac{1}{2} \sqrt{\kappa_{\mathrm{M}, j} \kappa_\mathrm{PT}} \, \cos \mleft( \frac{2\pi |x_j-x_\mathrm{P}|}{\lambda_0} \mright) , \\
\gamma_\mathrm{P} &= \frac{1}{2} \mleft( \kappa_\mathrm{PL} + \kappa_\mathrm{PT} \mright) .
\end{aligned}
\end{equation}

In addition to the waveguide-induced dissipation derived above, each YIG sphere also experiences intrinsic magnon loss due to coupling to free space, with a rate $\alpha_{\mathrm{M}, j}$ for the $j$th mirror YIG sphere and a rate $\alpha_\mathrm{P}$ for the probe YIG sphere. We take this effect into account phenomenologically by adding on-site Lindblad terms to the master equation. Neglecting the quantum jump terms in the Lindblad dissipator $\mathcal{D}[\rho]$ [see \eqref{coherent_dis_2}] and these added phenomenological terms, the effective non-Hermitian Hamiltonian of the system takes the form
\begin{equation} 
\label{eq_Heff}
\begin{split}
H_\mathrm{eff} =&\: \hbar \omega_0 m_\mathrm{P}^\dag m_\mathrm{P} + \hbar \sum_{j = 1}^{N} \omega_0 m_{\mathrm{M}, j}^\dag m_{\mathrm{M}, j} \\
& - \frac{i\hbar}{2} \mleft( \kappa_\mathrm{PL} + \kappa_\mathrm{PT} + \alpha_\mathrm{P} \mright) m_\mathrm{P}^\dag m_\mathrm{P} - \frac{i\hbar}{2} \sum_{j = 1}^{N} \mleft( \kappa_{\mathrm{M}, j} + \alpha_{\mathrm{M}, j} \mright) m_{\mathrm{M}, j}^\dag m_{\mathrm{M}, j}  \\
& + \hbar \sum_{\substack{j, l = 1 \\ j \neq l}}^{N} J_{j, l} m_{\mathrm{M}, j}^\dag m_{\mathrm{M}, l} + \hbar \sum_{j = 1}^{N} J_{j, \mathrm{P}} \mleft( m_\mathrm{P}^\dag m_{\mathrm{M}, j} + \text{H.c.} \mright) \\
& - i\hbar \sum_{\substack{j, l = 1 \\ j \neq l}}^{N} \gamma_{j, l} m_{\mathrm{M}, j}^\dag m_{\mathrm{M}, l} - i\hbar \sum_{j = 1}^{N} \gamma_{j, \mathrm{P}} \mleft( m_\mathrm{P}^\dag m_{\mathrm{M}, j} + \text{H.c.} \mright) .
\end{split}
\end{equation}
%


\subsection{Dual-waveguide scattering theory}

With the master equation and the non-Hermitian Hamiltonian in \eqref{eq_Heff} in place, we now turn to input-output theory for our system. 
In this and all following sections, we set $\hbar = 1$ for notational convenience and simplicity. 
Since there is no direct coupling between the two waveguides, the spatial input-output relations for the propagating fields~\cite{PhysRevA.88.043806, 29th-t4ks} can be expressed as
\begin{align}
	a_{\mathrm{out}, \alpha}(x) &= a_{\mathrm{in}, \alpha}(x) - i\sqrt{\frac{\kappa_\mathrm{M}}{2}} \sum_{j = 1}^{N} m_{\mathrm{M}, j} e^{i \eta_\alpha k_0 (x-x_j)} - i\sqrt{\frac{\kappa_\mathrm{PT}}{2}} m_\mathrm{P} e^{i \eta_\alpha k_0 (x-x_\mathrm{P})} , \\
	b_{\mathrm{out}, \alpha}(y) &= b_{\mathrm{in}, \alpha}(y) - i\sqrt{\frac{\kappa_\mathrm{PL}}{2}} m_\mathrm{P} e^{i \eta_\alpha k_0 (y-y_\mathrm{P})} ,
	\label{Ltransmission}
\end{align}
where $k_0$ is the wavenumber of the propagating photon field at frequency $\omega_0$. Here, for simplicity, we assume that all mirror YIG spheres couple to the transverse waveguide with an identical radiative decay rate, denoted $\kappa_\mathrm{M}$. The operators $a_{\mathrm{in/out}, \alpha}(x)$ and $b_{\mathrm{in/out}, \alpha}(y)$ denote the input and output field operators for photons propagating in the $\alpha \in \{\mathrm{F, B}\}$ direction in the transverse and longitudinal waveguides, respectively.

In our experiments on this magnonic system, we often use a weak coherent input field incident from Port~3 of the longitudinal waveguide (see \figref{system}). With our notation convention, input at this port corresponds to the forward-propagating direction ($\alpha = \mathrm{F}$), so the input field takes the form of a plane wave $b_{\mathrm{in}, \mathrm{F}}(y) = \varepsilon e^{i k_\mathrm{in} y}$, with frequency $\omega_\mathrm{p}$ and wavenumber $k_\mathrm{in}$. 
This field acts as a weak driving term for the collective magnon excitations, which can be expressed as
\begin{equation}
	\varepsilon V_\mathrm{drive} = \varepsilon \sqrt{\frac{\kappa_\mathrm{PL}}{2}} \mleft( m_\mathrm{P}^\dag + m_\mathrm{P} \mright).
	\label{drive}
\end{equation}
We define the reference ground state of all the magnons as $\ket{G} = \prod_{j = 1}^{N + 1} \ket{g_j}$, where $\ket{g_j}$ denotes the local ground state of the magnon mode in the $j$th YIG sphere. As discussed below \eqref{rhofinal}, because we are considering coherent dynamics governed by a bilinear Hamiltonian, this state acts as an effective vacuum state relative to the room-temperature thermal equilibrium background. In the weak-driving regime, the total state of the system can be written as
\begin{equation}
	\ket{\psi} = \ket{G} + \ket{\delta \psi} ,
\end{equation}
where $\ket{\delta \psi}$ represents a perturbative excitation induced by $\varepsilon V_\mathrm{drive}$. Substituting this ansatz into the Schr\"odinger equation
\begin{equation}
	i \partial_t \ket{\psi} = \mleft( H_\mathrm{eff} + \varepsilon V_\mathrm{drive} \mright) \ket{\psi},
\end{equation}
where $H_\mathrm{eff}$ is in the rotating frame at frequency $\omega_0$, and retaining terms up to first order in $\varepsilon$, we obtain
\begin{equation}
	i \partial_t \ket{\delta \psi} = H_\mathrm{eff} \ket{\delta \psi} + \varepsilon V_\mathrm{drive} \ket{G} ,
\end{equation}
where we assume the unperturbed state satisfies $H_\mathrm{eff} \ket{G} = 0$. 

By adopting the monochromatic steady-state ansatz $\ket{\delta \psi(t)} = \ket{\delta \psi} e^{-i \omega_\mathrm{p} t}$, the time derivative reduces to $i \partial_t \ket{\delta \psi} = \omega_\mathrm{p} \ket{\delta \psi}$. Thus the steady-state correction is given by
\begin{equation}
	\ket{\delta \psi} = \mleft( \omega_\mathrm{p} - H_\mathrm{eff} \mright)^{-1} \varepsilon V_\mathrm{drive} \ket{G} .
\end{equation}
Therefore, the steady-state excitation amplitude of the probe YIG sphere evaluates to
\begin{equation}
	\langle\psi|m_P|\psi\rangle = \varepsilon \sqrt{\frac{\kappa_\mathrm{PL}}{2}} \bra{G} m_P \mleft(\omega_\mathrm{p} - H_\mathrm{eff}\mright)^{-1} m_P^\dag \ket{G} .
	\label{eq:mP_expectation_initial}
\end{equation}
Because we operate in the weak-driving regime and retain only first-order corrections, the dynamics are strictly confined to the single-excitation subspace, which is spanned by exactly $N+1$ collective magnon modes. Consequently, we expand the resolvent operator using the biorthogonal right ($\ket{\psi_\mu^\mathrm{R}}$) and left ($\bra{\psi_\mu^\mathrm{L}}$) eigenstates of $H_\mathrm{eff}$. With the biorthogonality relation $\braket{\psi_\mu^\mathrm{L}}{\psi_\nu^\mathrm{R}} = \delta_{\mu\nu}$, the expansion takes the form
\begin{equation}
	\mleft( \omega_\mathrm{p} - H_\mathrm{eff} \mright)^{-1} = \sum_{\mu = 1}^{N + 1} \frac{\ketbra{\psi_\mu^\mathrm{R}}{\psi_\mu^\mathrm{L}}}{\omega_\mathrm{p} - \omega_0 - E_\mu},
\end{equation}
where $\omega_\mathrm{p} - \omega_0 = v \mleft( k_\mathrm{in} - k_0 \mright)$ and $E_\mu$ denotes the $\mu$th complex eigenvalue of $H_\mathrm{eff}$. 

In this single-excitation subspace, the state $m_P^\dag \ket{G}$ represents the condition where only the probe YIG sphere is excited. We can map this to an effective driving vector $\bm{V} = \mleft( 0,\, \dots, 1,\, \dots,\, 0 \mright)^\top$, whose only nonzero element corresponds to the position of the probe YIG sphere. Inserting the resolvent expansion into \eqref{eq:mP_expectation_initial}, the excitation amplitude can be expressed in matrix form as
\begin{equation}
	\langle\psi|m_P|\psi\rangle = \varepsilon\sqrt{\frac{\kappa_\mathrm{PL}}{2}}
	\sum_{\mu = 1}^{N + 1} \frac{ \bm{V}^\dag  \ketbra{\psi_\mu^\mathrm{R}}{\psi_\mu^\mathrm{L}} \bm{V} }{\omega_\mathrm{p} - \omega_0 - E_\mu} .
	\label{eq:mP_expectation_final}
\end{equation}
Substituting this result into the input-output relation \eqref{Ltransmission} and taking the expectation value, we obtain the photon transmission amplitude $S_{43}$ from Port~3 to Port~4 in the longitudinal waveguide:
\begin{equation}
	S_{43}  = 1 -  \frac{i \kappa_\mathrm{PL}}{2} \sum_{\mu = 1}^{N + 1} \frac{ \bm{V}^\dag  \ketbra{\psi_\mu^\mathrm{R}}{\psi_\mu^\mathrm{L}} \bm{V} }{ \omega_\mathrm{p} - \omega_0 - E_\mu } .
	\label{S43_anti}
\end{equation}

%
To elucidate the physical information extracted from \eqref{S43_anti}, it is instructive to evaluate a physically relevant special case. In the Bragg cavity, the probe magnon (PM) mode couples exclusively to a single collective dark supermode when $N$ is even, with zero coupling to any other modes; see Fig.~\ref{fig:bragg_N}.

The transmission amplitude can be written using the resolvent operator:
\begin{equation}
	S_{43} = 1 - \frac{i \kappa_\mathrm{PL}}{2} \bm{V}^\dag \mleft( \omega_\mathrm{p} - H_\mathrm{eff} \mright)^{-1} \bm{V} ,
\end{equation}
Assuming that the PM solely couples to a single dark cavity mode $\ket{\mathrm{D_1}}$ with coupling strength $J_\mathrm{D}$, we project the system onto the effective subspace spanned by $\{\ket{\mathrm{PM}}, \ket{\mathrm{D_1}}\}$. In this basis, the effective Hamiltonian truncates to
\begin{equation}
	H_{\mathrm{sub}} = 
	\begin{pmatrix}
		\omega_0 - i\Gamma_\mathrm{PM} & J_\mathrm{D} \\
		J_\mathrm{D} & \omega_\mathrm{D} - i\Gamma_\mathrm{D}
	\end{pmatrix} .
\end{equation}
Here, $\omega_\mathrm{D}$ and $\Gamma_\mathrm{D}$ denote the resonance frequency and the total decay rate of the uniquely coupled dark mode $\ket{\mathrm{D_1}}$, respectively. The term $\Gamma_\mathrm{PM} = \kappa_\mathrm{PL}/2 + \kappa_\mathrm{PT}/2 + \alpha_\mathrm{P}/2$ is the total decay rate of the PM.

By substituting $H_{\mathrm{sub}}$ into the resolvent operator and evaluating the matrix element corresponding to the PM mode, we obtain the exact analytical form:
\begin{equation}
	S_{43, \rm Bragg} = 1 + \frac{\kappa_\mathrm{PL} / 2}{i \mleft( \omega_\mathrm{p} - \omega_0 \mright) - \Gamma_\mathrm{PM} + \frac{J_\mathrm{D}^2}{ i \mleft( \omega_\mathrm{p} - \omega_\mathrm{D} \mright) - \Gamma_\mathrm{D} }} .
	\label{S_43_derived}
\end{equation}
Equation~\eqref{S_43_derived} is mathematically identical to the standard expression derived from phenomenological coupled-mode theory~\cite{Rao_PRL_2023}. This confirms that under the dominant two-mode coupling approximation, our multi-mode expansion strictly reduces to $S_{43, \rm Bragg}$, giving a clear physical picture of the transmission spectroscopy.

\subsection{Bound states in the continuum}

We now focus on the bare cavity formed by the mirror array of YIG spheres to analyze the photonic distribution therein, neglecting the probe YIG sphere and the longitudinal waveguide. Specifically, we calculate below the spatial distributions of the relative photonic amplitudes associated with distinct supermodes. We then reintroduce the probe YIG sphere to investigate the relationship between the probe-cavity coupling and these photonic distributions. Crucially, through this spatial analysis, we reveal that the cavity naturally hosts bound states in the continuum (BICs) perfectly confined within the system.

In real space, the Hamiltonian of the transverse waveguide is given by
\begin{equation}
H_\mathrm{wg}^{(x)} = -i v \int dx \mleft[ a_\mathrm{F}^\dag(x) \partial_x a_\mathrm{F}(x) - a_\mathrm{B}^\dag(x) \partial_x a_\mathrm{B}(x) \mright],
\end{equation}
where $a_\mathrm{F(B)}^\dag(x)$ and $a_\mathrm{F(B)}(x)$ are the creation and annihilation operators, respectively, of forward- (backward-) propagating photons at position $x$.

The bare Hamiltonian for the mirror YIG-sphere array is $H_\mathrm{MM}^{(x)} = \omega_0 \sum_{j = 1}^N m_{\mathrm{M}, j}^\dag m_{\mathrm{M}, j}$. The real-space interaction Hamiltonian, which describes the local coupling between the propagating waveguide photons and the discrete magnon modes of the mirror array, is given by
\begin{equation}
H_\mathrm{I}^{(x)} = \sum_{j = 1}^{N} \int dx \, \delta(x-x_j) g_\mathrm{M} \mleft[ a_\mathrm{F}^\dag(x) m_{\mathrm{M}, j} + a_\mathrm{B}^\dag(x) m_{\mathrm{M},j} + \mathrm{H.c.} \mright] .
\end{equation}
Consequently, the full Hamiltonian for this cavity system is given by $H^{(x)} = H_\mathrm{wg}^{(x)} + H_\mathrm{MM}^{(x)} + H_\mathrm{I}^{(x)}$.

Here, we consider collective excitations comprising two components: magnons in the mirror YIG spheres and photons in the waveguide. The total-excitation-number operator of the system is
\begin{equation}
N_E = \sum_{j = 1}^N m_{\mathrm{M}, j}^\dag m_{\mathrm{M},j} + \sum_{\alpha = \mathrm{F,B}} \int dx \, a_\alpha^\dag(x) a_\alpha(x) .
\end{equation}
Under the rotating-wave approximation and in the weak-driving limit, multi-excitation processes can be neglected. Therefore, the system dynamics are well described within the single-excitation subspace. Focusing strictly on the bare mirror-YIG array to analyze its intrinsic BIC modes, we define the reference ground state as
$\ket{\mathrm{vac}, G} = \ket{\mathrm{vac}} \otimes \prod_{j = 1}^N \ket{g_j}$,
where $\ket{\mathrm{vac}}$ denotes the vacuum state of the photonic modes in the waveguide. As discussed below~\eqref{rhofinal}, this combined reference state serves as an effective vacuum relative to the thermal equilibrium background. Accordingly, a general eigenstate can be expressed as
\begin{equation}
\ket{\Psi} = \sum_{\alpha = \mathrm{F,B}} \int dx \, \phi_\alpha(x) a_\alpha^\dag(x) \ket{\mathrm{vac}, G} + \sum_{j = 1}^N c_j m_{\mathrm{M}, j}^\dag \ket{\mathrm{vac}, G} ,
\end{equation}
where $\phi_\alpha(x)$ ($\alpha = \mathrm{F,B}$) is the spatial amplitude of a forward- or backward-propagating photon in the waveguide, and $c_j$ denotes the excitation amplitude of the $j$th mirror magnon mode.

To investigate the photonic distribution corresponding to a specific supermode labeled by $n$, we first express its magnon sector using a collective mode operator:
\begin{equation}
\ket{\psi_n} = M_n^\dag \ket{\mathrm{vac}, G} ,
\qquad
M_n^\dag = \sum_{j = 1}^N c_{n, j} m_{\mathrm{M}, j}^\dag .
\end{equation}
Here, $\ket{\psi_n}$ is the pure magnonic state of this specific $n$th supermode, $M_n^\dag$ is the corresponding collective magnon creation operator, and $c_{n, j}$ denotes the excitation amplitude of the $j$th mirror magnon within this supermode.
The corresponding single-excitation state of the $n$th supermode can then be written as
\begin{equation}
\ket{\Psi_n} = \sum_{\alpha = \mathrm{F,B}} \int dx \, \phi_{n, \alpha}(x) a_\alpha^\dag(x) \ket{\mathrm{vac}, G} + \ket{\psi_n} .
\end{equation}

We substitute the single-excitation state $\ket{\Psi_n}$ into the stationary Schr\"odinger equation,
\begin{equation}
H^{(x)} \ket{\Psi_n} = E_n \ket{\Psi_n} ,
\end{equation}
where $E_n$ is the complex eigenenergy corresponding to the $n$th collective supermode of the bare mirror-YIG-sphere array. By projecting this equation into the real-space basis of the waveguide, we obtain separate coupled constitutive equations for the forward- and backward-propagating photonic amplitudes:
\begin{align}
-i v \partial_x \phi_{n, \mathrm{F}}(x) + g_\mathrm{M} \sum_{j = 1}^N c_{n, j} \delta(x - x_j) &= E_n \phi_{n, \mathrm{F}}(x) , \label{eq:1st_F} \\
+i v \partial_x \phi_{n, \mathrm{B}}(x) + g_\mathrm{M} \sum_{j = 1}^N c_{n, j} \delta(x - x_j) &= E_n \phi_{n, \mathrm{B}}(x) . \label{eq:1st_B}
\end{align}
To decouple these expressions, we define the total photonic wave function as $\phi_n(x) = \phi_{n, \mathrm{F}}(x) + \phi_{n, \mathrm{B}}(x)$. Taking the spatial derivatives of Eqs.~(\ref{eq:1st_F}) and (\ref{eq:1st_B}) and combining them, we eliminate the individual directional components to obtain a second-order differential equation:
\begin{equation}
\mleft( \partial_x^2 + \frac{E_n^2}{v^2} \mright) \phi_n(x) = \frac{2 E_n g_\mathrm{M}}{v^2} \sum_{j = 1}^N c_{n, j} \delta(x - x_j) .
\label{eq:2ndOrderDE}
\end{equation}

In the narrow-band limit typical for magnonic cavity systems, the collective radiative frequency shifts and decay rates contained within the complex eigenvalue $E_n$ are orders of magnitude smaller than the bare resonance frequency of the YIG spheres. This allows us to apply the Markovian approximation, replacing $E_n$ with the bare quantum energy $\omega_0$ in the differential coefficients ($E_n \approx \omega_0$). Under this approximation, \eqref{eq:2ndOrderDE} simplifies to its final effective form governed by the bare wave vector $k_0 = \omega_0 / v$:
\begin{equation}
\mleft( \partial_x^2 + k_0^2 \mright) \phi_n(x) = \frac{2k_0 g_\mathrm{M}}{v} \sum_{j = 1}^N c_{n, j} \delta(x - x_j) .
\end{equation}

As a consequence, the photonic field in the waveguide can be expressed as a superposition of waves emitted from each scattering center. To facilitate a systematic comparison of the spatial distributions across different collective excitations, we define a normalized relative amplitude $\tilde{\phi}_n(x)$ [often denoted simply as $\tilde{\phi}(x)$ without the mode index] for the $n$th supermode:
\begin{equation}
\tilde{\phi}_n(x) = A \sum_{j = 1}^N c_{n, j} e^{i k_0 |x - x_j|} ,
\end{equation}
where the normalization constant $A$ is chosen such that the maximum amplitude of the profile is scaled to unity ($\max |\tilde{\phi}_n(x)| = 1$). This relative spatial profile provides a clear visualization of the photonic distribution.

\begin{figure*}
\includegraphics[width=\linewidth]{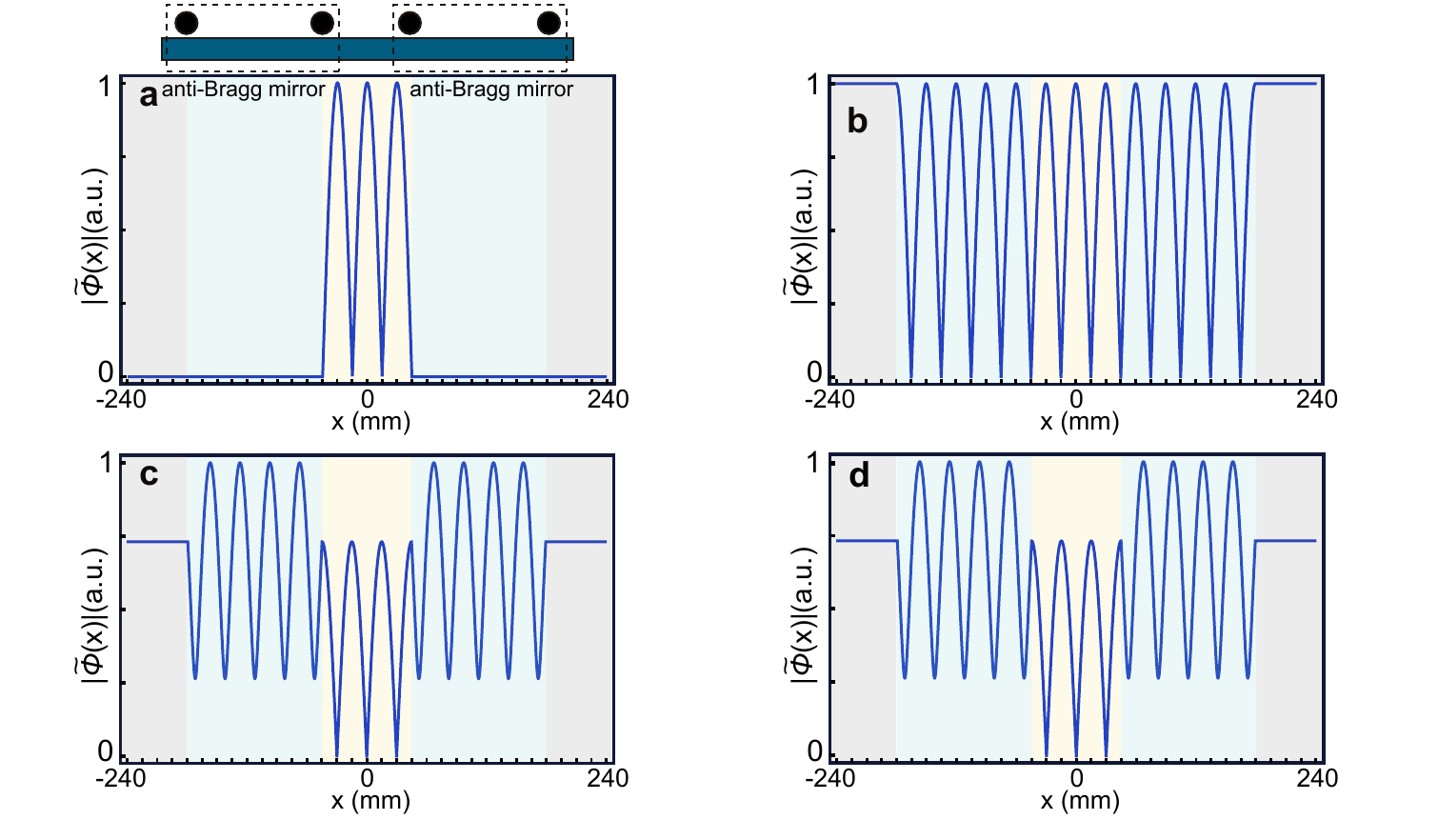}
\caption{Spatial distribution of the relative photonic amplitude $\tilde{\phi}(x)$ in an anti-Bragg magnonic cavity with $N = 4$ mirror YIG spheres along the transverse waveguide, for
(a) the dark supermode,
(b) the bright supermode,
and (c,~d) the two subradiant supermodes.
The model configurations are the same as those in Figs.~4(e)--(f) in the main text, and the parameters used here are 
$\kappa_\mathrm{M}/2\pi = \unit[9.2]{MHz}$, 
$\alpha_\mathrm{M}/2\pi = \unit[6.2]{MHz}$.
The collective response in the spectrum, detected by adding a probe YIG sphere, is shown in Figs.~4(e)--(f) in the main text.
The gray, cyan, and yellow shaded regions denote the areas outside the cavity, within the mirrors formed by the YIG spheres, and inside the cavity, respectively. This background color convention is consistent across all photonic amplitude profiles.
}
\label{fig:anti-bragg_photonic_amplitude}
\end{figure*}

To demonstrate this visualization explicitly, we present in \figref{fig:anti-bragg_photonic_amplitude} the calculated spatial distributions of the relative photonic amplitudes for the four supermodes in an $N = 4$ anti-Bragg magnonic cavity, obtained using the same cavity configuration and parameters as in Fig.~4 in the main text.
Specifically, \figpanels{fig:anti-bragg_photonic_amplitude}{a}{b} show the relative photon distributions for the dark supermode and the bright supermode, respectively, while \figpanels{fig:anti-bragg_photonic_amplitude}{c}{d} display the profiles for the two subradiant supermodes. Notably, the spatial profile of the dark supermode reveals that its photonic amplitude is strictly confined within the internal waveguide region of the cavity, whereas all other supermodes exhibit non-vanishing distributions outside the cavity. Because the collective excitation of this dark supermode is strictly localized within the intra-cavity waveguide, completely decoupling it from the external radiative channels, it constitutes a bound state in the continuum (BIC), protected from waveguide dissipation.


\subsection{Scaling and positional behavior of coupling rates}
	
One of the central results of this work is the tunability of the coupling rates between the probe YIG sphere and a particular dark supermode by changing the total number of YIG spheres in the Bragg or anti-Bragg mirrors, as done in Figs.~1(g) and 2(e) in the main text, and by changing the position of the probe YIG sphere, as done in Fig.~3 in the main text. In this section, we provide a detailed theoretical explanation for all these behaviors observed in our experiments.
For convenience, we eliminate the photonic degrees of freedom and consider magnonic components of supermodes.


\subsubsection{Bragg mirrors}   
\label{Bragg lattice}

We start with the case of Bragg mirrors, and first focus on the properties of such a Bragg lattice. The effective non-Hermitian Hamiltonian of a cavity constructed by $N$ mirror YIG spheres can be written as [see \eqref{eq_Heff}; here we have transformed to a frame rotating at frequency $\omega_0$, and neglected both the intrinsic magnon loss to free space and the presence of the probe YIG sphere]
%
\begin{equation} 
\label{eq_H}
\begin{split}
H_\mathrm{eff} = \sum_{j, l = 1}^N A_{j, l} m_{\mathrm{M}, j}^\dag m_{\mathrm{M}, l}
\end{split},
\end{equation}
with $A_{j, l} = -i \sqrt{\kappa_{\mathrm{M}, j} \kappa_{\mathrm{M}, l}} \exp(i \phi_{j, l}) / 2$ and $\phi_{j, l} = 2\pi |x_j - x_l| / \lambda_0$. In the Bragg-lattice case, the distance between neighboring YIG spheres is either $\lambda_0 / 2$ or $3 \lambda_0 / 2$, resulting in a phase of $\phi_{j, j+1} = \pi \pmod{2\pi}$. Assuming that all the YIG spheres couple to the waveguide with the same strength $\kappa_\mathrm{M}$, \eqref{eq_H} reduces to
\begin{equation}
\label{eq_HB}
H_\mathrm{eff} = -\frac{i \kappa_\mathrm{M}}{2} \sum_{j, l = 1}^N (-1)^{|j - l|} m_{\mathrm{M}, j}^\dag m_{\mathrm{M}, l} .
\end{equation}
Equation~(\ref{eq_HB}) can be analytically solved: it hosts a bright supermode
\begin{equation} 
\label{eq_psiB}
\ket{B}_N = \frac{1}{\sqrt{N}} \sum^N_{j = 1} (-1)^j m_{\mathrm{M}, j}^\dag \ket{G},
\end{equation}
with eigenvalue $-i N \kappa_\mathrm{M} / 2$ and $N - 1$ degenerate dark supermodes $\ket{D_n}_N$ with eigenvalue 0:
\begin{equation}
\ket{D_n}_N = \sum^N_{j = 1} c_{n, j} m_{\mathrm{M}, j}^\dag \ket{G} , \quad (n = 1, \ldots, N - 1) ,
\end{equation}
where the coefficients satisfy
\begin{equation}
\sum^N_{j = 1}(-1)^j c_{n, j} = 0, \quad \forall n .
\end{equation}

We now include the probe YIG sphere. The effective Hamiltonian then becomes
\begin{equation} 
\label{eq_Hfull}
H_\mathrm{eff} = \sum_{j, l = 1}^N A_{j, l} m_{\mathrm{M}, j}^\dag m_{\mathrm{M}, l}   + \sum_{j = 1}^N A_{j, \mathrm{P}} \mleft( m_{\mathrm{M}, j}^\dag m_\mathrm{P} + \text{H.c.} \mright) - i \gamma_\mathrm{P} m_\mathrm{P}^\dag m_\mathrm{P} ,
\end{equation}
with $A_{j, \mathrm{P}} = -i \gamma_\mathrm{MP} \exp(i \phi_{j, \mathrm{P}}) / 2$, $\gamma_\mathrm{MP} = \sqrt{\kappa_\mathrm{M} \kappa_\mathrm{PT}}$, and $\phi_{j, \mathrm{P}} = 2\pi |x_j - x_\mathrm{P}| / \lambda_0$. To obtain the effective coupling between the probe YIG sphere and the bright and dark supermodes of the Bragg lattice, we write \eqref{eq_Hfull} in the basis of these supermodes:
\begin{equation}
b^\dag = \frac{1}{\sqrt{N}} \sum^N_{j = 1} (-1)^j m_{\mathrm{M}, j}^\dag, \qquad d_n^\dag = \sum^N_{j = 1} c_{n, j} m_{\mathrm{M}, j}^\dag ,
\end{equation}
which yields
\begin{equation}
H_\mathrm{eff} = - \frac{i \sqrt{N} \kappa_\mathrm{M}}{2} b^\dag b + G_\mathrm{BP} \mleft( b^\dag m_\mathrm{P} + \text{H.c.} \mright) + \sum_{n = 1}^{N - 1} G_{\mathrm{DP}, n} \mleft( d^\dag_n m_\mathrm{P} + \text{H.c.} \mright)- i \gamma_\mathrm{P} m_\mathrm{P}^\dag m_\mathrm{P} ,
\end{equation}
where
\begin{align}
G_\mathrm{BP} &= \frac{1}{\sqrt{N}} \sum^N_{j = 1} (-1)^j A_{j, \mathrm{P}} , \\
G_{\mathrm{DP}, n} &= \sum^N_{j = 1} c^*_{n, j} A_{j, \mathrm{P}}.
\end{align}

We first consider the case in which $N$ is even and the probe YIG sphere is placed between the two Bragg mirrors. In this case, the probe YIG sphere couples symmetrically to the two Bragg lattices:
\begin{equation}
A_{j, \mathrm{P}} = \mleft\{ 
\begin{array}{cc} 
- \frac{1}{2} \gamma_\mathrm{MP} (-1)^{j + N / 2} & (j \leq N / 2) \\ 
- \frac{1}{2} \gamma_\mathrm{MP} (-1)^{j + N / 2 + 1} & (j > N / 2) 
\end{array} 
\mright. ,
\end{equation}
satisfying $A_{j, \mathrm{P}} = A_{(N + 1 - j), \mathrm{P}}$. 
Additionally, since $A_{j, \mathrm{P}}$ and the eigenvector coefficients $c_{n,j}$ are real in this configuration, $G_{\mathrm{DP, n}}$ is real.
To evaluate the coupling between the probe YIG sphere and the subspace of dark supermodes, we can separate the dark subspace into a dark supermode that is phase-coherent with $A_{j, \mathrm{P}}$,
\begin{equation} 
\label{eq_psiD}
\ket{D_1}_N = \frac{1}{\sqrt{N}} \mleft( \sum^{N / 2}_{j = 1} (-1)^{j + N / 2} m_{\mathrm{M}, j}^\dag + \sum^N_{j = N / 2 + 1} (-1)^{j + N / 2 + 1} m_{\mathrm{M}, j}^\dag \mright) \ket{G} ,
\end{equation}
and a subspace spanned by $\ket{D_{2, \cdots, N - 1}}_N$ that is orthogonal to $\ket{D_1}_N$ and therefore has $G_{\mathrm{DP, n}} = 0$. We then obtain
\begin{equation}
G^{\rm even}_{\mathrm{DP}, 1} = J^{\rm even} = - \frac{\sqrt{N}}{2} \gamma_\mathrm{MP} .
\end{equation}
This yields
\begin{equation}
|2 J^{\rm even}| = 2 J_\mathrm{D} = \sqrt{N \kappa_\mathrm{PT} \kappa_\mathrm{M}} ,
\end{equation}
in agreement with the results in Fig.~1(g) in the main text.

When $N$ is odd, the bright supermode is still given by \eqref{eq_psiB}, and the coupling between the probe YIG sphere and the bright supermode now becomes non-zero: $G^{\rm odd}_\mathrm{BP} = \pm \gamma_\mathrm{MP} / (2 \sqrt{N})$. The dark subspace in this case is the one for the $N - 1$ case plus an additional mode
\begin{equation}
\ket{D_{N - 1}}_N = \frac{1}{\sqrt{N}} \mleft( \ket{B}_{N - 1} \otimes \ket{g_N} - \sqrt{N - 1} \ket{G}_{N - 1} \otimes \ket{e_N} \mright) ,
\end{equation}
where $\ket{B}_{N - 1}$ is the bright supermode of the Bragg mirrors with $N - 1$ YIG spheres, and $\ket{g_N}$ and $\ket{e_N}$ are the ground and first excited states, respectively, of the added $N$th YIG sphere. With this choice of basis, the only two dark supermodes that couple to the probe YIG sphere are $\ket{D_{N - 1}}_N$, with a coupling strength of $G_{\mathrm{DP}, N - 1} = - \sqrt{N - 1} \gamma_\mathrm{MP} / (2 \sqrt{N})$, and $\ket{D_1}_N = \ket{D_1}_{N - 1}\otimes \ket{g_N}$ given by \eqref{eq_psiD}, with a coupling strength of $G_{\mathrm{DP}, 1}  = - \sqrt{N - 1} \gamma_\mathrm{MP} / 2$. We can now choose a different basis
\begin{align}
\ket{D'_1}_N &= \frac{1}{\sqrt{N + 1}} \mleft( \sqrt{N} \ket{D_1}_N - \ket{D_{N - 1}}_N \mright) , \\
\ket{D'_{N - 1}}_N &= \frac{1}{\sqrt{N + 1}} \mleft( \ket{D_1}_N + \sqrt{N} \ket{D_{N - 1}}_N \mright) ,
\end{align}
such that the probe YIG sphere only couples to $\ket{D'_1}_N$ with strength
\begin{equation}
G^{\rm odd}_{\mathrm{DP}, 1} =  J^{\rm odd} = - \frac{\sqrt{N^2 - 1}}{2 \sqrt{N}} \gamma_\mathrm{MP} .
\end{equation}
This yields $|2 J^{\rm odd}| = \sqrt{(N^2 - 1) \kappa_\mathrm{PT} \kappa_\mathrm{M}/N}$, which is smaller than the $\sqrt{N}$ scaling. We confirm this result experimentally in Fig.~1(g) of the main text.

\begin{figure*}
\includegraphics[width=\linewidth]{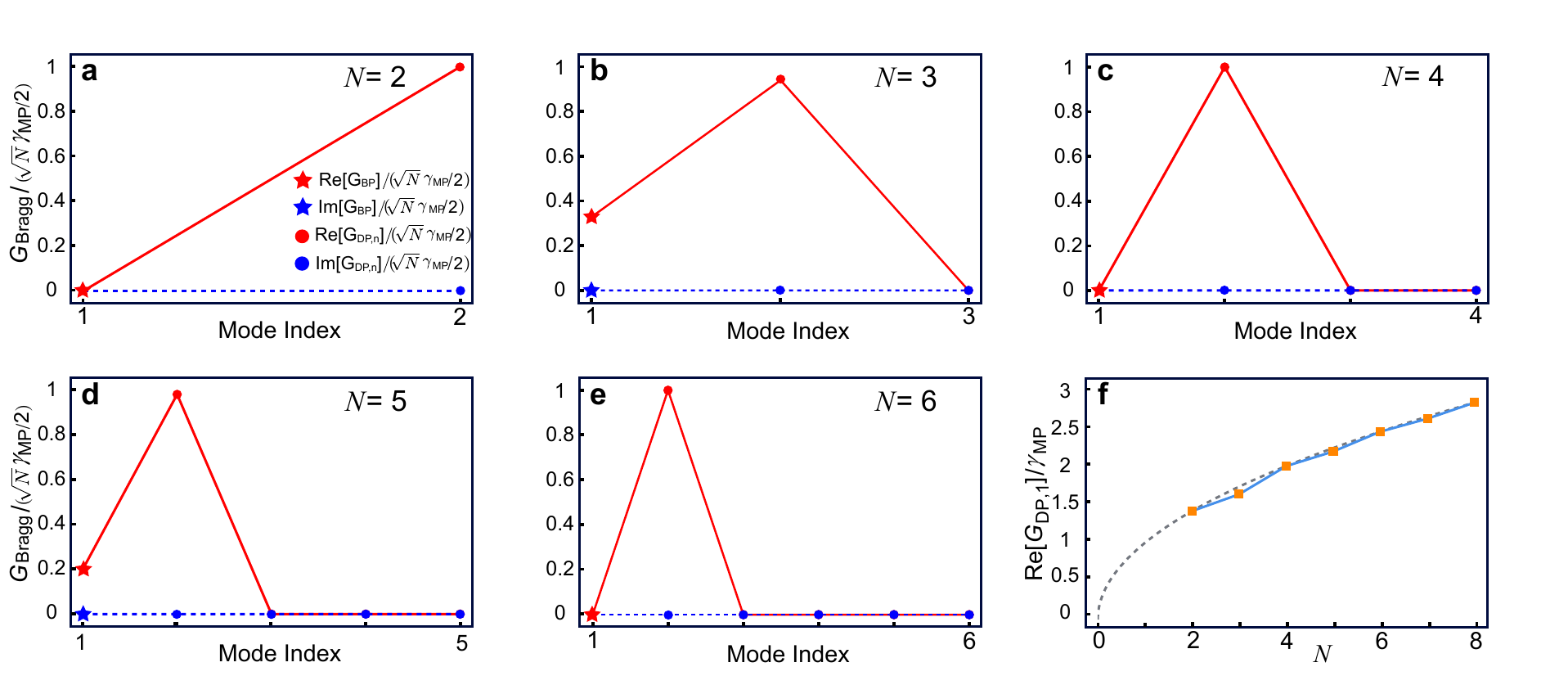}
\caption{(a)--(e) Coupling strengths $G_{\mathrm{Bragg}}$ between the probe YIG sphere and different supermodes for $N =$ 2--6 mirror YIG spheres.
The red (blue) markers correspond to the real (imaginary) part of the couplings.
The couplings marked with pentagrams (circles) correspond to the coupling $G_\mathrm{BP}$ ($G_{\mathrm{DP, n}}$) between the probe YIG sphere and the bright (dark) supermode(s).
(f) Coherent coupling $G_{\mathrm{DP}, 1}$ between the probe YIG sphere and the dark supermode [the supermode index is in panels (a)--(e)] as a function of the number $N$ of mirror YIG spheres. The parameters used here are 
$\kappa_\mathrm{PT} / 2\pi = \unit[8.5]{MHz}$,
$\kappa_\mathrm{PL} / 2\pi = \unit[10]{MHz}$,
$\kappa_\mathrm{M} / 2\pi = \unit[8.0]{MHz}$,
$\alpha_\mathrm{M} / 2\pi = \unit[6.2]{MHz}$,
and $\alpha_\mathrm{P} / 2\pi = \unit[7.0]{MHz}$.} 
\label{fig:bragg_N}
\end{figure*}

We verify the above observations numerically in \figref{fig:bragg_N}. In \figpanels{fig:bragg_N}{a}{e}, we show the real and imaginary parts of the couplings $G_\mathrm{BP}$ and $G_{\mathrm{DP, n}}$ for different numbers $N$ of mirror YIG spheres. In \figpanel{fig:bragg_N}{f}, we demonstrate the characteristic $\sqrt{N}$ scaling of $G_{\mathrm{DP, 1}}$, with negative deviations when $N$ is odd.
	
\begin{figure*}
\includegraphics[width=\linewidth]{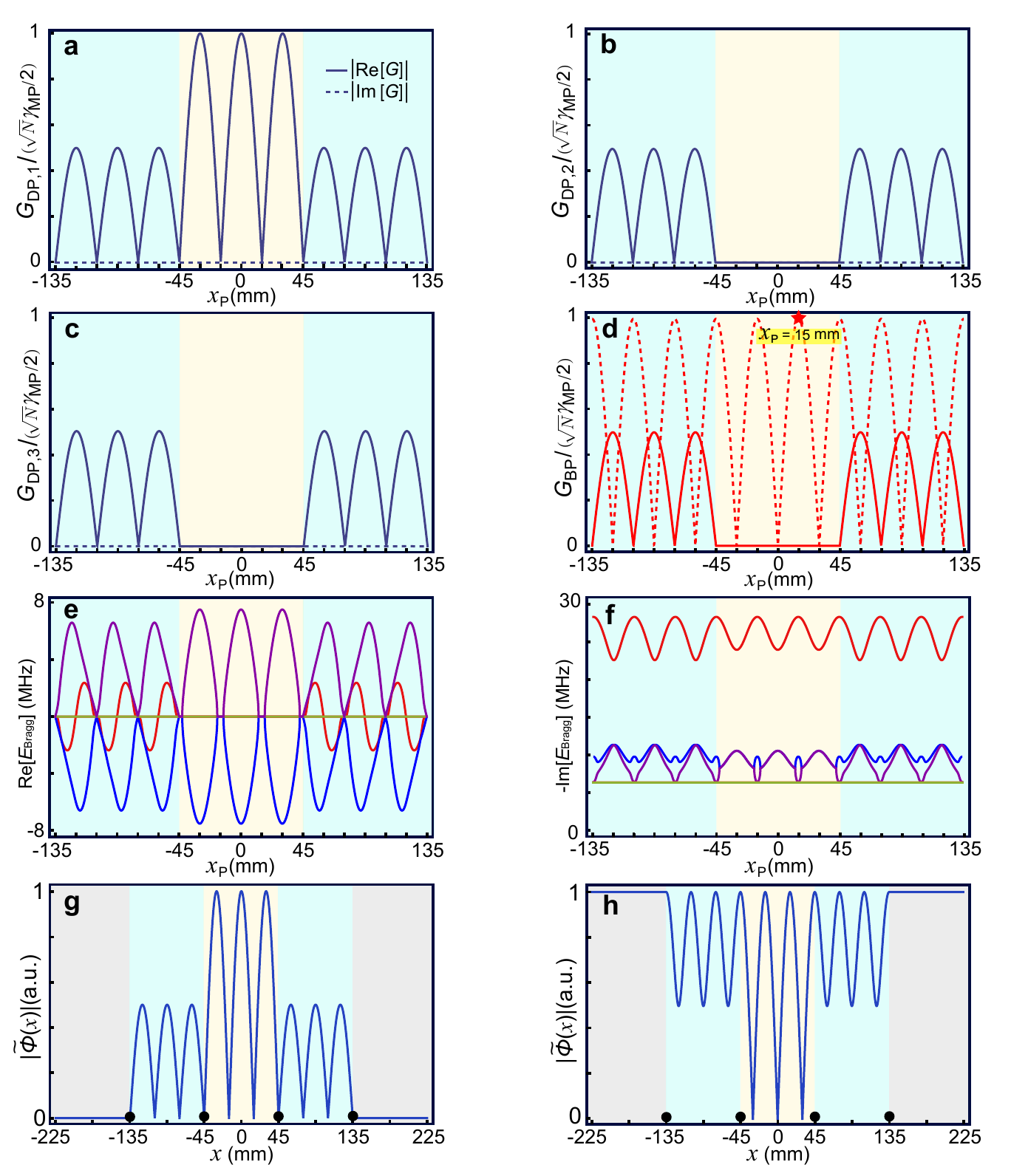}
\caption{Coupling strengths, energy levels, decay rates, and photonic amplitudes for a probe YIG sphere moved across a Bragg cavity formed by $N = 4$ YIG spheres.
(a)--(c) Couplings $G_\mathrm{DP, n}$ between the probe YIG sphere and the three dark supermodes as a function of the position of the probe YIG sphere. 
(d) Coupling $G_\mathrm{BP}$ between the probe YIG sphere and the bright supermode as a function of the position of the probe YIG sphere. The solid (dashed) curves in (a)--(d) denote coherent (dissipative) coupling.
(e) Energy levels and (f) decay rates of the hybridized modes formed by the probe magnon mode (PM) and the collective supermodes, as a function of the position of the probe YIG sphere. Lines sharing the same color denote the same supermode in the two panels. 
(g,~h) Spatial distributions of the relative photonic amplitudes of the dark supermode $\ket{D_1}_N$ and the bright supermode $\ket{B}_N$.
The parameters used in this figure are 
$\kappa_\mathrm{PT} / 2\pi = \unit[8.5]{MHz}$,
$\kappa_\mathrm{PL} / 2\pi = \unit[10]{MHz}$,
$\kappa_\mathrm{M} / 2\pi = \unit[8.0]{MHz}$,
$\alpha_\mathrm{M} / 2\pi = \unit[6.2]{MHz}$,
and $\alpha_\mathrm{P} / 2\pi = \unit[7.0]{MHz}$.
The corresponding experimental data are shown in Fig.~3 and Figs.~4(a)--(c) in the main text. 
}
\label{fig:moving_bragg}
\end{figure*}
	
To further investigate the coupling between the probe YIG sphere and the magnonic Bragg cavity, we move the probe YIG sphere along the transverse waveguide and analyze how its position affects the coupling strength. We illustrate this scenario for $N = 4$ in \figref{fig:moving_bragg}.
\figpanelshead{fig:moving_bragg}{a}{c} and \figpanel{fig:moving_bragg}{d} display the couplings of the probe YIG sphere to the three dark supermodes and to the bright supermode, respectively. Inside the cavity, the probe YIG sphere exhibits coherent coupling only to a single dark supermode, while its interaction with the bright supermode is purely dissipative~\cite{Mirhosseini2019}. As the probe YIG sphere is translated along the transverse waveguide, the coupling strengths to the cavity modes display a periodic oscillation, consistent with the behavior observed in the experiments in Figs.~3 and 4 in the main text. 
Notably, at $x_\mathrm{P} = \unit[15]{mm}$, the PM is dissipatively coupled solely to the superradiant state, allowing the effective Hamiltonian to be expressed in a reduced form:
\begin{equation}
	H_{\mathrm{eff}} = 
	\begin{pmatrix}
		-i\Gamma_\mathrm{B} & G_{\mathrm{BP}} \\
		G_{\mathrm{BP}} & -i\Gamma_{\mathrm{PM}}
	\end{pmatrix}.
\end{equation}
Here, $\Gamma_\mathrm{B}$ denotes the decay rate of the bright supermode, and $G_{\mathrm{BP}} = i\sqrt{N \kappa_{\mathrm{PT}} \kappa_{\mathrm{M}}}/2$ represents the effective dissipative coupling strength. By diagonalizing this reduced Hamiltonian, the eigenvalues of the two hybridized modes are analytically obtained as:
\begin{equation}
	\lambda_{\pm} = -\frac{i}{2} \mleft( \Gamma_\mathrm{B} + \Gamma_{\mathrm{PM}} \pm \sqrt{(\Gamma_\mathrm{B} - \Gamma_{\mathrm{PM}})^2 - 4 G_{\mathrm{BP}}^2} \mright).
\end{equation}
In this regime, the slow decay mode is dominantly localized at the PM, with its decay rate reduced compared to that of the bare PM into the longitudinal waveguide in the absence of the cavity. This demonstrates that at $x_\mathrm{P} = \unit[15]{mm}$, the dissipative coupling between the PM and the superradiant supermode strongly suppresses the decay rate of the PM into the longitudinal waveguide, directly accounting for the linewidth narrowing observed in the transmission spectrum.
	
Outside the cavity, the probe YIG sphere couples to all dark supermodes, but these dark supermodes do not introduce any dissipative coupling. The probe YIG sphere also interacts with the bright supermode, but because the bright supermode rapidly decays into the waveguide, its contribution to the effective probe-cavity coupling is negligible. Therefore, whether inside or outside the cavity, the dark supermodes dominate the interaction between the probe YIG sphere and the cavity.
	
In \figpanels{fig:moving_bragg}{e}{f}, we plot the real and imaginary parts, respectively, of the eigenvalues of the dual-waveguide probe-cavity system. The purple and blue branches in the spectra correspond to the mode splittings originating from the probe YIG sphere's coupling to the bright supermode and the dark supermode of the cavity. In addition, we observe the emergence of exceptional points located between the two oscillation periods inside the cavity. Between these two exceptional points, the real parts of the eigenvalues become degenerate while their imaginary parts split, such that no mode-splitting can be observed in the transmission spectrum. This behavior is also confirmed by the measurements in Fig.~4 in the main text.
	
In \figpanels{fig:moving_bragg}{g}{h}, we show the spatial distributions of the relative photonic amplitudes for the states $\ket{D_1}_N$ and $\ket{B}_N$, respectively. We find that the spatial profile of the dark supermode is in good agreement with the experimental results obtained from the moving-probe-YIG-sphere measurements of the probe-cavity coupling in Fig.~4 in the main text. This agreement indicates that the bound state in the continuum can be detected using the technique with a moving probe YIG sphere.

    
\subsubsection{Anti-Bragg mirrors}

\begin{figure*}
\includegraphics[width=0.7\linewidth]{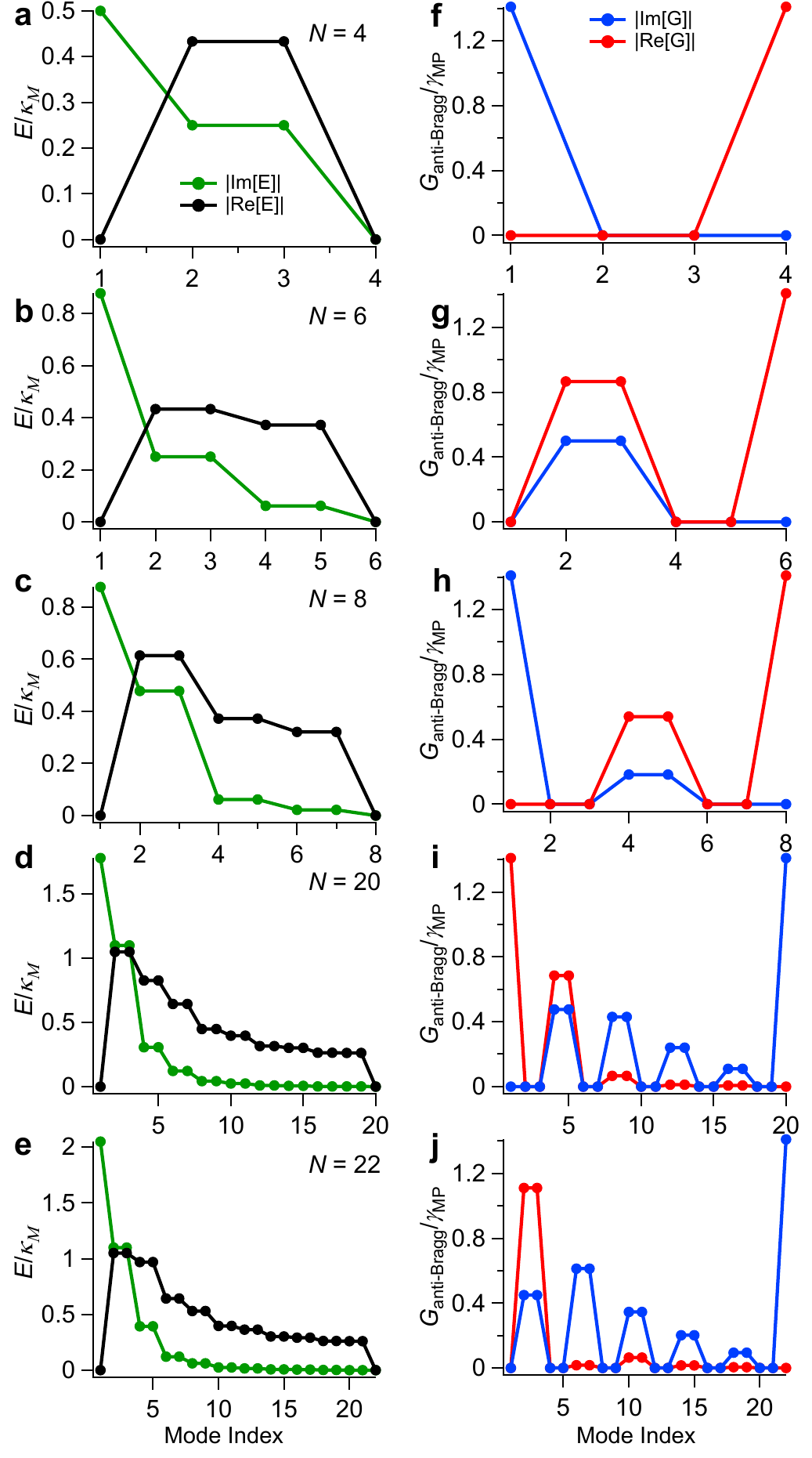}
\caption{(a)--(e) Eigenvalues of the modes of the anti-Bragg mirrors. The real part (black) gives the energy and the imaginary part (green) gives the decay rate.
(f)--(j) Coupling strengths (real parts in red, imaginary parts in blue) between the probe YIG sphere and the cavity eigenmodes for different numbers $N$ of mirror YIG spheres in the anti-Bragg mirrors.
The parameters used here are $\kappa_\mathrm{PT}$=0.8$\kappa_\mathrm{M}$, and $\alpha_\mathrm{M}=0$.}
\label{fig:anti-bragg_N}
\end{figure*}

For the case of an anti-Bragg cavity, we consider an even number of YIG spheres symmetrically distributed on the two anti-Bragg lattices: $N = 2 N_\mathrm{L} = 2 N_\mathrm{R}$. In this case, analytically solving the effective Hamiltonian in \eqref{eq_H} is no longer straightforward, since the phase difference corresponding to $3 \lambda_0 / 4$ is $3 \pi / 2$, which yields $\exp(i \phi_{j, j+1}) = -i$. We therefore numerically diagonalize the effective Hamiltonian, and show its spectrum for different values of $N$ in \figpanels{fig:anti-bragg_N}{a}{e}. Unlike the Bragg-cavity case, there is now a unique dark supermode,
\begin{equation}
\ket{D}_N = \frac{1}{\sqrt{2}} \mleft( m^\dag_{\mathrm{M},N / 2} + m^\dag_{\mathrm{M},N / 2 + 1} \mright) \ket{G} ,
\end{equation}
which resides on the two sites closest to the center of the setup, and does not change its position or strength when new YIG spheres are introduced. The brightest supermode in this case is also not superradiant; its decay rate has a sub-linear scaling with $N$.

The couplings $G_{\rm anti-Bragg}$ between the probe YIG sphere and these modes are shown in \figpanels{fig:anti-bragg_N}{f}{j}. Since the dark supermode does not change with $N$, its coupling to the probe YIG sphere is also a constant:
$G_\mathrm{DP} = \sqrt{2} \gamma_\mathrm{MP}/2$. However, the coupling to the bright supermodes is no longer vanishing in this case. In fact, the bright-probe coupling is of the same order of magnitude as the dark-probe coupling. It exhibits diverse collective effects and intricate interference patterns. These effects strongly modify the anti-crossing in $\abs{S_{43}}$. We see that as $N$ increases, more bright supermodes will be coupled to the probe, but with decaying coupling strengths, resulting in progressively smaller modification of the anti-crossing. We thus expect that the anti-crossing size should approach a constant as $N \to \infty$.

\begin{figure*}
\includegraphics[width=0.85\linewidth]{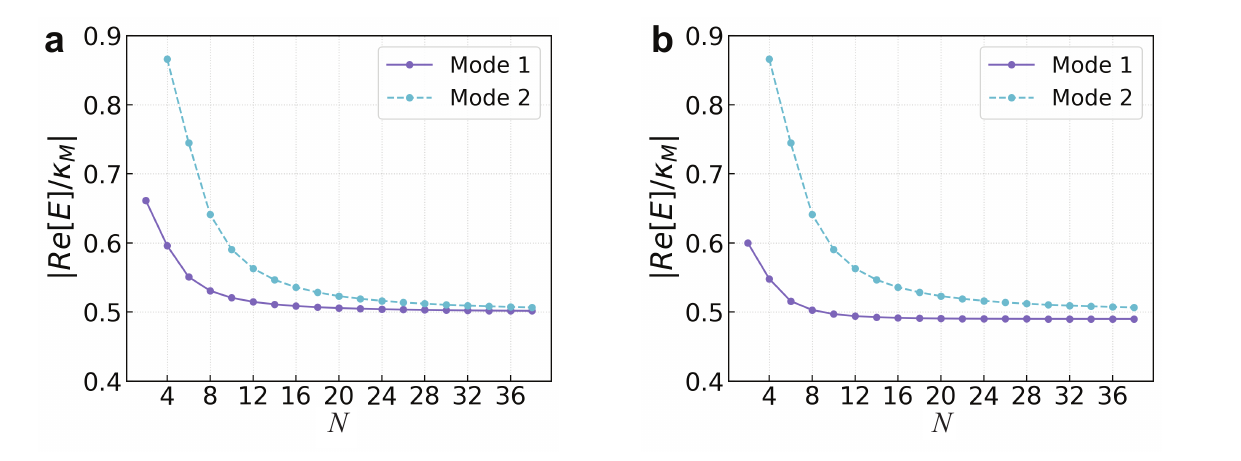}
\caption{The two smallest absolute real eigenvalues of modes having non-zero distribution on the probe YIG sphere in the anti-Bragg cavity. 
(a) $\kappa_\mathrm{PT} = \kappa_\mathrm{M}$, $\kappa_\mathrm{PL} = 0$, $\alpha_\mathrm{M}=\alpha_\mathrm{P}=0$. 
In (b), $\kappa_\mathrm{PT} = 0.8 \kappa_\mathrm{M}$ instead.}
\label{fig:anti-bragg_gap}
\end{figure*}
    
To verify this prediction, we compute the eigenvalue difference by diagonalizing the effective Hamiltonian for the probe YIG sphere and the anti-Bragg mirrors. 
In \figref{fig:anti-bragg_gap}, modes~1 and~2 are the two hybrid modes---formed by the coupling between the probe magnon mode and the anti-Bragg supermodes---with the smallest absolute real eigenvalues; these eigenvalues, measured in the rotating frame of $\omega_0$, correspond to the frequencies (shifted by $-\omega_0$) at which the main and additional anti-crossings occur in Fig.~2(d) of the main text.
The eigenfrequency of mode~1 also implies the existence of a negative eigenfrequency in the rotating frame. 
The difference between the two eigenvalues gives the anti-crossing splitting observed in the transmission spectrum.
We observe that, as expected, both anti-crossings approach a constant frequency as $N\to\infty$. 
Specifically, in the ideal case $\kappa_\mathrm{PT} = \kappa_\mathrm{M}$ and $\kappa_\mathrm{PL} = 0$, both eigenvalues approach $\kappa_\mathrm{M}/2$ as $N \to \infty$ [\figpanel{fig:anti-bragg_gap}{a}]. 
This value is smaller than the effective coupling $J_\mathrm{D}=\sqrt{2\kappa_\mathrm{PT}\kappa_\mathrm{M}}/2 = \sqrt{2}\kappa_\mathrm{M}/2$, reflecting the influence of the bright supermode in the ideal case, and is qualitatively consistent with the experimental results in Fig.~2(e) of the main text.
Decreasing $\kappa_\mathrm{PT}$ reduces the eigenvalue splitting [\figpanel{fig:anti-bragg_gap}{b}], indicating a corresponding reduction in $J_D$, as expected.

\begin{figure*}
 \includegraphics[width=0.7\linewidth]{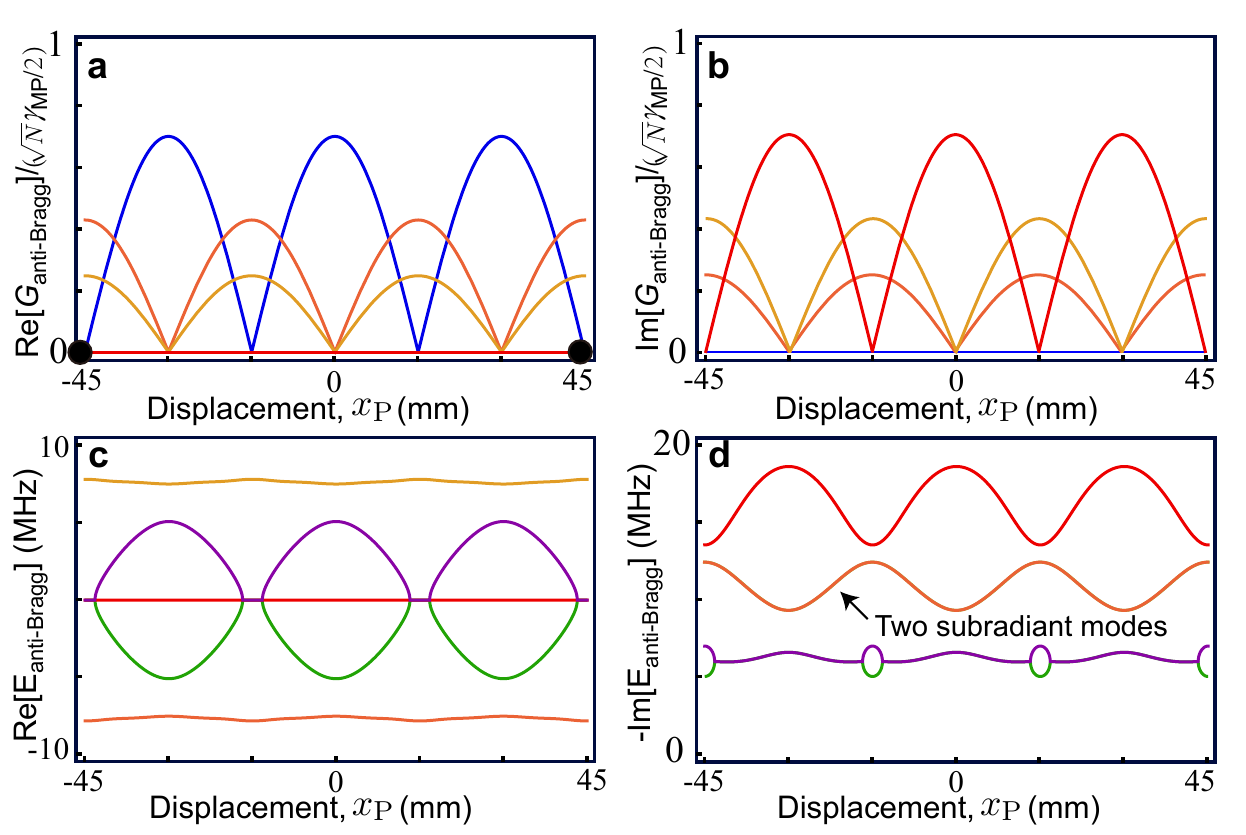}
\caption{Couplings, dissipation, and exceptional points in a magnonic cavity with anti-Bragg mirrors consisting of $N = 4$ mirror YIG spheres.
(a) Coherent and (b) dissipative couplings between the probe YIG sphere and the supermodes of the anti-Bragg cavity, as a function of the position of the probe YIG sphere. 
The blue line represents the dark supermode, the red line denotes the bright (superradiant) supermode, while the orange and yellow lines correspond to two subradiant modes.
Black circles indicate the locations of the nearest mirror YIG spheres; the other two mirror YIG spheres are not shown.
(c) Energy levels and (d) decay rates of the supermodes between the probe YIG sphere and the collective modes in the magnonic anti-Bragg cavity, as a function of the position of the probe YIG sphere.
The parameters used in (c,~d) are 
$\kappa_\mathrm{PT}/2\pi = \unit[9.2]{MHz}$, 
$\kappa_\mathrm{PL}/2\pi = \unit[11]{MHz}$, 
$\kappa_\mathrm{M}/2\pi = \unit[9.2]{MHz}$, 
$\alpha_\mathrm{M}/2\pi = \unit[6.2]{MHz}$, 
and $\alpha_\mathrm{P}/2\pi = \unit[7.0]{MHz}$.
The corresponding experimental data are shown in Figs.~4(e)--(f) in the main text.} 
\label{fig:moving_anti_bragg}
\end{figure*}

We now turn to investigating the spatial dependence of the coupling between the probe YIG sphere and the anti-Bragg cavity.
\figpanelshead{fig:moving_anti_bragg}{a}{b} illustrate how the coherent and dissipative couplings between the probe YIG sphere and the four supermodes vary as the probe is moved to different positions $x_\mathrm{P}$ inside the cavity.  Within the cavity, we find that all collective modes participate in the interaction with the probe YIG sphere. Specifically, the probe YIG sphere couples coherently only to the dark supermode (blue), dissipatively to the bright supermode (red), and exhibits both coherent and dissipative couplings to the other two subradiant modes (orange and yellow).

\begin{figure*}
\includegraphics[width=0.8\linewidth]{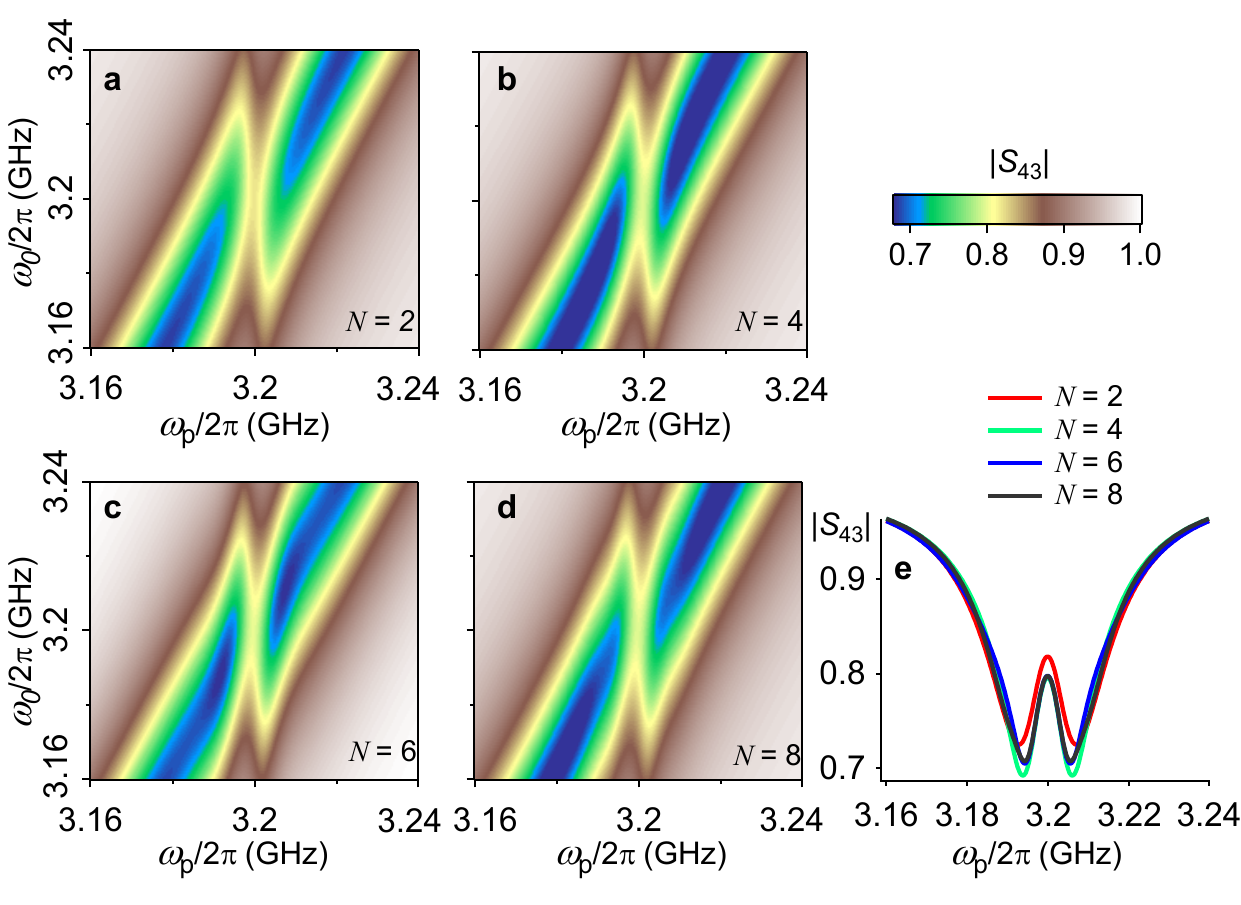}
\caption{Simulations of $|S_{43}|$ spectra for anti-Bragg cavities.
(a)--(d) Theoretical transmission spectra for different numbers $N$ of mirror YIG spheres, as a function of the probe magnon mode (PM) resonance frequency ($\omega_0$) and the probe frequency ($\omega_\mathrm{p}$).
(e) Line cuts of (a)--(d) at the resonance frequency of $\omega_0 / 2\pi = \unit[3.2]{GHz}$.
The simulations are based on \eqref{S43_anti} and use the parameters $\kappa_\mathrm{PT} / 2\pi = \unit[8]{MHz}$, $\kappa_\mathrm{PL} / 2\pi = \unit[10]{MHz}$, $\kappa_\mathrm{M} / 2\pi = \unit[10]{MHz}$, $\alpha_\mathrm{M} / 2\pi = \unit[6.7]{MHz}$, and $\alpha_\mathrm{P} / 2\pi = \unit[13.1]{MHz}$.
Variations in the individual $\kappa_{\mathrm{M}, j}$ and $\alpha_{\mathrm{M}, j}$ values lead to deviations between the simulation and experimental data.
Note that the parameters used here are different from the ones used in \figpanels{fig:moving_anti_bragg}{c}{d} because the former is obtained by fitting the experimental data of the probe YIG sphere fixed at the center of the anti-Bragg cavity, while the latter is obtained by fitting the experimental data of the moving probe YIG sphere in the anti-Bragg cavity.
}
\label{fig:anti_bragg_simulation}
\end{figure*}

Furthermore, in \figpanels{fig:moving_anti_bragg}{c}{d}, we present the real and imaginary parts of the eigenvalues of the anti-Bragg cavity as the probe YIG sphere is moved across different positions. 
Because all eigenmodes participate in the interaction with the probe YIG sphere inside the anti-Bragg cavity, the eigenvalues of the eigenmodes exhibit position-dependent variations in both their real and imaginary components. 
In particular, at certain probe positions inside the cavity, we identify a pair of eigenmodes whose real parts become degenerate while their imaginary parts split, indicating two exceptional points located at the two sides of the interval.
Within this exceptional-point interval, the real parts remain degenerate and the linewidths are split, and accordingly no mode splitting can be observed in the transmission spectrum, consistent with the experimental findings in the main text. 
In \figref{fig:anti_bragg_simulation}, we display simulations of transmission spectra obtained for anti-Bragg cavities with different numbers $N$ of mirror YIG spheres, while keeping the probe YIG sphere fixed at $x_\mathrm{P} = \unit[0]{mm}$. 
As $N$ increases, the observed splitting gradually diminishes.

\begin{figure*}
\includegraphics[width=\linewidth]{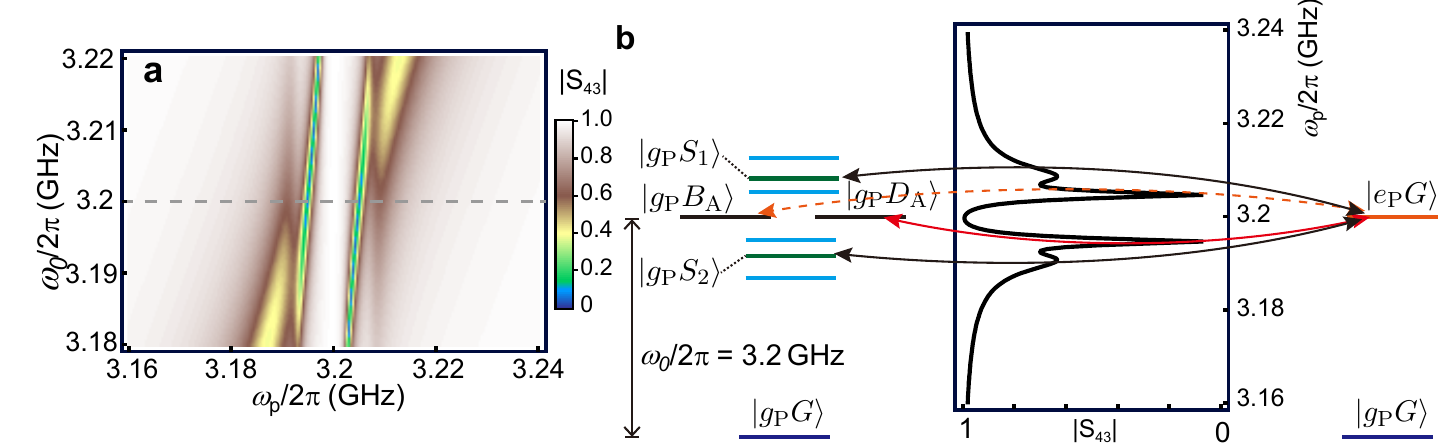}
\caption{Transmission spectrum and energy levels of a magnonic cavity with anti-Bragg mirrors each having $N_\mathrm{L} = N_\mathrm{R} = 4$ YIG spheres.
(a) Theoretical transmission spectrum as a function of $\omega_0$ and $\omega_\mathrm{p}$ in the absence of free-space radiative losses. 
(b) Left: energy diagram of the $N = 8$ anti-Bragg cavity. The notation (e.g., $\ket{g_\mathrm{P}B_\mathrm{A}}$) is explained in Fig.~2 of the main text.
Right: a line cut of (a) at the resonance frequency $\omega_0 / 2\pi = \unit[3.2]{GHz}$.
The simulations here are based on \eqref{S43_anti} and use the parameters $\kappa_\mathrm{PT} / 2\pi = \unit[8]{MHz}$, $\kappa_\mathrm{PL} / 2\pi = \unit[10]{MHz}$, $\kappa_\mathrm{M} / 2\pi = \unit[10]{MHz}$, $\alpha_\mathrm{M} / 2\pi \approx 0$, and $\alpha_\mathrm{P} / 2\pi \approx 0$.
}
\label{fig:anti_bragg_N=8}
\end{figure*}

In \figpanel{fig:anti-bragg_N}{h}, we plot the coherent and dissipative couplings between the probe YIG sphere and the eight eigenmodes of an anti-Bragg cavity with $N = 8$. 
We find that, in addition to coupling to the bright and dark supermodes, the probe YIG sphere also couples to two other collective modes. 
These additional couplings are directly responsible for the extra pair of splittings in \figpanel{fig:anti_bragg_N=8}{a}, where the full eigenvalue spectrum of the probe-cavity system is shown. 
A corresponding line cut of these splittings is provided in \figpanel{fig:anti_bragg_N=8}{b}, clearly illustrating how the probe YIG sphere couples to all four relevant eigenmodes and how these interactions manifest in the spectral features.        


\section{Experimental setup}

\begin{figure*}
\includegraphics[width=0.7\linewidth]{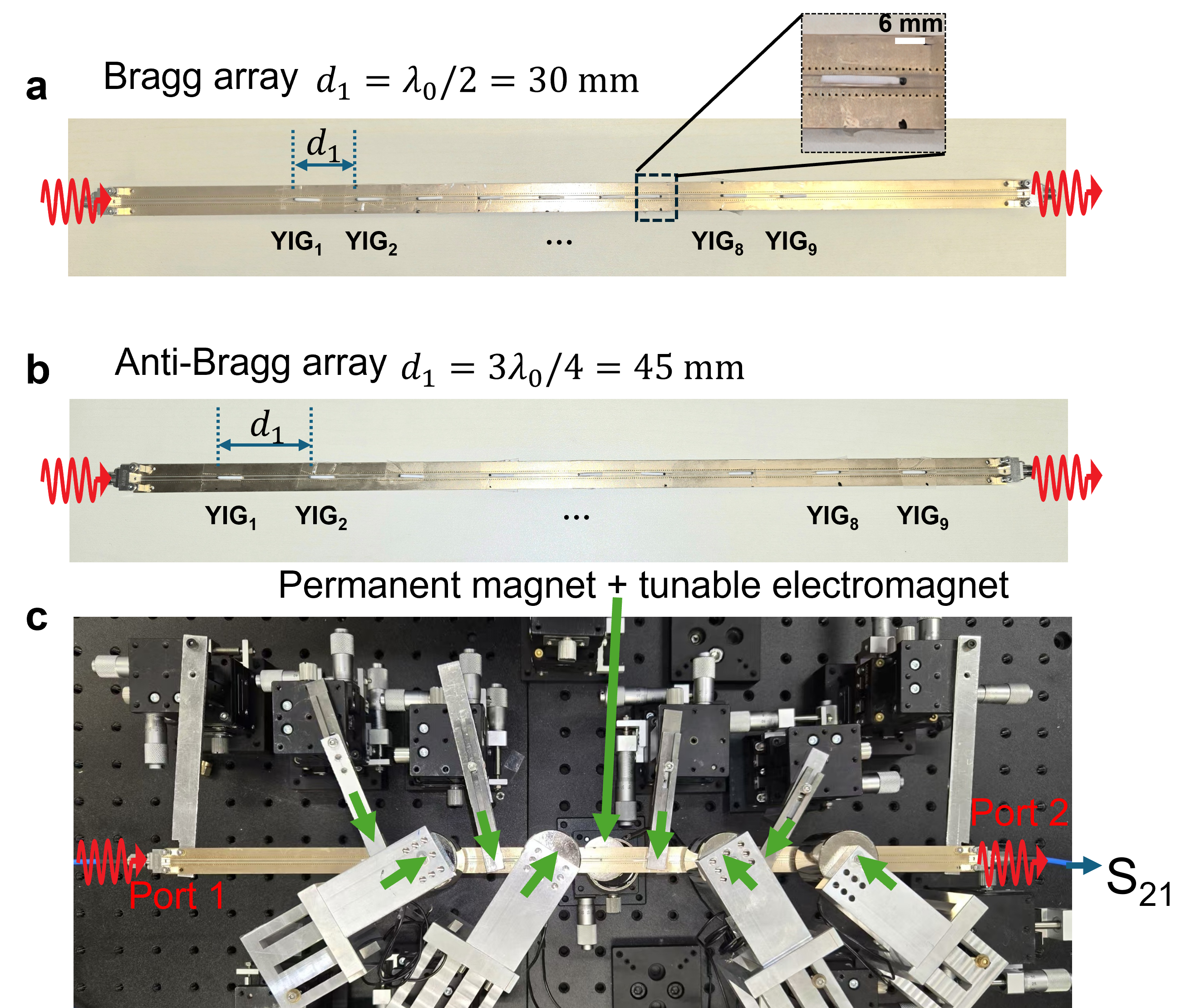}
\caption{Experimental setup for Bragg and anti-Bragg arrays.
Nine YIG spheres, numbered sequentially from left to right, are positioned on a common CPW.
Each sphere supports a KM that is inductively coupled to microwave photons propagating in the waveguide.
(a) The Bragg-array (Bragg-mirror) configuration, with an inter-sphere spacing of $d_1 = \unit[30]{mm}$.
(b) The anti-Bragg-array (anti-Bragg-mirror) configuration, with an inter-sphere spacing of $d_1 = \unit[45]{mm}$.
(c) The experimental configuration for transmission coefficient ($S_{21}$) measurements, shown with the Bragg array from panel (a) installed. This setup forms a two-port microwave network. Local magnetic flux sources (indicated by arrows) are used to tune the resonance frequencies of the KMs. The transmission spectrum is measured by detecting microwave photons sent in through Port 1 and exiting through Port 2 using a vector network analyzer (VNA).}
\label{fig:Bragg and anti-Bragg setup}
\end{figure*}

We study multiple magnon modes interacting with microwave photons in a one-dimensional (1D) open coplanar waveguide (CPW). 
The CPW, shown in \figref{fig:Bragg and anti-Bragg setup}, is designed to be $\approx \unit[450]{mm}$ long. 
It was manufactured by JX Quantum (www.jx-quantum.com), with a center conductor width of $\unit[1.5]{mm}$, a gap of $\unit[0.17]{mm}$ between the center conductor and the two ground planes, a dielectric thickness of $\unit[1.524]{mm}$, and a characteristic impedance of $Z_0 \approx \unit[50]{\Omega}$. 

We place highly polished YIG spheres on the CPW in two configurations: a Bragg array [\figpanel{fig:Bragg and anti-Bragg setup}{a}] and an anti-Bragg array [\figpanel{fig:Bragg and anti-Bragg setup}{b}]. 
In the Bragg array, the distance between adjacent YIG spheres is $d_1 = \lambda_0 / 2 = \unit[30]{mm}$, while in the anti-Bragg array it is $d_1 = 3 \lambda_0 / 4 = \unit[45]{mm}$.
The YIG spheres are held by ceramic rods, which are glued onto the CPW such that each sphere sits on the center conductor. These ceramic rods are transparent to microwave photons.

\figpanelhead{fig:Bragg and anti-Bragg setup}{c} shows the YIG spheres aligned perpendicular to the external magnetic bias field. The KM in each YIG sphere can be tuned by moving a permanent magnet using a mechanical translation stage. We probe the system's transmission coefficient from Port 1 to Port 2 using a VNA at a low probe power of $\unit[-30]{dBm}$.

\begin{figure*}
\includegraphics[width=0.7\linewidth]{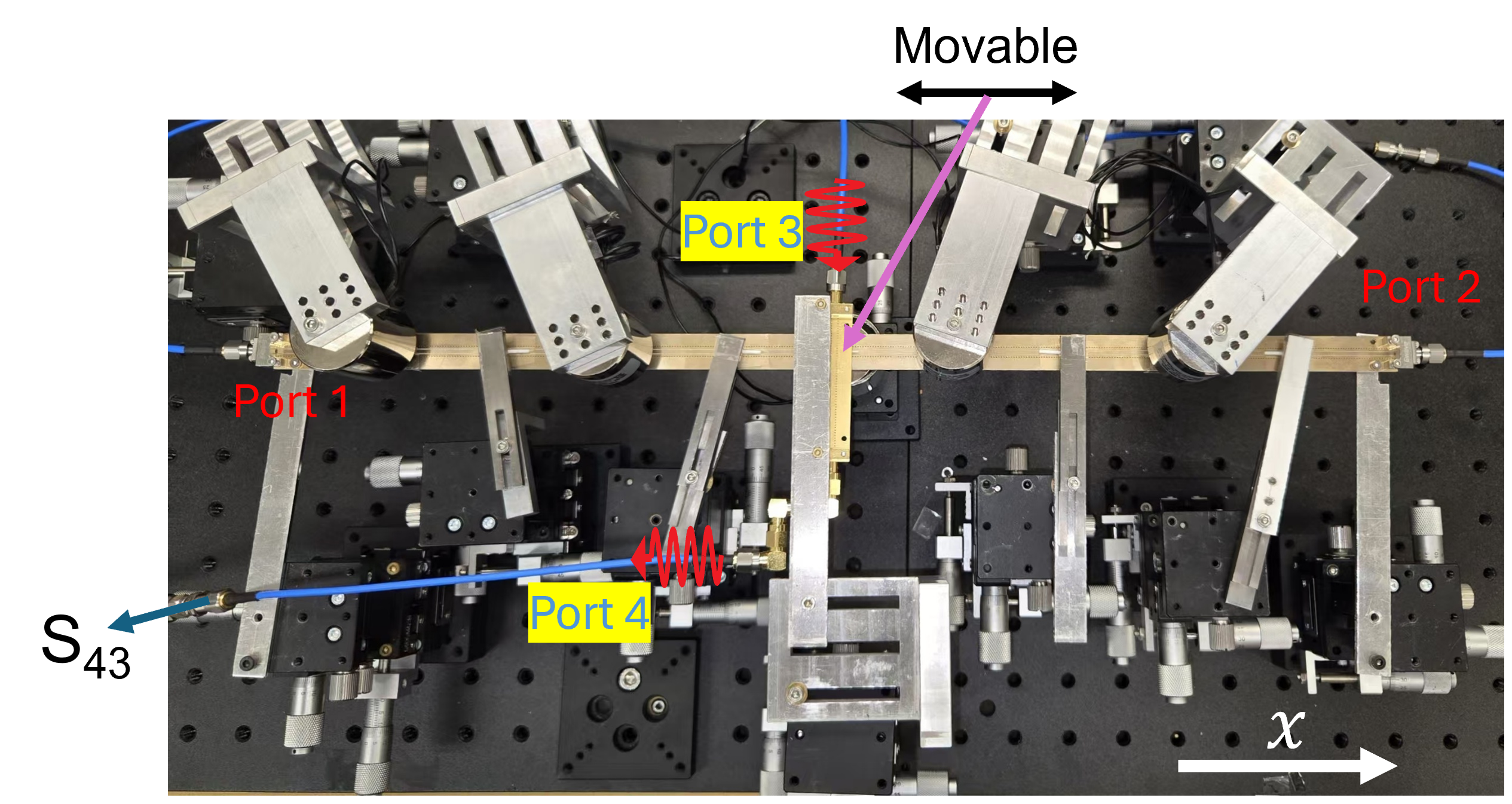}
\caption{Dual-waveguide setup for magnon-cavity interaction.
This setup builds upon the configuration shown in \figpanel{fig:Bragg and anti-Bragg setup}{c}. A second, longitudinal waveguide is positioned directly above the original transverse waveguide.
The arrangement of the YIG spheres has been reconfigured to construct a cavity. 
A probe YIG sphere is placed between the two waveguides, and its KM is referred to as the probe magnon mode (PM).
On the transverse waveguide, periodic arrays of YIG spheres are arranged on either side of the probe YIG sphere. These arrays act as Bragg (or anti-Bragg) mirrors, thereby defining a magnonic Bragg (or anti-Bragg) cavity. The configuration depicted in this figure is specifically an anti-Bragg cavity.
For transmission coefficient $S_{43}$ measurements, Ports 1 and 2 of the transverse waveguide are terminated with $\unit[50]{\Omega}$ loads.
We measure the $S_{43}$ between Port 3 and Port 4 of the longitudinal waveguide to detect the interaction between the PM and the magnonic cavity. This experimental configuration can be conceptually described by the schematic shown in Fig. 1(b) of the main text.
The longitudinal waveguide for the Bragg cavity is a CPW with a $\unit[0.813]{mm}$ dielectric thickness, a $\unit[0.63]{mm}$ center conductor width, and a $\unit[0.44]{mm}$ gap.
The longitudinal waveguide for the anti-Bragg cavity is a microstrip with a $\unit[0.813]{mm}$ dielectric thickness and a $\unit[1.83]{mm}$ center conductor width.
}
\label{fig:cavity_setup}
\end{figure*}

To further probe our systems using a probe YIG sphere, we employ the dual-waveguide setup shown in \figref{fig:cavity_setup}. When the spacing between YIG spheres in an array satisfies the Bragg (anti-Bragg) condition, we refer to the configuration as a Bragg (anti-Bragg) mirror, and the constituent YIG spheres are termed mirror YIG spheres. Two such mirrors, separated by $d_0 = 3 \lambda_0 / 2$, form a Bragg (or anti-Bragg) magnonic cavity. At the center of the cavity, we place a single YIG sphere, which we term the probe YIG sphere. A second, longitudinal waveguide is positioned above the probe YIG sphere to detect its interaction with the magnonic cavity by measuring the transmission coefficient, $S_{43}$, from Port 3 to Port 4 (see Figs.~\ref{system} and \ref{fig:cavity_setup}). 
Under resonance conditions, the radiation emitted by the probe YIG sphere excites the KMs in the mirror YIG spheres, which then interfere with each other, yielding observable features in $S_{43}$.

In the experiment, the resonance frequency of the probe magnon mode (PM) in the probe YIG sphere is tuned by varying the current in an electromagnet, while an additional permanent magnet ensures a large frequency tuning range. The resonance frequencies of the KMs in the mirror YIG spheres are fixed at \unit[3.2]{GHz} using individually applied permanent magnets. We find that sweeping the electromagnet current causes magnetic crosstalk in our setup. However, this crosstalk only affects the two mirror YIG spheres close to the probe YIG sphere. We attribute this relatively local effect to the electromagnet current being swept over only a small range around a fixed bias value. Thus, the permanent magnets far from the probe YIG sphere are unaffected, ensuring that the magnon resonance frequencies of these far-away YIG spheres remain unchanged. Based on our results, we conclude that this crosstalk has negligible effects.

The experimental setup shown in \figref{fig:cavity_setup} also enables further exploration of the properties of the Bragg and anti-Bragg magnonic cavities, through the detection of magnon-photon interactions at various locations of the probe YIG sphere. 
By mounting the longitudinal waveguide and the attached probe YIG sphere onto a high-precision displacement platform, we can translate the assembly along the transverse waveguide in the horizontal ($\pm x$) direction. 
This spatial scanning technique allows us to map the coupling strength and resonant behavior of the PM as it interacts with the magnonic cavity at different positions.

\begin{table}[]
\centering
\begin{tabular}{cccccccccc}
\hline
Parameter & YIG$_1$ & YIG$_2$ & YIG$_3$ & YIG$_4$ & YIG$_5$ & YIG$_6$ & YIG$_7$ & YIG$_8$ & YIG$_9$ \\ \hline
$\kappa / 2\pi$~(MHz) & 12.8 & 9.4 & 11.2 & 9.0 & 9.1 & 9.1 & 8.3 & 9.5 & 4.4 \\
$\Gamma / 2\pi$~(MHz) & 9.0 & 6.6 & 8.6 & 7.2 & 6.2 & 6.9 & 6.3 & 7.2 & 3.7 \\
$\alpha / 2\pi$~(MHz) & 5.2 & 3.8 & 6.0 & 5.4 & 3.3 & 4.7 & 4.3 & 4.9 & 3.0 \\
\hline
\end{tabular}
\caption{Decay parameters at the KM resonant frequency of \unit[3.2]{GHz} for each YIG sphere in the Bragg array. The average $\kappa_\mathrm{M} / 2\pi$ is \unit[9.2]{MHz} with a standard deviation of \unit[2.3]{MHz}.
}
\label{tab:Bragg_array_parameter}
\end{table}

\begin{table}[]
\centering
\begin{tabular}{cccccccccc}
\hline
Parameter & YIG$_1$ & YIG$_2$ & YIG$_3$ & YIG$_4$ & YIG$_5$ & YIG$_6$ & YIG$_7$ & YIG$_8$ & YIG$_9$ \\ \hline
$\kappa / 2\pi$~(MHz) & 9.0 & 10.0 & 9.8 & 6.1 & 10.4 & 12.4 & 8.0 & 9.3 & 8.4 \\
$\Gamma / 2\pi$~(MHz) & 7.0 & 6.8 & 7.4 & 4.9 & 9.4 & 9.1 & 6.2 & 6.7 & 6.3 \\
$\alpha / 2\pi$~(MHz) & 5.0 & 3.6 & 5.0 & 3.7 & 8.4 & 5.8 & 4.4 & 4.1 & 4.2 \\ 
\hline
\end{tabular}
\caption{Decay parameters at the KM resonant frequency of \unit[3.2]{GHz} for each YIG sphere in the anti-Bragg array. The average $\kappa_\mathrm{M} / 2\pi$ is \unit[9.3]{MHz} with a standard deviation of \unit[1.7]{MHz}.}
\label{tab:anti-Bragg_array_parameter}
\end{table}

\begin{table}[]
\centering
\begin{tabular}{ccccccccc}
\hline
Parameter & YIG$_1$ & YIG$_2$ & YIG$_3$ & YIG$_4$ & YIG$_5$ & YIG$_6$ & YIG$_7$ & YIG$_8$ \\ \hline
$\kappa / 2\pi$~(MHz) & 12.3 & 8.7 & 12.8 & 8.5 & 7.9 & 9.3 & 9.2 & 8.0 \\
$\Gamma / 2\pi$~(MHz) & 9.5 & 6.4 & 9.7 & 6.8 & 6.3 & 7.2 & 7.1 & 6.3 \\
$\alpha / 2\pi$~(MHz) & 6.7 & 4.1 & 6.6 & 5.1 & 4.7 & 5.1 & 5.0 & 4.6 \\
\hline
\end{tabular}
\caption{Decay parameters at the KM resonant frequency of \unit[3.2]{GHz} for each YIG sphere in the Bragg cavity. The average $\kappa_\mathrm{M} / 2\pi$ is \unit[9.6]{MHz} with a standard deviation of \unit[1.9]{MHz}.}
\label{tab:Bragg_cavity_parameter}
\end{table}

\begin{table}[]
\centering
\begin{tabular}{ccccccccc}
\hline
Parameter & YIG$_1$ & YIG$_2$ & YIG$_3$ & YIG$_4$ & YIG$_5$ & YIG$_6$ & YIG$_7$ & YIG$_8$ \\ \hline
$\kappa / 2\pi$~(MHz) & 8.3 & 9.2 & 11.5 & 8.5 & 5.8 & 7.5 & 13.3 & 7.9 \\
$\Gamma / 2\pi$~(MHz) & 7.0 & 6.4 & 9.8 & 7.4 & 5.2 & 7.6 & 10.0 & 7.4 \\
$\alpha / 2\pi$~(MHz) & 5.7 & 4.4 & 8.1 & 6.3 & 4.6 & 7.7 & 6.7 & 6.9 \\
\hline
\end{tabular}
\caption{Decay parameters at the KM resonant frequency of \unit[3.2]{GHz} for each YIG sphere in the anti-Bragg cavity. The average $\kappa_\mathrm{M} / 2\pi$ is \unit[9.0]{MHz} with a standard deviation of \unit[2.4]{MHz}.}
\label{tab:anti-Bragg_cavity_parameter}
\end{table}

We also measure how the KM in each YIG sphere absorbs microwaves at its own resonance frequency. 
During this measurement, we make sure the modes in all the other YIG spheres are set to far-detuned frequencies so they do not interfere. 
From this type of measurement, we can extract the radiative decay rate ($\kappa$) and the total decoherence rate ($\Gamma$) for each YIG sphere's KM.
For data taken at low probe power ($\unit[-30]{dBm}$), we employ a circle-fitting routine to fit the measured complex transmission coefficient of the magnon response~\cite{wu2024microwaveinterferencespinensemble}. This fit gives direct access to the rates $\kappa$ and $\Gamma$, as well as the intrinsic loss rate, $\alpha$.
We summarize, in four tables, the extracted KM parameters for each YIG sphere of diameter \unit[1.2]{mm} in the various configurations: the Bragg array (\tabref{tab:Bragg_array_parameter}), the anti-Bragg array (\tabref{tab:anti-Bragg_array_parameter}), the Bragg cavity (\tabref{tab:Bragg_cavity_parameter}), and the anti-Bragg cavity (\tabref{tab:anti-Bragg_cavity_parameter}).
The average $\kappa / 2\pi$ values determined at a resonance frequency of \unit[3.2]{GHz} are: $\kappa_\mathrm{M} = \unit[9.2 \pm 2.3]{MHz}$ for the Bragg array, $\unit[9.3 \pm 1.7]{MHz}$ for the anti-Bragg array, $\unit[9.6 \pm 1.9]{MHz}$ for the Bragg cavity, and $\unit[9.0 \pm 2.4]{MHz}$ for the anti-Bragg cavity.


\section{Additional results for Bragg and anti-Bragg arrays}

\begin{figure*}
\includegraphics[width=\linewidth]{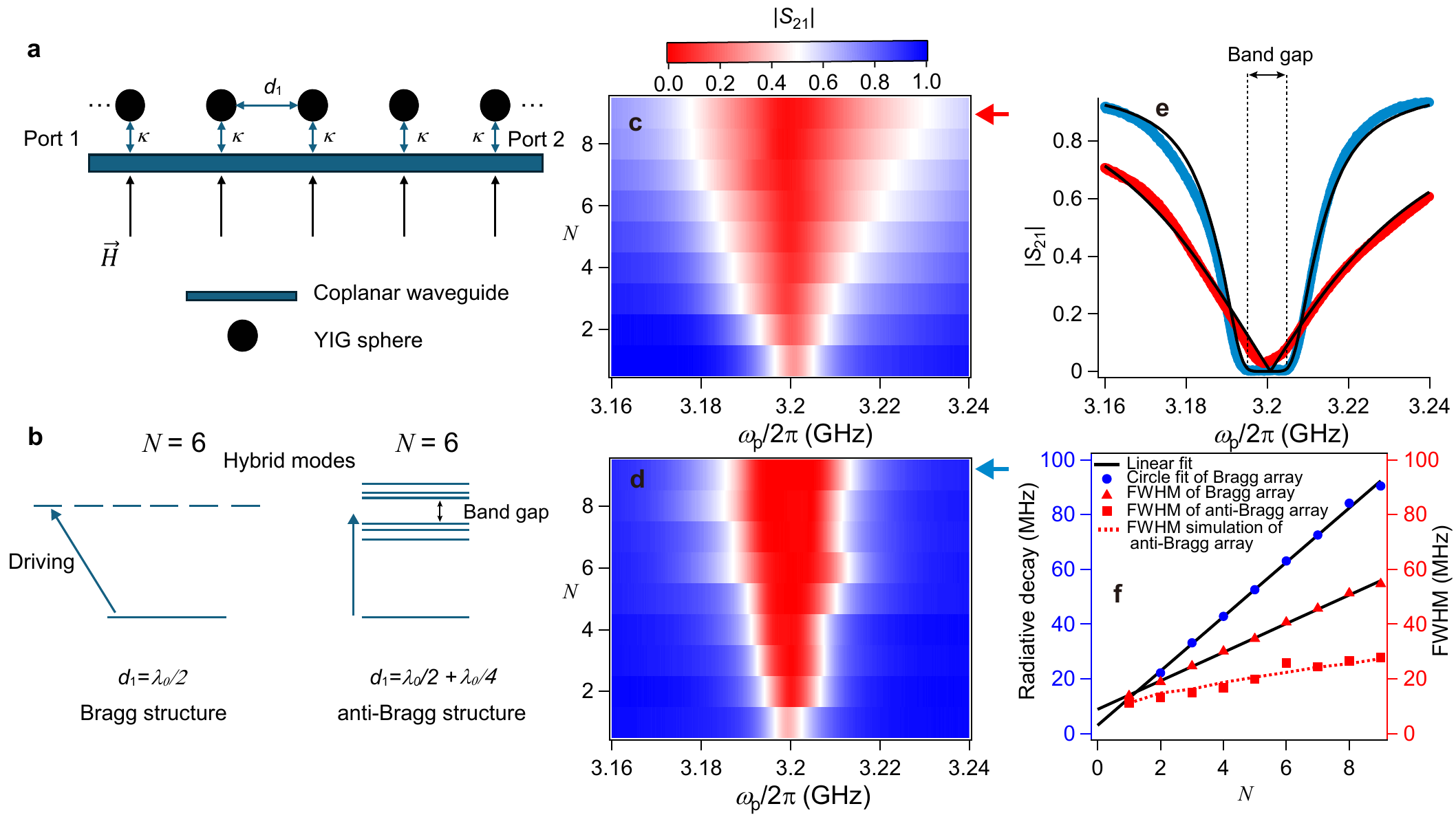}
\caption{Theory and experimental results for Bragg and anti-Bragg arrays.
(a) A simplified schematic of the experimental setup depicted in \figpanel{fig:Bragg and anti-Bragg setup}{c}. 
Each YIG sphere's KM couples to the waveguide with a radiative decay rate of $\kappa$.
(b) Eigenfrequency diagrams of the Bragg and anti-Bragg arrays.
For the Bragg array (left), the $N$ magnon modes hybridize into degenerate levels, forming one superradiant mode (which can be excited through the waveguide, as indicated by the blue arrow) and $N - 1$ subradiant modes (decoupled from the waveguide).
For the anti-Bragg array (right), the $N$ magnon modes hybridize into distinct levels.
In the limit of infinite $N$, the discrete polaritons form two continuous polaritonic branches, separated by a band gap~\cite{Brehm2021_1000130266}.
(c) Measured $|S_{21}|$ spectra for the Bragg array, plotted as a function of $N$ and probe frequency $\omega_\mathrm{p}$. The KMs of the YIG spheres are successively tuned into resonance at \unit[3.2]{GHz} as $N$ increases.
(d) Same as (c), but for the anti-Bragg array.
(e) Line cuts for $N = 9$, taken at the positions indicated by the colored arrows in (c) and (d). The red and blue data correspond to the Bragg and anti-Bragg arrays, respectively. The black curve overlaying the red data is a fit using the circle-fit method to extract parameters (e.g., radiative decay), which is suitable for the Lorentzian line shape.
The black curve for the blue data is a fit based on the transfer-matrix method.
(f) Extracted parameters as a function of the number of YIG spheres ($N$).
The blue dots show the radiative decay (left y axis) of the visible superradiant mode.
The red triangles and squares represent the full width at half maximum (FWHM, right y axis) for the data in (c) and (d), respectively.
The black curves are linear fits, while the dashed curve shows the simulated FWHM obtained from the transfer-matrix method.
}
\label{fig:Bragg and anti-Bragg mirror_result}
\end{figure*}

In this section, we provide further experimental and theoretical insights into the mode properties of Bragg and anti-Bragg arrays in the low-excitation limit (relative to the thermal equilibrium) and their dependence on the number of magnon modes $N$. Throughout this section, magnon modes are tuned to a common resonance frequency of \unit[3.2]{GHz}. \figpanelhead{fig:Bragg and anti-Bragg mirror_result}{a} provides a simplified schematic of the experimental setup used, which was shown in full in \figpanel{fig:Bragg and anti-Bragg setup}{c}.

For the Bragg array, solving \eqref{eq_Heff} without a probe YIG sphere shows that the collective modes become degenerate [see the left part of \figpanel{fig:Bragg and anti-Bragg mirror_result}{b}], forming a single bright mode and $N - 1$ dark modes. The bright mode can be observed in the $S_{21}$ spectrum, while the dark modes are ideally decoupled from the waveguide. In contrast, for the anti-Bragg array [see the right panel of \figpanel{fig:Bragg and anti-Bragg mirror_result}{b}], the modes are non-degenerate, forming multiple energy levels separated by a frequency gap.

\figpanelhead{fig:Bragg and anti-Bragg mirror_result}{c} shows experimental results for the Bragg array. The transmission spectrum ($|S_{21}|$) shows the bright mode manifesting as a broadened transmission dip (red), whose linewidth increases with $N$. The resonant dip exhibits a Lorentzian line shape, as seen in the corresponding line cut (red) in \figpanel{fig:Bragg and anti-Bragg mirror_result}{e}. A circle fit [black curve in \figpanel{fig:Bragg and anti-Bragg mirror_result}{e}] reveals that the bright mode of the nine-magnon-mode array has a collective decay rate of $\kappa_\mathrm{B} = \unit[90.4]{MHz}$, approximately nine times the average value of a single magnon mode ($\kappa_\mathrm{M} / 2\pi = 9.2 \pm \unit[2.3]{MHz}$). 

By fitting the resonance dips for different $N$, we extract $\kappa_\mathrm{B}$. As plotted in \figpanel{fig:Bragg and anti-Bragg mirror_result}{f}, the extracted $\kappa_\mathrm{B}$ confirms that the collective decay rate increases linearly with $N$ (black curve and blue data points, left y axis). For the anti-Bragg array, the transmission spectrum exhibits a flat-bottomed stopband as $N$ increases [\figpanel{fig:Bragg and anti-Bragg mirror_result}{d}], a clear signature of an emerging polariton band gap. 

We use a consistent definition of the full width at half maximum (FWHM) to compare the spectral features of the Bragg and anti-Bragg arrays. As depicted in \figpanel{fig:Bragg and anti-Bragg mirror_result}{f}, the FWHMs for both configurations exhibit a linear dependence on $N$, but with distinct slopes (red markers, right y axis). It should be noted that this FWHM includes the intrinsic loss of each magnon mode. Although the FWHM in the Bragg array is numerically different from the radiative decay rate ($N \kappa_\mathrm{M}$), both quantities share the same linear dependence on $N$. We also simulate the FWHM's dependence on $N$ for the anti-Bragg array using the transfer-matrix method~\cite{Brehm2021_1000130266} with the parameters $\kappa / 2\pi = \unit[9.3]{MHz}$, $\Gamma / 2\pi = \unit[7.1]{MHz}$, and $\alpha / 2\pi = \unit[4.9]{MHz}$. The resulting theoretical curve [dashed line in \figpanel{fig:Bragg and anti-Bragg mirror_result}{f}] shows excellent agreement with the extracted data.

\begin{figure*}
\includegraphics[width=0.8\linewidth]{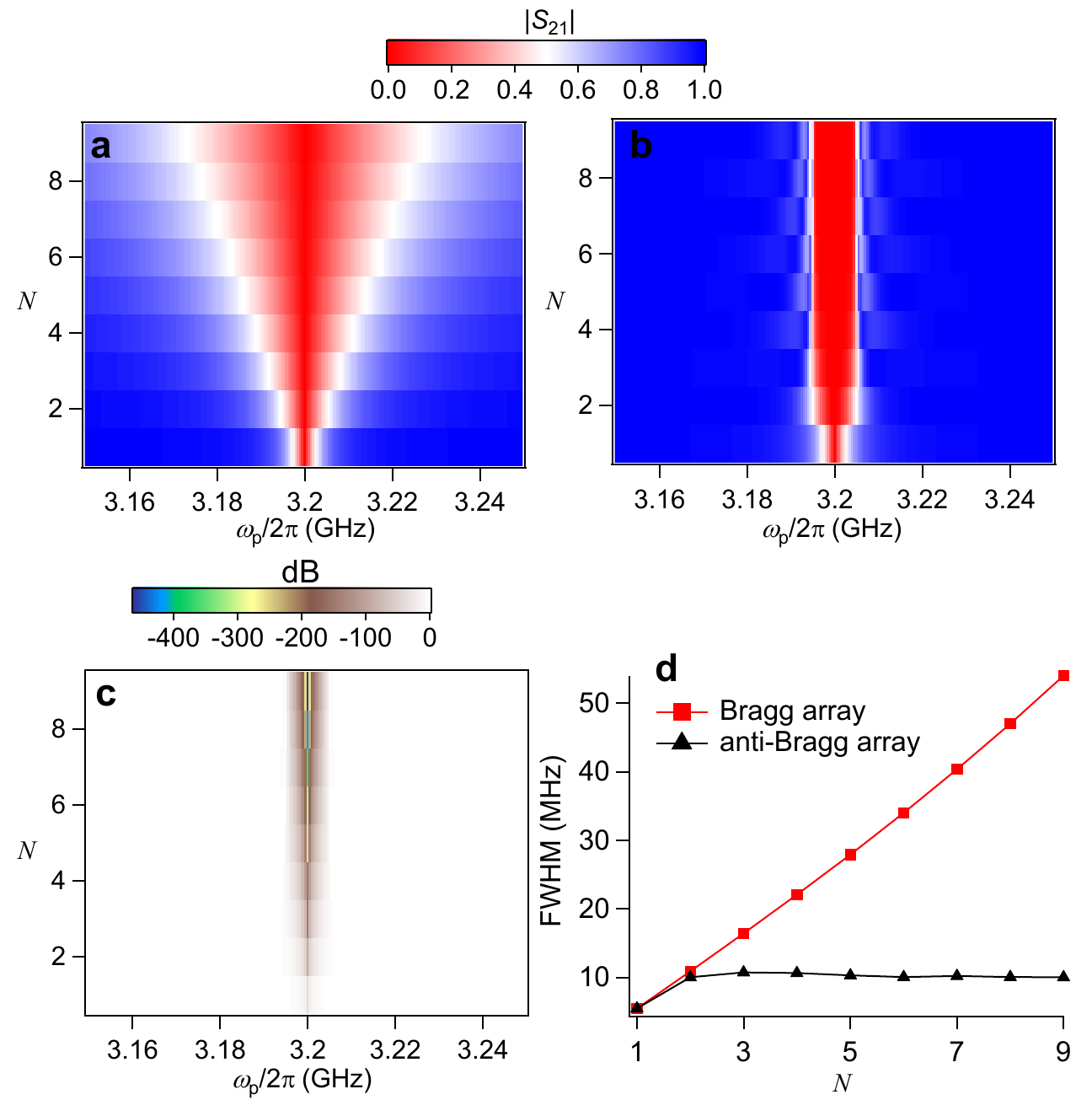}
\caption{Simulation results for low intrinsic loss.
(a) Transmission spectrum of the Bragg arrays, simulated using the transfer-matrix model, as a function of the probe frequency $\omega_\mathrm{p}$ and the number of YIG spheres, $N$. For these simulations, each KM is assigned a radiative decay rate of \unit[9.3]{MHz} and a negligible intrinsic loss of \unit[0.05]{MHz}, resulting in a total decoherence rate of \unit[4.7]{MHz}.
(b) Same as (a), for anti-Bragg arrays. 
(c) Same as (b), but plotted with a dB scale for the transmission instead of a linear scale.
(d) The FWHM of the Bragg and anti-Bragg arrays as a function of $N$, extracted from the simulated spectra in (a) and (b).
}
\label{fig:Bragg and anti_Bragg sim}
\end{figure*}

We further perform transfer-matrix simulations for both types of arrays as a function of $N$, assuming that the intrinsic loss is close to zero.
Specifically, we set the radiative decay rate to $\kappa / 2\pi = \unit[9.3]{MHz}$, the total decay rate to $\Gamma / 2\pi = \unit[4.7]{MHz}$, and use an intrinsic loss of $\alpha / 2\pi = \unit[0.05]{MHz}$. For the Bragg array, the simulated results in \figpanel{fig:Bragg and anti_Bragg sim}{a} show that the absorption dips broaden as $N$ increases. In contrast, for the anti-Bragg array, the simulations reveal that the frequency gap ceases to expand once $N > 2$ [\figpanel{fig:Bragg and anti_Bragg sim}{b}], indicating the formation of a band gap. Instead, for larger $N$, side peaks emerge on either side of the band gap. These side peaks correspond to polaritons, which are associated with the real parts of the eigenfrequencies of the effective Hamiltonian~\cite{Brehm2021_1000130266}. The width of the non-expanding band gap is $\approx \unit[9.5]{MHz}$ [see \figpanel{fig:Bragg and anti_Bragg sim}{b}], which is approximately equal to the radiative decay rate of a single magnon mode. However, in the dB-scale spectrum [\figpanel{fig:Bragg and anti_Bragg sim}{c}], the dip around the \unit[3.2]{GHz} resonance becomes progressively deeper as $N$ increases. This is because the anti-Bragg condition leads to destructive interference in the forward direction of each KM's radiation.


\section{Additional results for Bragg and anti-Bragg cavities}
\label{Additional results}

This section provides experimental and fitting data to support the results on Bragg and anti-Bragg cavities presented in the main text. 


\subsection{Coupling between the probe magnon mode and a Bragg cavity}

\begin{figure*}
\includegraphics[width=\linewidth]{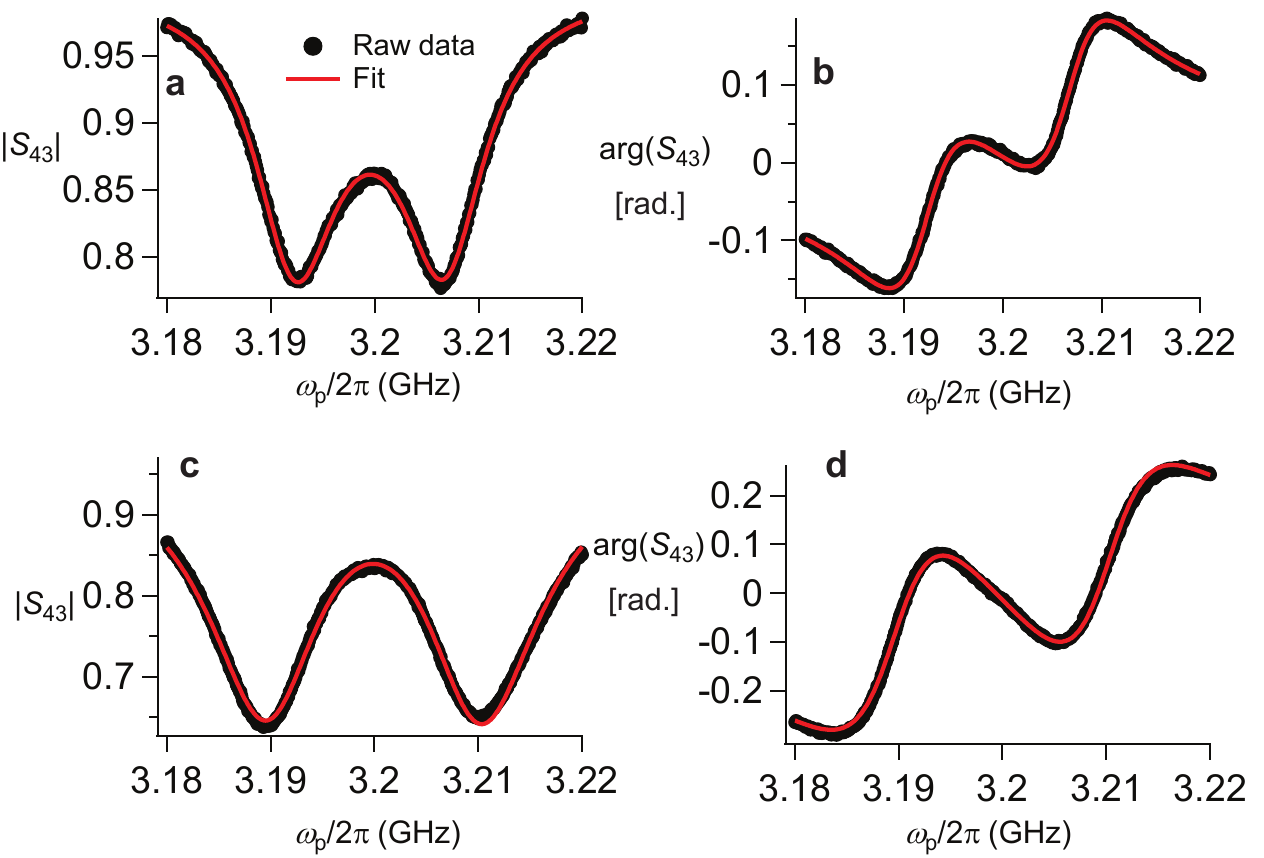}
\caption{Transmission spectra and extracted coupling strengths for a probe magnon mode (PM) coupled to a Bragg cavity.
(a) Line cut (black) from Fig.~1(e) in the main text and (b) the corresponding phase response, taken when all KMs in the constituent mirror YIG spheres are tuned into resonance. 
The solid red curves are fits to the experimental data using \eqref{S_43_derived}.	
(c,~d) Same as panels (a) and (b), respectively, but for a line cut from Fig.~1(f) in the main text instead.}
\label{fig:on_resonant_linecut}
\end{figure*}

\figpanelshead{fig:on_resonant_linecut}{a}{b} show horizontal cross-sections of the transmission magnitude ($|S_{43}|$) and phase response at the resonance condition depicted in Fig.~1(e) in the main text. 
This corresponds to the case where each mirror consists of a single mirror YIG sphere ($N_\mathrm{L} = N_\mathrm{R} = 1$). Here, a probe YIG sphere with a diameter of \unit[0.5]{mm} is used to probe the exchange interaction with the dark supermode. The solid red lines in \figpanels{fig:on_resonant_linecut}{a}{b} are fitted to the data using \eqref{S_43_derived} with the parameters $\kappa_\mathrm{PM} / 2\pi = \unit[4.0]{MHz}$, $\Gamma_\mathrm{PM} / 2\pi = \unit[5.5]{MHz}$, $\Gamma_\mathrm{C} / 2\pi = \unit[2.7]{MHz}$, $\omega_0/2\pi=\omega_D/2\pi=\unit[3.2]{GHz}$, and $2 J_\mathrm{D} / 2\pi = \unit[6.8]{MHz}$. The good agreement between the fitted curve and the experimental data indicates that the system can be well described by the coupled-mode model.

Next, we analyze the horizontal cross-section at the resonance condition of Fig.~1(f) in the main text (where $N_\mathrm{L} = N_\mathrm{R} = 4$, such that we have eight mirror YIG spheres and one \unit[1.0]{mm} probe YIG sphere), as shown in \figpanels{fig:on_resonant_linecut}{c}{d}. From the fitting, we extract the parameters $\kappa_\mathrm{PM} / 2\pi = \unit[10]{MHz}$, $\Gamma_\mathrm{PM} / 2\pi = \unit[10.1]{MHz}$, $\Gamma_\mathrm{C} / 2\pi = \unit[4.7]{MHz}$, and $2 J_\mathrm{D} / 2\pi = \unit[20.4]{MHz}$. The coherent coupling strength is thus enhanced by approximately a factor of three compared to the previous case. This enhancement occurs because the larger probe YIG sphere provides a greater number of spins, resulting in a larger $\kappa_\mathrm{PT}$, while an increased number of mirror YIG spheres provides a $\sqrt{N}$ collective enhancement effect, thereby strengthening the overall coupling.


\subsection{Transmission spectra with the mirror magnons far detuned}

\begin{figure*}
\centering
\includegraphics[width=0.8\linewidth]{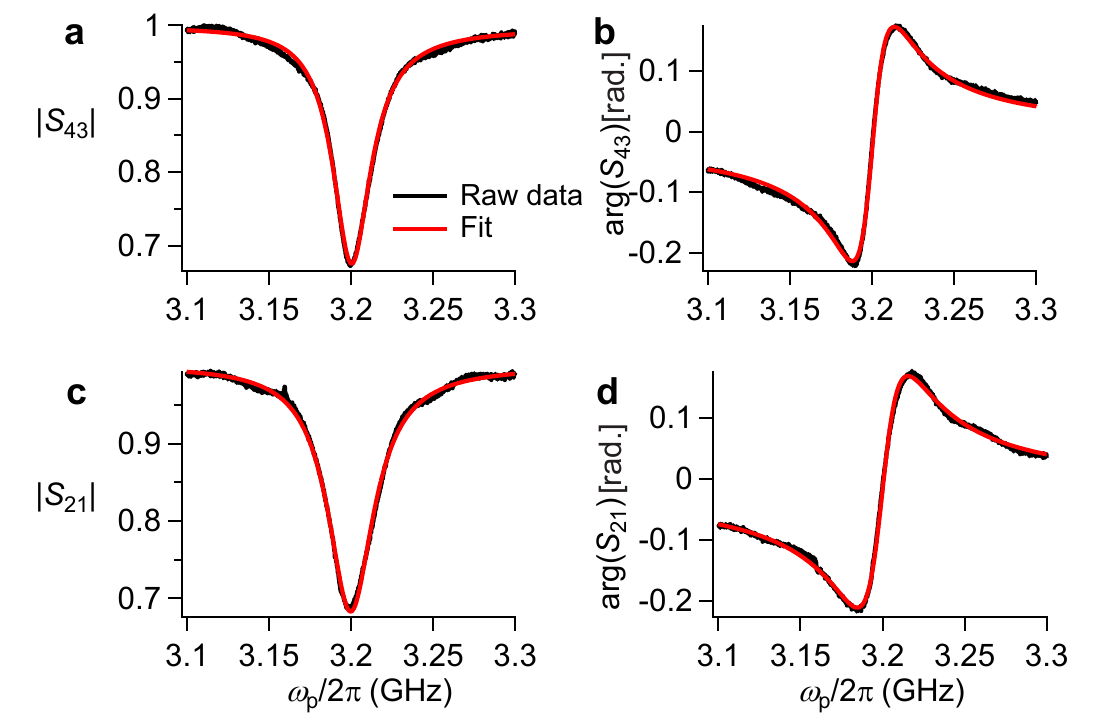}
\caption{Transmission spectra for the PM through both waveguides. 
(a) Transmission magnitude $|S_{43}|$ and (b) corresponding phase response arg$(S_{43})$ as a function of probe frequency $\omega_\mathrm{p}$.
(c,~d) Same as panels (a) and (b), respectively, but for $S_{21}$ instead of $S_{43}$.
Black points are data and red lines are fits using the circle-fit method.
The mirror magnon modes are far detuned from the PM.
The probe YIG sphere has a diameter of \unit[1.2]{mm}.
\label{fig:anti-Bragg_probe_top_bottom}}
\end{figure*}

\figrefhead{fig:anti-Bragg_probe_top_bottom} shows the transmission spectra of the PM in a $\unit[1.2]{mm}$ probe sphere coupling to both the transverse and longitudinal waveguides when the mirror magnon modes are far detuned.
The longitudinal waveguide is the microstrip, and the transverse waveguide is the CPW (see \figref{fig:cavity_setup}).
The parameters extracted from the $S_{43}$ data using the circle-fit method~\cite{ProbstReview} are $\kappa_\mathrm{PL} / 2\pi = \unit[10.0]{MHz}$ and $\Gamma_\mathrm{PM} / 2\pi = \unit[15.5]{MHz}$ [\figpanels{fig:anti-Bragg_probe_top_bottom}{a}{b}]. 
The parameters extracted from the $S_{21}$ data are $\kappa_\mathrm{PT} / 2\pi = \unit[11.9]{MHz}$ and $\Gamma_\mathrm{PM} / 2\pi = \unit[18.7]{MHz}$ [\figpanels{fig:anti-Bragg_probe_top_bottom}{c}{d}].

In the main text, we use \eqref{S43_anti} to fit the $S_{43}$ data in Fig.~2(g) when the mirror magnon modes are on resonance with the PM. 
From this fitting, we obtain $\kappa_\mathrm{PL} / 2\pi = \unit[10.0]{MHz}$ and $\kappa_\mathrm{PT} / 2\pi = \unit[8.0]{MHz}$.  
We observe that the $\kappa_\mathrm{PL}$ value obtained in this way is comparable to the value obtained when the mirror magnon modes are far detuned, but the $\kappa_\mathrm{PT}$ value is clearly different.
We attribute this discrepancy in $\kappa_\mathrm{PT}$ to the modification of the electromagnetic environment in the transverse waveguide. 
Specifically, resonant interactions among magnon modes and radiation interference alter this environment, thereby affecting $\kappa_\mathrm{PT}$.


\subsection{Dissipative coupling}

\begin{figure*}
\includegraphics[width=0.9\linewidth]{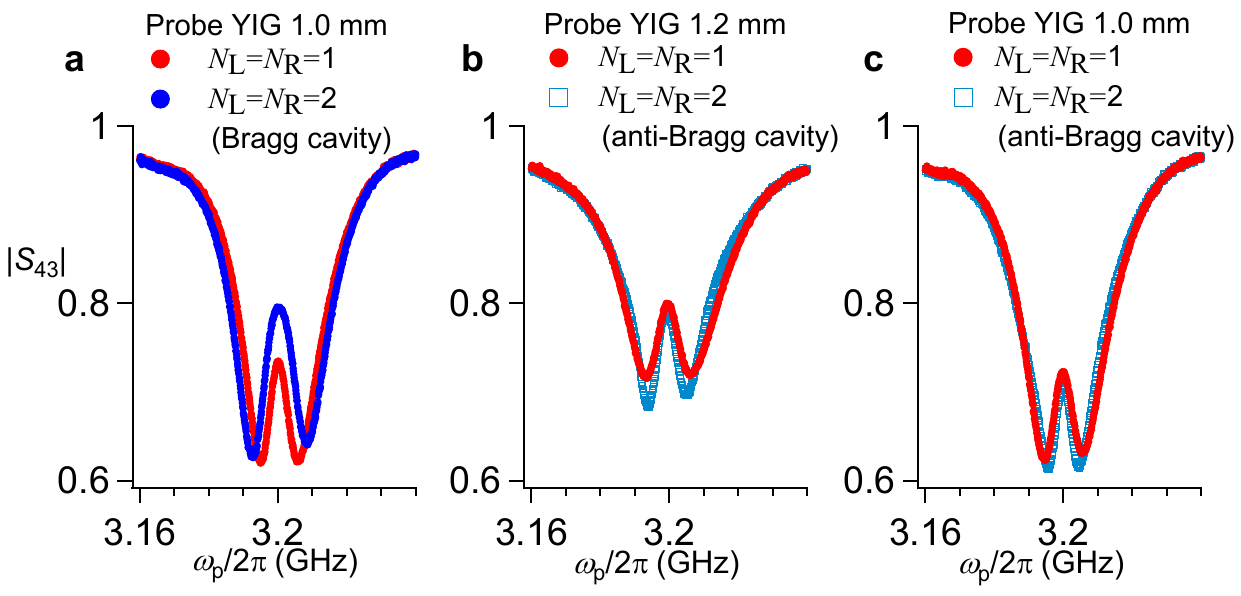}
\caption{Measured transmission spectra of Bragg and anti-Bragg cavities with one or two YIG spheres in each mirror.
(a) Measured transmission coefficient $|S_{43}|$ for $N_\mathrm{L} = N_\mathrm{R} =1$ (red) and 2 (blue) in a Bragg cavity with a \unit[1.0]{mm} probe YIG sphere.
All magnon modes are resonant at \unit[3.2]{GHz}.
(b) Measured $|S_{43}|$ for $N_\mathrm{L} = N_\mathrm{R} = 1$ (red) and 2 (teal) in an anti-Bragg cavity with a \unit[1.2]{mm} probe YIG sphere.
(c) Same as (b), but with a \unit[1.0]{mm} probe YIG sphere.}
\label{fig:dissipative_coupling}
\end{figure*}

Next, we discuss the distinct behavior of Bragg and anti-Bragg cavities when the number of left and right mirror YIG spheres is set to $N_\mathrm{L} = N_\mathrm{R} = 2$, with a particular focus on the emergence of dissipative coupling.
For the Bragg cavity shown in \figpanel{fig:dissipative_coupling}{a}, the two dips in the transmission spectrum for the $N_\mathrm{L} = N_\mathrm{R} = 2$ case (blue) are shallower than those for the $N_\mathrm{L} = N_\mathrm{R} = 1$ case (red). In stark contrast, for the anti-Bragg cavity [\figpanel{fig:dissipative_coupling}{b}], the two transmission dips are observed to be deeper in the $N_\mathrm{L} = N_\mathrm{R} = 2$ configuration (teal) when the PM is resonant with the cavity.
We interpret this deepening of the transmission dips as a signature of dissipative coupling between the PM and the anti-Bragg cavity's bright supermode. Here, ``dissipative coupling'' refers to the $\gamma_{j, \mathrm{P}}$ terms in the effective Hamiltonian \eqref{eq_Heff}, whose explicit form is given in \eqref{dissipation_rates}: $\gamma_{j, \mathrm{P}} \propto \cos(2\pi |x_j - x_\mathrm{P}| / \lambda_0)$. In contrast to the coherent exchange coupling $J_{j, \mathrm{P}} \propto \sin(2\pi |x_j - x_\mathrm{P}| / \lambda_0)$, which causes level repulsion, a non-zero $\gamma_{j, \mathrm{P}}$ induces level attraction between the coupled modes.
The underlying mechanism involves two steps: first, the dissipative coupling synchronizes the phases of the two modes to cause level attraction~\cite{Wang_2020_JAP}; second, the bright supermode of the anti-Bragg cavity exhibits enhanced radiative decay into the transverse waveguide [see how $\kappa_\mathrm{M}$ of mode index 1 varies as $N$ increases in \figpanels{fig:anti-bragg_N}{a}{e}]. 
To rule out the possibility that the observed dip deepening is an artifact tied to the specific probe geometry---such as an accidental size-dependent resonance or a direct geometrical interference effect---we repeated the experiment using a probe YIG sphere with a different diameter (\unit[1.0]{mm}); see \figpanel{fig:dissipative_coupling}{c}.
The \unit[1.0]{mm} probe exhibits the same qualitative behavior: deeper transmission dips in the $N_\mathrm{L} = N_\mathrm{R} = 2$ configuration compared to $N_\mathrm{L} = N_\mathrm{R} = 1$.
This rules out probe-specific artifacts and confirms that the dip deepening is an intrinsic consequence of the dissipative coupling mechanism $\gamma_{j, \mathrm{P}}$ in the anti-Bragg cavity.

\begin{figure}
\centering
\includegraphics[width=\linewidth]{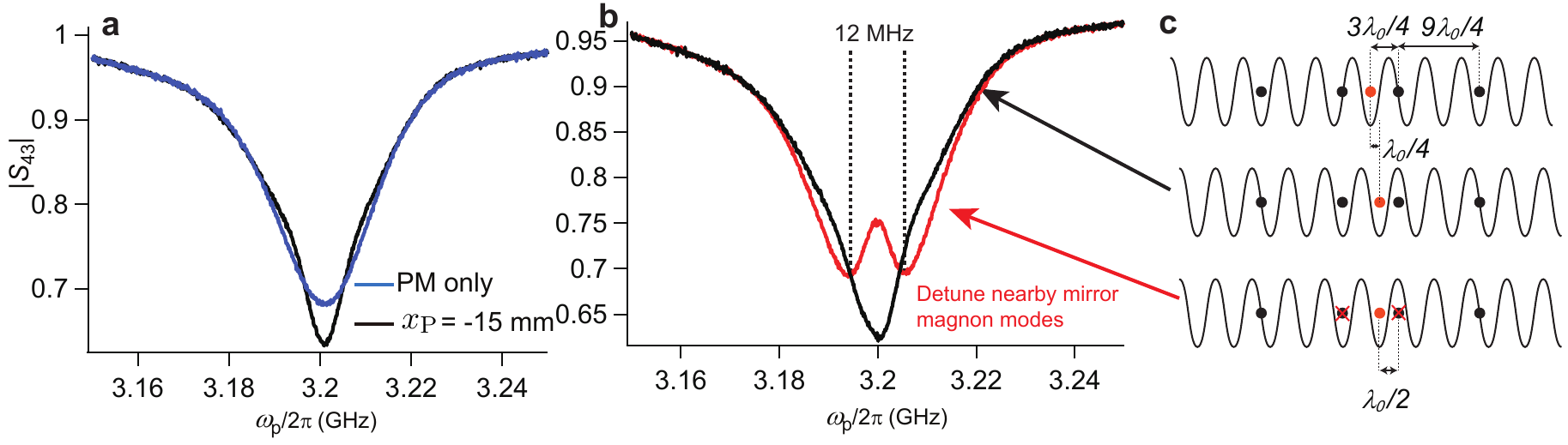}
\caption{Effect of probe position and mirror detuning in an anti-Bragg cavity.
(a) The transmission magnitude response $|S_{43}|$ as a function of probe frequency $\omega_\mathrm{p}$ for the probe YIG sphere at $x_\mathrm{P} = \unit[-15]{mm}$ and the bare PM response.
(b) $|S_{43}|$ as a function of probe frequency $\omega_\mathrm{p}$. 
(c) Schematic diagrams of the anti-Bragg cavity, illustrating different probe YIG sphere positions and conditions. 
Top: The probe YIG sphere (orange sphere) is positioned at the center of the cavity.
Middle: The probe YIG sphere is moved $\lambda_0 / 4$ ($x_\mathrm{P} = \unit[15]{mm}$) to the right of the center of the cavity. 
Bottom: The magnon modes of adjacent mirror YIG spheres are far detuned.
The black and red arrows link the spectral features in (b) to their corresponding physical configurations in (c). When the magnon modes of the adjacent mirror YIG spheres are far detuned (bottom), an anti-crossing reappears in the transmission spectrum.
}
\label{fig:anti_bragg_far_detune}
\end{figure}

Moreover, for the anti-Bragg cavity, when the probe sphere is located at $x_\mathrm{P} = \unit[-15]{mm}$, the PM decouples from the dark and bright supermodes but couples dissipatively and coherently to other subradiant supermodes; see \figpanels{fig:moving_anti_bragg}{a}{b}.
This leads to the observation of a reduced transmission dip in \figpanel{fig:anti_bragg_far_detune}{a} compared to the bare PM response.
Furthermore, the exchange interaction $J_{j, \mathrm{P}}$---the coherent coupling between the probe and the $j$th mirror YIG sphere introduced in \eqref{coherent_dis_1}---can be recovered by detuning the nearby mirror magnon modes far from resonance; see the red curve in \figpanel{fig:anti_bragg_far_detune}{b} for the probe YIG sphere at $x_\mathrm{P} = \unit[15]{mm}$.
Because the distances between the probe YIG sphere and the distant mirror YIG spheres are $13\lambda_0/4$ and $11\lambda_0/4$ [the bottom part of \figpanel{fig:anti_bragg_far_detune}{c}], the propagating microwave between the probe YIG sphere and the mirror YIG spheres mediates the exchange interaction $J_{j, \mathrm{P}}$ only when the magnon modes of the adjacent mirror YIG spheres are far detuned.
This recovery is impossible in the Bragg cavity at the same probe-YIG-sphere position, since the distance between the probe YIG sphere and the mirror YIG spheres is $n \lambda_0 / 2$ ($n$ is an integer), resulting in only dissipative coupling.


\section{Comparison with other cavity-based and free-space platforms}

\begin{table}
\centering
\begin{tblr}{
width = \linewidth,
colspec = {X[0.75,m] X[1.5,m] X[1.5,m] X[1.5,m] X[1.5,m] X[1.5,m]},  
cell{1-6}{1} = {font=\bfseries},
hlines,
vlines
}

Setup 
& Dual-waveguide configuration 
& 3D metal cavity 
& {Single 1D \\ waveguide (CPW)}
& Fabry--P\'erot cavity 
& {Photonic crystal \\ slab (PhC slab)} \\  

Field modes 
& Continuum modes 
& {Single-mode \\ standing wave} 
& Continuum modes 
& {Single-mode \\ standing wave} 
& Continuum modes (TE/TM-like) \\   

{Coupling \\ objects}
& YIG spheres
& A single YIG sphere
& {Transmons \\ (superconducting qubits)~\cite{Mirhosseini2019}, \\ YIG spheres~\cite{Wang2026-aea6000}}
& {Atom array \\ (trapped by \\ optical tweezers)}
& {Square lattice \\ of holes in the \\ PhC slab} \\

Coupling mediator
& Traveling photons
& Cavity photons
& Traveling photons
& Cavity photons
& Traveling photons \\

{Coupling \\ charac- \\ teristics \\ and key \\ findings}

& {We implement two distinct magnonic cavities within the dual-waveguide setup by precisely arranging multiple YIG spheres. \\
We demonstrate collective magnon-photon $\sqrt{N}$ enhancement in the Bragg cavity [Fig.~1(g)]. \\
In the anti-Bragg cavity, the $\sqrt{N}$ scaling breaks down [Fig.~2(f)]. \\
The anti-Bragg cavity behaves as a multi-level cavity [Fig.~2(d,e)]. \\
We probe field properties of BICs in both cavity types via a moving probe YIG sphere, a unique advantage of this platform [Fig.~3, Fig.~4]. \\
}

& {Magnon-photon coupling is enhanced by physically reducing the cavity size~\cite{ZhangPhysRevLett} or by using a larger YIG sphere (increasing the number of interacting spins)~\cite{TabuchiPhysRevLett}.}

& {In Ref.~\cite{Mirhosseini2019}: \\
Two transmons separated by $\lambda / 2$ act as resonant mirrors to form an atomic cavity. \\
A probe qubit at the center exhibits strong exchange coupling with this atomic cavity. \\ \\
In Ref.~\cite{Wang2026-aea6000}: \\
A reflectionless exceptional point is realized by engineering collective states in an anti-Bragg magnonic mirror array. \\
An extension to a multi-emitter cavity demonstrating $\sqrt{N}$ scaling is currently absent.}

& {Experimentally tests light-matter interaction by verifying the $\sqrt{N}$ scaling with the number of atoms in the array. \\
Atoms are confined at the cavity antinodes. \\
Other array configurations remain unexplored.}

& {Experimentally realizes a single-resonance parametric BIC in the PhC slab. \\
Due to the absence of a suitable probe to interact with the BIC, the spatial field distribution of the trapped-state standing wave remains experimentally unverified.}
 \\   

References 
& This work 
& Refs.~\cite{TabuchiPhysRevLett, ZhangPhysRevLett} 
& Refs.~\cite{Mirhosseini2019, Wang2026-aea6000} 
& Ref.~\cite{LiuPhysRevLett} 
& Ref.~\cite{Hsu2013} \\

\end{tblr}
\caption{Comparison of our dual-waveguide system with other representative experimental platforms for light-matter interaction.}
\label{tab:comparison}
\end{table}

To compare our dual-waveguide magnonic system with other representative experimental platforms, we summarize the key distinctions in \tabref{tab:comparison}. 
In the dual-waveguide configuration, the spatial arrangement of mirror YIG spheres along the transverse waveguide plays a crucial role in engineering collective magnon-photon interactions with the probe YIG sphere. 
This configuration offers two primary advantages over traditional cavity-based and open-waveguide systems: positional flexibility and the ability to explore richer collective light-matter interactions.

Unlike conventional cavity magnonic systems~\cite{TabuchiPhysRevLett, ZhangPhysRevLett}, our scheme does not require modification of a 3D metallic cavity to achieve strong magnon-photon coupling. 
In such cavity systems, the cavity frequency is fixed once the structure is fabricated. 
In contrast, in our platform, the interaction frequency can be flexibly selected by positioning the YIG spheres at appropriate locations within the Bragg cavity, where the distance between adjacent YIG spheres satisfies $n \lambda_0 / 2$. 
Collective enhancement of the magnon-photon coupling can then be achieved by increasing the number of mirror YIG spheres (see Fig.~1 in the main text).
In the anti-Bragg cavity configuration, the collective supermodes possess different eigenfrequencies and exhibit both coherent and dissipative coupling with the PM. 
In this regime, the coherent coupling does not experience collective enhancement. Instead, the dissipative coupling between the bright supermode and the PM leads to level attraction. 
Experimentally, we observe that the avoided-crossing gap decreases as more mirror YIG spheres are added to the anti-Bragg cavity (Fig.~2 in the main text).

Furthermore, the magnonic cavity in our experiments arises from the existence of BICs in the transverse waveguide. 
By moving the probe sphere together with the longitudinal waveguide, we can flexibly probe a variety of light-matter-interaction phenomena, including the detection of BICs (Fig.~4 in the main text).
Additionally, when the mirror YIG spheres form a spatially asymmetric configuration (different numbers of YIG spheres in the two mirrors) around the probe in the Bragg cavity, we observe a slight reduction in the level repulsion [Fig.~1(g) in the main text]. 
In the anti-Bragg cavity, at specific probe positions, we find that detuning the mirror magnon modes enables switching the distant coherent coupling on and off in the frequency domain, as the field distribution of the magnonic cavity is modified.
This interaction is described by Eqs.~(\ref{eq:final_master})--(\ref{eq_Heff}).

The ability to manipulate and detect collective magnon-photon interactions in the dual-waveguide configuration provides a versatile and scalable approach for engineering reconfigurable magnonic networks. 
Moreover, the scanning-probe framework enables the exploration of various exotic states~\cite{Liu2017, Atala2013}, e.g., emitter-photon bound states~\cite{Liu2017}.


\normalem
\bibliography{References_SM}